\documentclass[aps,showpacs,pre,amsmath,amsfonts,amssymb,superscriptaddress,nofootinbib,notitlepage,a4paper]{revtex4}

\usepackage{cancel}

\usepackage{graphicx}
\usepackage{psfrag}
\usepackage[naturalnames]{hyperref}

\newcommand{\beq}{\begin{equation}} 
\newcommand{\eeq}{\end{equation}}
\newcommand{\bem}{\begin{multline}}
\newcommand{\bes}{\begin{split}} \newcommand{\ees}{\end{split}} 
\newcommand{\bea}{\begin{eqnarray}} \newcommand{\eea}{\end{eqnarray}}

\def\s{{\sigma}}
\def\us{{\underline{\sigma}}}
\def\ut{{\underline{t}}}
\def\di{{\partial i}}
\def\dimj{{\partial i \setminus j}}
\def\tmin{{\theta_{\rm min}}}
\def\tminz{{\theta_{\rm min,0}}}
\def\tmino{{\theta_{\rm min,1}}}
\def\tr{{\theta_{\rm r}}}
\def\tc{{\theta_{\rm c}}}
\def\td{{\theta_{\rm d}}}
\def\muc{{\mu_{\rm c}}}
\def\mud{{\mu_{\rm d}}}
\def\ms{{m_{\rm s}}}
\def\ys{{y_{\rm s}}}
\def\ind{{\mathbb{I}}}
\def\lmin{{\underset{l}{\min}}}
\def\lmunmin{{\underset{l-1}{\min}}}

\def\ziter{{z_{\rm iter}}}
\def\zsite{{z_{\rm site}}}
\def\zedge{{z_{\rm edge}}}
\def\Fsite{{F_{\rm site}}}
\def\Fedge{{F_{\rm edge}}}
\def\Ziter{{{\cal Z}_{\rm iter}}}
\def\Zsite{{{\cal Z}_{\rm site}}}
\def\Zedge{{{\cal Z}_{\rm edge}}}
\def\hlambda{{\widehat{\lambda}}}

\def\hu{{\widehat{u}}}
\def\hv{{\widehat{v}}}
\def\hg{{\widehat{g}}}

\def\hP{{\widehat{P}}}
\def\hG{{\widehat{G}}}
\def\hziter{{\widehat{z}_{\rm iter}}}
\def\hzsite{{\widehat{z}_{\rm site}}}
\def\hzedge{{\widehat{z}_{\rm edge}}}
\def\hZiter{{\widehat{\cal Z}_{\rm iter}}}
\def\hZsite{{\widehat{\cal Z}_{\rm site}}}
\def\hZedge{{\widehat{\cal Z}_{\rm edge}}}
\def\etasite{{\eta_{\rm site}}}
\def\hetasite{{\widehat{\eta}_{\rm site}}}
\def\dd{{\rm d}}
\def\la{{\langle}}
\def\ra{{\rangle}}

\def\P{{\cal P}}
\def\S{{\cal S}}
\def\tlambda{{\widetilde{\lambda}}}
\def\tx{{\widetilde{x}}}
\def\txr{{\widetilde{x}_{\rm r}}}
\def\tv{{\widetilde{v}}}
\def\tu{{\widetilde{u}}}
\def\tp{{\widetilde{p}}}
\def\tq{{\widetilde{q}}}

\begin{document}

\title{Minimal contagious sets in random regular graphs}

\author{Alberto Guggiola}
\author{Guilhem Semerjian}
\affiliation{LPTENS, Unit\'e Mixte de Recherche (UMR 8549) du CNRS et
 de l'ENS, associ\'ee \`a l'UPMC Univ Paris 06, 24 Rue Lhomond, 75231
 Paris Cedex 05, France.}

\begin{abstract}
The bootstrap percolation (or threshold model) is a dynamic process modelling the propagation of an epidemic on a graph, where inactive vertices become active if their number of active neighbours reach some threshold. We study an optimization problem related to it, namely the determination of the minimal number of active sites in an initial configuration that leads to the activation of the whole graph under this dynamics, with and without a constraint on the time needed for the complete activation. This problem encompasses in special cases many extremal characteristics of graphs like their independence, decycling or domination number, and can also be seen as a packing problem of repulsive particles. We use the cavity method (including the effects of replica symmetry breaking), an heuristic technique of statistical mechanics many predictions of which have been confirmed rigorously in the recent years. We have obtained in this way several quantitative conjectures on the size of minimal contagious sets in large random regular graphs, the most striking being that 5-regular random graph with a threshold of activation of 3 (resp. 6-regular with threshold 4) have contagious sets containing a fraction $1/6$ (resp. $1/4$) of the total number of vertices. Equivalently these numbers are the minimal fraction of vertices that have to be removed from a 5-regular (resp. 6-regular) random graph to destroy its 3-core. We also investigated Survey Propagation like algorithmic procedures for solving this optimization problem on single instances of random regular graphs.

\end{abstract}

\maketitle

\section{Introduction}

Models of epidemic spreadings as dynamical processes occurring on a graph appear in various contexts besides epidemiology~\cite{Hethcote_review,Dorogovtsev_review,Newman_review,Boccaletti_review,Barrat_book}; for instance social sciences study viral marketing campaigns aimed at propagating new social trends, and in economy it is crucial to understand cascading effects potentially leading to the bankrupt of financial institutions. In these models individual agents are located on the vertices of a graph, and their state (healthy or contaminated for instance) evolve in time according to the state of their neighbours, the edges of the graph representing the contacts between agents that can possibly transmit the illness from one contaminated agent to an healthy one.

There is a great diversity in the details of these models: the dynamics can occur in continuous (asynchronous) or discrete time, according to deterministic or random rules, the state of an agent can be boolean (healthy or contaminated) or describe several levels of contamination, and finally the dynamics can be monotonous or not. To precise this last point, a dynamics is said monotonous if the states of an agent always occur in the same order in time, for instance in the Susceptible-Infected-Recovered (SIR) model the only allowed transitions are S$\to$I and I$\to$R, a Recovered individual being immune forever, whereas in the SIS model an agent can become infected several times in a row. In this paper we will concentrate on a simple monotonous dynamics, that evolve deterministically in discrete time, with inactive (Susceptible) variables becoming active (Infected) when their number of active neighbours reach some threshold, and then remain active for ever. For this reason it is called the threshold model, see~\cite{thr_socio} for a version introduced in sociology with an underlying complete graph, and~\cite{bootstrap} for its first appearance in physics under the name of bootstrap percolation (on random regular graphs).

Given one specific dynamical model there are many different questions that can be asked. The first, a priori simplest, issue concerns the time evolution of the system from a random initial condition, taking the initial state of each agent as an independent random variable. For monotonous dynamics a stationary state is reached after some time, and one can wonder whether the epidemic has invaded the whole graph (in other words whether it percolates) in this final state. The probability of this event obviously depends on the fraction of infected vertices in the initial condition, and this may lead to phase transitions for certain class of graphs; see~\cite{AiLe88,Ho03,BaBoetal12} for such a study of the bootstrap percolation on finite-dimensional lattices, and~\cite{bootstrap,BaPi07,KaNe10,Lelarge2012,janson2012,ShMo14,SIR_Bohman,SIR_Janson} for various type of dynamics on random graphs. In particular one finds for the bootstrap percolation on random regular graphs a phase transition at some initial critical density $\tr$ (dependent on the degree of the graph and the threshold of activation): with high probability initial conditions with a fraction $\theta$ of active vertices (without correlations between the sites) are percolating if and only if $\theta > \tr$.

Besides these studies of the ``forward'' (or ``direct'') time evolution, which are somehow simplified by the independence assumption for the initial state variables, one can also formulate more difficult inference and optimization questions. An example of the former type is to infer some information on the initial state given a snapshot of the epidemic after some time evolution~\cite{inference_Shah,inference_Pinto,inference_Torino,inference_Orsay}; this ``inverse problem'' is particularly relevant in epidemiology in the search of the ``zero patient'' who triggered the spreading of an illness. For what regards the latter type of questions, the design of an efficient vaccination campaign can indeed be seen as an optimization problem: find the smallest set of nodes (to minimize the economical and social cost) whose vaccination will prevent the epidemic to reach a given fraction of the population~\cite{vaccination_Torino}. We shall actually consider in this paper the somehow reverse optimization problem, namely targeting a small set of initially active sites that lead to the largest possible propagation of the contagion. This obviously makes more sense in the perspective of viral marketing, in which it was first considered~\cite{Kempe} than in the epidemiological one; the initial adopters of a new product, that can be financially incited to do so, are expected to convince most of their acquaintances and progressively the largest possible part of the population. From this point of view the additional constraint that the propagation should be as fast as possible is also a relevant one.

More precisely, one can define two versions of this optimization problem: (i) given a fixed number of initially active agents, choose them in order to maximize the number of active agents at some fixed later time, or in the final state of the propagation; (ii) find the minimal number of initially active agents such that all the agents are active, again after some time or in the final state. We will concentrate on the latter version of the problem but part of our analysis applies to both. These optimization problems are known to be hard from a (worst-case) computational complexity point of view~\cite{Kempe,Chen,Dreyer09}, even to approximate. Exhibiting minimal percolating sets for bootstrap percolation on finite dimensional lattices is relatively easy thanks to their regular structures, but more refined extremal problems are also relevant in this case, see for instance~\cite{Morris09,maxtime2d}. The understanding of these optimization problems seems less advanced in the case of sparse random graphs. There exist upper and lower bounds on the size of minimal contagious sets~\cite{Dreyer09,Ackerman10,bounds12}, some based in particular on the expansion properties of such graphs~\cite{bounds_Amin}. One particular case of the optimization problem (when the threshold of activation is equal to the degree of the vertex minus one) is actually equivalent to the decycling number problem of graph theory~\cite{decycling_Beineke} (also known as minimal Feedback Vertex Set), which was settled rigorously for 3-regular random graphs in~\cite{decycling} (this paper also contains bounds for higher degrees). As this last point unveils the notion of minimal contagious sets is connected in some special cases to many other problems in graph theory; one way to see this connection is to picture the inactive sites of the initial condition as particles to be put on the graph. One wants to pack as many as possible of them (to obtain a contagious set of minimal size), yet they do have some kind of repulsive interactions because of the constraint of complete percolation at a later time. This is particularly clear when the threshold of activation is equal to the degree for all vertices: the problem is then exactly equivalent to the hard-core particle model, also known as independent set or vertex cover.

The strategy we shall follow to determine the minimal size of contagious sets of sparse random graphs will be the same as in~\cite{Torino1,Torino2}, namely a reformulation under the form of a statistical mechanics model which can be treated with the so-called cavity method~\cite{cavity,cavity_T0,MeZe,MeMo_book}. This (heuristic) method yields predictions for any interacting model defined on a sparse random graph; its use in the context of random constraint satisfaction problems led to the discovery of a very rich phenomenology of phase transitions~\cite{MeZe,KrMoRiSeZd}, with many of these predictions later confirmed rigorously~\cite{MoraMezard05b,clus_rig_Fede,AC08,molloy_col_freezing,amin_ksat,amin_cond_col}. Let us emphasise in particular the determination of the maximal size of independent sets of random regular graphs (which as we saw is a problem related to the present one), for which the predictions of the cavity method (see~\cite{is_japan} and references therein) have been recently rigorously confirmed (for graphs of large enough but finite degree) in~\cite{is_Sly}. Another example in the context of graph theory is the study of matchings in random graphs, where the cavity method~\cite{ZdMe_matchings} has also been proved to be correct~\cite{BoLeSa_matchings}. The main originality of our contribution with respect to~\cite{Torino1,Torino2} is the use of a more refined version of the cavity method (i.e. incorporating the effects of replica symmetry breaking), and an analytical study of the limit where the time at which the complete activation is required is sent to infinity.

The rest of the article is organized as follows.
In Sec.~\ref{sec_def_and_res} we define precisely the dynamics under study, recall briefly some known results for random initial conditions, formulate the optimization problem and propose various interpretations of it, and for the convenience of the reader we summarize the main results to be obtained in the following. 
In Sec.~\ref{sec_cavity} we derive the cavity method equations, both at the replica symmetric and one step of replica symmetry breaking level. 
The solution of these equations for random regular graphs is presented in Sec.~\ref{sec_results_rrg}, which contains the main analytical results of this work.
Sec.~\ref{sec_single_sample} is devoted to single sample numerical experiments, where we confront the analytical predictions with the optimized initial configurations obtained with two kind of algorithms (a simple greedy one and a more involved procedure based on message passing).
We finally draw our conclusions and present perspectives for future work in Sec.~\ref{sec_conclu}. The most technical parts of the computations are deferred to two Appendices.

\section{Definitions and main results}
\label{sec_def_and_res}

\subsection{Definition of the dynamics}
\label{sec_def_dyn}

Let us consider a graph on $N$ vertices (or sites), $G=(V,E)$, with the vertices labelled as $V=\{1,\dots,N\}$, and the number of edges denoted $|E|=M$. The dynamical process under study concerns the evolution of variables $\s_i^t$ on the vertices, $\s_i^t=0$ (resp. $1$) if the vertex $i$ is inactive (resp. active) at time $t$. We shall denote $\us^t=(\s_1^t,\dots,\s_N^t)$ the global configuration at time $t$. The latter is determined by the initial condition $\us$ at the initial time, $\us^0=\us$, and then evolves subsequently in a deterministic and parallel way, in discrete time, according to the rules:
\beq
\s_i^{t} = \begin{cases} 1 & \text{if} \ \s_i^{t-1}=1 \\
1 & \text{if} \ \s_i^{t-1}=0 \ \text{and} \ \underset{j \in \di}{\sum} \s_j^{t-1} \ge l_i \\
0 & \text{otherwise} \end{cases} \ ,
\label{eq_dynrules}
\eeq
where $\di$ is the set of neighbours of $i$ on the graph, and $l_i$ is a fixed threshold for each vertex; we will also use $d_i = |\di|$ to denote the degree of vertex $i$. The dynamics is monotonous (irreversible), an active site remaining active at all later times, an inactive site $i$ becoming active if its number of active neighbours at the previous time crosses the threshold $l_i$. Note that the configuration $\us^t$ at time $t$ is a deterministic function of the initial condition $\us=\us^0$, and that by monotonicity one can define the final configuration $\us^{\rm f}=\underset{t \to \infty}{\lim} \us^t$, this stationary configuration being reached in a finite number of steps for all finite graphs.

It turns out that the final configuration $\us^{\rm f}$ is also the one reached by a sequential dynamics in which at each time step only one site $i$ with at least $l_i$ active neighbours is activated; a moment of thought reveals the independence of the final configuration with respect to the order of the updates. $\us^{\rm f}$ is indeed the smallest configuration (considering the partial order $\us \le \us'$ if and only if $\s_i \le \s'_i$ for all vertices) larger than the initial condition $\us$, such that no further site can be activated. It will sometimes be useful in the following to think of this process in a dual way, corresponding to the original presentation of bootstrap percolation in~\cite{bootstrap}, namely to consider that inactive sites are sequentially removed if they have less than a certain number of inactive neighbours. An equivalent definition of $\us^{\rm f}$ is thus given by the inactive sites it contains, that form the largest set (with respect to the inclusion partial order) contained in the set of inactive sites of $\us$, and such that in their induced graph the degree of site $i$ is larger or equal than $d_i-l_i+1$; they form thus a (generalized inhomogeneous version of the) core of the initially inactive sites.

\subsection{Reminder of the behaviour for random initial conditions on random regular graphs}
\label{sec_reminder_random}
To put in perspective the optimization problem to be studied in this paper it is instructive to first recall briefly some well-known results for the evolution from a random initial configuration~\cite{bootstrap,BaPi07}. For the sake of simplicity let us consider $G$ to be a $k+1$-random regular graph (i.e. a graph drawn uniformly at random among all graphs in which every vertex has degree $k+1$), with a uniform threshold for activation set to $l_i=l$ for all vertices. Suppose that the states of the vertices in the initial condition are chosen randomly, independently and identically for each vertex, with a probability $\theta$ (resp. $1-\theta$) for a vertex to be active (resp. inactive). The probability for one vertex $i_0$ to be active at some time $t+1$, denoted $x_{t+1}$, can be computed from the following equation:
\beq
x_{t+1} = \theta + (1-\theta) \sum_{p=l}^{k+1} \binom{k+1}{p} \tx_t^p (1-\tx_t)^{k+1-p} \ .
\label{eq_random_x}
\eeq
Indeed such a vertex was either active in the initial condition, or has seen at least $l$ of its neighbours activate themselves before time $t$, and without the participation of $i_0$. The probability $\tx_t$ of this last event obeys the recursive equation
\beq
\tx_{t+1} = \theta + (1-\theta) \sum_{p=l}^k \binom{k}{p} \tx_t^p (1-\tx_t)^{k-p} \ ,
\label{eq_random_tx}
\eeq
with a number of participating neighbours reduced from $k+1$ to $k$ as $i_0$ has to be supposed inactive here. The initial condition for these equations is $x_0=\tx_0=\theta$. In the limit $t \to \infty$ of large times $\tx_t \to \tx_\infty(\theta)$, the smallest fixed-point in $[0,1]$ of the recursion (\ref{eq_random_tx}). For each $k\ge 2$ and $l$ with $2\le l \le k$ there exists a threshold $\tr(k,l)$ such that $\tx_\infty(\theta)$ is equal to 1 for $\theta > \tr$, strictly smaller than 1 for $\theta<\tr$. From Eq.~(\ref{eq_random_x}) one realizes that the same statement applies to $x_\infty(\theta)$, hence $\tr$ is the threshold for complete activation (percolation) from a Bernouilli random initial condition with probability $\theta$ for each active site. Studying more precisely Eq.~(\ref{eq_random_tx}) one realizes that for $l=k$ the transition is continuous ($x_\infty(\theta_{\rm r}^-)=1$), with an explicit expression for the threshold, $\tr(k,k)=\frac{k-1}{k}$. For $2\le l \le k-1$ the transition is discontinuous ($x_\infty(\theta_{\rm r}^-)<1$), the threshold $\tr$ is obtained as the solution of the equations:
\beq
\begin{cases}
\txr= \tr + (1-\tr) \overset{k}{\underset{p=l}{\sum}} \binom{k}{p} \txr^p (1-\txr)^{k-p} \\
1 = (1-\tr) l \binom{k}{l} \txr^{l-1} (1-\txr)^{k-l}
\end{cases} \ ,
\label{eq_tandx_r}
\eeq
where $\txr=\tx_\infty(\theta_{\rm r}^-)$ is the value of the fixed-point of (\ref{eq_random_tx}) at the bifurcation where it disappears discontinuously. For $l=2$ these equations can be solved explicitly and yield
\beq
\tr(k,l=2) = 1- \frac{(k-1)^{2k-3}}{k^{k-1}(k-2)^{k-2}}\ .
\eeq
For generic values of the parameters $k,l$ there is no explicit expression of $\tr$, as (\ref{eq_tandx_r}) are algebraic equations of arbitrary degree; some numerical values of $\tr$ will be given in Table \ref{table_res_Tinfty}. For a given value of $k$ the threshold $\tr(k,l)$ is growing with $l$: if an initial condition leads to complete activation for some parameter $l$ it will also be activating under the less constrained dynamics with $l'<l$.

The relevant range for the threshold parameter $l$ in this study of random initial conditions is $2\le l \le k$. Indeed for $l=0$ after one step the configuration is completely active regardless of $\us^0$, for $l=1$ a single active site (per connected component) in the initial configuration is enough to activate the whole graph, hence in these two cases $\tr=0$. On the other hand if $l=k+1$ one has $\tr=1$: any pair of adjacent inactive sites in the initial condition will remain inactive for ever, and the number of such pairs is linear in $N$ as soon as $\theta<1$.

Note that the recursion equations (\ref{eq_random_x},\ref{eq_random_tx}) are exact if the neighbourhood up to distance $t$ of the vertex $i_0$ is a regular tree of degree $k+1$. The limit $t\to\infty$ can be taken in this way only if the graph considered is an infinite regular tree. A rigorous proof that this reasoning is in fact correct also for the large size limit of random regular graphs (that converge locally to regular trees) can be found in~\cite{BaPi07}.

\subsection{Definition of the optimization problem over initial conditions}
\label{sec_def_optimization}

Let us now come back to a general graph $G$ with some thresholds $l_i$ for vertex activation, and consider the minimal fraction of active vertices in an initial configuration that activates the whole graph, i.e.
\beq
\tmin(G,\{l_i\}) = \frac{1}{N}\min_\us \left\{ \sum_{i=1}^N \s_i \ | \ \s_i^{\rm f} = 1 \ \forall i \right\} \ .
\eeq
This corresponds to the minimal size of a contagious (or percolating) set, divided by the total number of vertices. Following~\cite{Torino1,Torino2} it will turn out useful to introduce another parameter $T$ (a positive integer) in this optimization problem, and impose now that the fully active configuration is reached within this time horizon $T$:
\beq
\tmin(G,\{l_i\},T) = \frac{1}{N}\min_\us \left\{ \sum_{i=1}^N \s_i \ | \ \s_i^T = 1 \ \forall i \right\} \ .
\eeq
Obviously for any finite graph $\tmin(G,\{l_i\},T)$ decreases when $T$ increases and has $\tmin(G,\{l_i\})$ as its limit for $T\to\infty$. To turn the computation of $\tmin$ into a form more reminiscent of statistical mechanics problems we shall introduce a probability measure over initial configurations:
\beq
\eta(\us) = \frac{1}{Z(G,\{l_i\},T,\mu,\epsilon)} 
e^{\underset{i=1}{\overset{N}{\sum}}[\mu \s_i - \epsilon (1-\s_i^T)]}
\ ,
\label{eq_eta_us}
\eeq
where $\us^T$ is as above the configuration obtained after $T$ steps of the dynamics starting from the configuration $\us=\us^0$, the $\mu$ and $\epsilon$ are for the time being arbitrary parameters, and the partition function $Z$ ensures the normalization of this law. The parameter $\mu$ is a ``chemical potential'' that controls the fraction of initially active vertices (if $\epsilon=0$ the measure $\eta$ reduces to the Bernouilli measure), while $\epsilon$ is the cost to be paid for each site $i$ inactive at the final time $T$. In particular if $\epsilon=+\infty$ one has
\beq
\eta(\us) = \frac{1}{Z(G,\{l_i\},T,\mu,\epsilon=+\infty)} 
e^{\mu \underset{i=1}{\overset{N}{\sum}} \s_i} \prod_{i=1}^N \ind(\s_i^T=1)
\ ,
\eeq
with $\ind(A)$ is the indicator function of the event $A$, the measure is thus supported by activating initial configurations (within the time horizon $T$). It is then obvious that the knowledge of $Z$ allows to deduce the sought-for minimal density $\tmin$, as
\beq
\tmin(G,\{l_i\},T) = \lim_{\mu \to -\infty}\frac{1}{\mu} \frac{1}{N} \ln Z(G,\{l_i\},T,\mu,\epsilon=+\infty) \ .
\eeq
Actually one can gain more information from the whole dependency of the partition function on $\mu$. Suppose indeed that the number of initial configurations with a fraction $\theta$ of active vertices that activate the whole graph in $T$ steps is, at the leading exponential order, $e^{N s(\theta)}$, with an entropy density $s(\theta)$ of order one with respect to $N$. Then this entropy density can be computed, in the large $N$ limit, as a Legendre transform of the logarithm of the partition function. More precisely, defining the free-entropy density $\phi$ as
\beq
\phi(G,\{l_i\},T,\mu,\epsilon=+\infty) = \frac{1}{N} \ln Z(G,\{l_i\},T,\mu,\epsilon=+\infty) \ ,
\eeq
the evaluation of the sum over configurations in the definition of $Z$ via the Laplace method yields in the large $N$ limit:
\beq
\phi(G,\{l_i\},T,\mu,\epsilon=+\infty) = \sup_{\theta \in [\tmin,1]} \,[ \mu \, \theta + s(\theta)] \ ,
\eeq
hence $s(\theta)$ can be obtained by an inverse Legendre transform of $\phi(\mu)$, with $s(\theta) = \phi(\mu) - \mu \, \theta$ and $\theta = \phi'(\mu)$.

For completeness let us also make a similar statement when $\epsilon$ is finite, i.e. when one does not impose strictly the constraint of complete activation at time $T$. Denoting $s(\theta,\theta')$ the entropy density of initial configurations that have a fraction $\theta$ of initially active vertices and that lead after $T$ steps of evolution to a configuration with a fraction $\theta'$ of active sites, one has
\beq
\phi(G,\{l_i\},T,\mu,\epsilon) = \frac{1}{N} \ln Z(G,\{l_i\},T,\mu,\epsilon)= \underset{\theta,\theta'}{\sup}\,[ \mu \, \theta - \epsilon \, (1-\theta') + s(\theta,\theta') ] \ .
\label{eq_Legendre_both}
\eeq
Varying the parameters $\mu$ and $\epsilon$ thus allows to reconstruct the function $s(\theta,\theta')$, and hence to solve the optimization problem denoted (i) in the introduction, namely for a fixed value of $\theta$ find the maximal reachable $\theta'$. We will mainly concentrate in the following of the paper on the optimization problem denoted (ii) in the introduction, that is imposing the full activation of the graph at time $T$ ($\theta'=1$), which as explained above can be studied via the computation of $s(\theta)=s(\theta,\theta'=1)$ from the inverse Legendre transform of the free-entropy with $\epsilon=+\infty$.

The definitions above were valid for any graph and any choice of the activation thresholds; we shall however be particularly interested in the case of large random regular graphs with uniform thresholds, we thus define
\beq
\tmin(k,l) = \lim_{N \to \infty} \mathbb{E}[\tmin(G,\{l_i=l\}) ] \ , \qquad
\tmin(k,l,T) = \lim_{N \to \infty} \mathbb{E}[\tmin(G,\{l_i=l\},T) ] \ ,
\eeq
where the average is over uniformly chosen regular graphs of degree $k+1$ on $N$ vertices, with the same threshold for activation $l$ on every vertex. The fact that the limit in the definition of $\tmin(k,l,T)$ exists could actually be proven rigorously using the method developed in~\cite{Gamarnik_interpolation}, and it is expected that $\tmin(G,\{l_i=l\},T)$ is self-averaging (i.e. concentrates around its average in the large $N$ limit). The existence of $\tmin(k,l)$ might be a more difficult mathematical problem that we shall not discuss further; it is a reasonable conjecture that it coincides with the limit of $\tmin(k,l,T)$ when $T\to\infty$, i.e. that the large size and large time limits commute. We will see in Sec.~\ref{sec_kequall_Tinfty} one argument in favour of this conjecture. Let us emphasize that $\tmin(k,l) < \tr(k,l)$, with a strict inequality. This is indeed a large-deviation phenomenon: even if most initial configurations with density smaller than $\tr$ do not activate the whole graph some very rare ones (with a probability exponentially small in $N$ in the Bernouilli measure of parameter $\theta<\tr$) are able to do so. Note also that $\tmin(k,l)$ is growing with $l$ at fixed $k$, for the same reasons as explained above in the discussion of $\tr$. The computations of $\tmin$ we shall present will follow the strategy explained above on an arbitrary graph, namely the computation of a free-entropy density, that we define in the case of random regular graphs as the quenched average over the graph ensemble,
\beq
\phi(k,l,T,\mu,\epsilon) = \lim_{N \to \infty} \frac{1}{N} \mathbb{E}[\ln Z(G,\{l_i=l\},T,\mu,\epsilon)] \ .
\label{eq_def_phi}
\eeq

\subsection{Equivalence with other problems and bounds}
\label{sec_connections}

As mentioned in the introduction the problem of minimal contagious sets can be related, for appropriate choices of the threshold parameters $l_i$, to other standard problems in graph theory.

Consider first the case of an arbitrary graph where the thresholds $l_i$ are equal to the degrees $d_i$ for all vertices. An inactive site in the initial configuration will be activated only if it is surrounded by active vertices, and it will do so in a single step. In other words in any percolating initial condition, whatever the time horizon $T$, the inactive vertices must form an independent set (no two inactive vertices are allowed to be neighbours). For regular random graphs one has thus $\tmin(k,k+1,T)=\tmin(k,k+1)$ for all $T$, and this quantity is equal to 1 minus the density of the largest independent sets of a $k+1$-regular random graph.

Another correspondance with previously studied models arises when $T=1$, for any choice of the thresholds $l_i$. Indeed in this case the vertex $i$ can be inactive in a percolating initial configuration only if its number of inactive neighbours is smaller than some value (namely, $\le d_i-l_i$). These generalized hard-core constraints (repulsion between inactive vertices) correspond exactly to the so-called Biroli-M\'ezard (BM) model~\cite{bm_prl,bm} (with the correspondance inactive vertex $\leftrightarrow$ vertex occupied by a particle in the BM model, and $d_i-l_i \leftrightarrow \ell_i$ of the BM model). Hence for $T=1$ the minimal density $\tmin$ is 1 minus the density of a close packing of the corresponding BM model. Further specializing this $T=1$ case by setting $l_i=1$ on each vertex leads to the constraint that every inactive site in a percolating initial configuration has to be adjacent with at least one active site, in other words that the active sites form a dominating set of $G$. The minimal density $\tmin$ is thus the domination number (divided by $N$) of $G$.

Consider now the thresholds of activation to be 1 less than the degrees, i.e. $l_i=d_i-1$ on all vertices, with no constraint on the time of activation ($T=\infty$). As explained at the end of Sec.~\ref{sec_def_dyn}, the inactive vertices in the final configuration form the 2-core of the inactive ones in the initial configuration. A percolating initial configuration must be such that this 2-core is empty, in other words the subgraph induced by the inactive sites of the initial configuration must be acyclic (a tree or a forest), i.e. the active sites have to form a decycling set~\cite{decycling_Beineke} (also known as a Feedback Vertex Set), and $N \tmin$ is the decycling number of $G$. This characterization leads to the following bound for every $k+1$-regular graph with thresholds $k$ of activation on every site,
\beq
\tmin(k,k) \ge \frac{k-1}{2 k} \ .
\label{eq_bound_kk}
\eeq
Indeed if $A$ denotes the number of active vertices in a percolating initial configuration, the $N-A$ other vertices induces a forest, the number of edges between inactive vertices is thus at most $N-A-1$. On the other hand this number is at least $\frac{k+1}{2} N - (k+1)A$ (the first term being the total number of edges, and the number of edges incident to at least one active site being at most $(k+1)A$). The decycling number of random regular graphs was studied in~\cite{decycling}, proving in particular that the bound (\ref{eq_bound_kk}) is actually tight for 3-regular large random graphs, i.e. $\tmin(2,2)=1/4$, and it was conjectured to be also the case for 4-regular ones (i.e. $\tmin(3,3)=1/3$). An asymptotic lowerbound on $\tmin(k,k)$ for large values of $k$ was worked out in~\cite{lb_klequal_rig} , we will come back on this result in Sec.~\ref{sec_kequall_Tinfty}. Note also that the decycling number of arbitrary sparse random graphs was studied with physics methods in~\cite{fvs_Zhou1,fvs_Zhou2}.

For general thresholds smaller than the degrees minus one the active sites of a percolating initial configuration must form a ``de-coring'' set instead of a ``de-cycling'' set (i.e. their removal has to provoke the disappearance of a $q$-core with $q>2$). A generalization of the lower bound (\ref{eq_bound_kk}) to any $k+1$-regular graph with uniform threshold $l$ was given in~\cite{Dreyer09}, and reads
\beq
\tmin(k,l) \ge \frac{2l-k-1}{2 l} \ .
\label{eq_bound_kl}
\eeq
Its proof goes as follows. Consider the sequential process explained at the end of Sec.~\ref{sec_def_dyn} in which at each time $t$ a single vertex gets activated, and denote $E(t)$ the number of edges between active and inactive vertices after $t$ steps of this process. By definition of the activation rule $E(t+1)-E(t) \le k+1-2l$. If as above $A$ denotes the number of active sites in a percolating initial configuration, by definition $E(N-A)=0$, hence $E(0)\ge (N-A) (2l-k-1)$. On the other hand $E(0) \le (k+1) A$, which gives the lower bound (\ref{eq_bound_kl}) on the possible values of $A$.

We should also mention an upper bound on the minimal sizes of contagious sets derived in~\cite{Ackerman10,bounds12} for graphs of arbitrary degree distributions, which yields in the case of $k+1$-regular graphs:
\beq
\tmin(k,l) \le \frac{l}{k+2} \ .
\eeq

To conclude this discussion let us mention that the ``de-coring'' perspective on the minimal contagious set problem is somehow reminiscent (even if not directly equivalent), to the Achlioptas processes~\cite{Achli_proc09,riordan2012} (more precisely of their offline version~\cite{Achli_offline}) where one looks for an extremal event avoiding the appearance of an otherwise typical structure (a giant component in the Achlioptas processes, a core in the minimal contagious set case).

\subsection{Main analytical results}

Let us draw here a more detailed plan of the rest of the paper to make its reading easier and faster for someone not interested in the technical details of the statistical mechanics method (who can browse quickly over the next section and jump to the results announced in Sec.~\ref{sec_results_rrg}). In order to compute the minimal density $\tmin$ of contagious sets we shall rephrase this problem as a statistical mechanics model and apply to it the cavity method. The latter is based on self-consistent assumptions of various degrees of sophistication, parametrized by the so-called level of replica symmetry breaking. We will study the first two levels of this hierarchy, named replica symmetric (RS) and one step of replica symmetry breaking (1RSB). These two approaches will lead to two predictions for $\tmin$, to be denoted respectively $\tminz(k,l,T)$ and $\tmino(k,l,T)$. From general bounds established in the context of disordered statistical mechanics models (first for the Sherrington-Kirkpatrick model~\cite{Gu03,Ta06,book_Panchenko} and later for some models on sparse random graphs~\cite{FrLe,FranzLeone03b,PaTa}) it is expected that the different levels of the cavity method provide improving lower bounds on $\tmin$, namely
\beq
\tminz(k,l,T) \le \tmino(k,l,T) \le \tmin(k,l,T) \ .
\eeq
Our computation of $\tminz(k,l,T)$ and $\tmino(k,l,T)$ relies on the resolution of a set of roughly $2T$ algebraic equations on $2T$ unknowns, explicit numbers will be given in Sec.~\ref{sec_results_rrg}. We managed to perform analytically the $T\to\infty$ limit and reduce the determination of $\tminz(k,l)$ and $\tmino(k,l)$ (their limit when $T\to\infty$) to a finite number of equations, that will also be presented along with numerical results in Sec.~\ref{sec_results_rrg}. We found four particular cases in which the predictions of the first two levels of replica symmetry breaking coincide when $T\to\infty$, which led us to conjecture that they are the exact ones, namely:
\beq
\tmin(2,2)=\frac{1}{4} \ , \qquad
\tmin(3,3)=\frac{1}{3} \ , \qquad
\tmin(4,3)=\frac{1}{6} \ , \qquad
\tmin(5,4)=\frac{1}{4} \ ,
\eeq
all these cases saturating the lower bounds of (\ref{eq_bound_kk},\ref{eq_bound_kl}). The first (resp. second) equality was actually proven (resp. conjectured) in~\cite{decycling}. We have also performed a large degree expansion of the decycling number of random regular graphs, yielding the conjecture
\beq
\tmin(k,k) = 1- \frac{2 \ln k}{k} - \frac{2}{k} + O\left(\frac{1}{k \ln k} \right) \ .
\eeq

\section{Cavity method treatment of the problem}
\label{sec_cavity}

\subsection{Factor graph representation}

As explained in Sec.~\ref{sec_def_optimization} the central quantity to compute is the free-entropy density defined from the partition function normalizing the probability law (\ref{eq_eta_us}), that for completeness we shall generalize to possibly site dependent chemical potentials $\mu_i$ and costs for non-activation $\epsilon_i$:
\beq
\eta(\us) = \frac{1}{Z(G,\{l_i\},T,\{\mu_i,\epsilon_i\})} 
e^{\, \underset{i=1}{\overset{N}{\sum}}[\mu_i \s_i - \epsilon_i (1-\s_i^T)]}
\ .
\label{eq_eta_us_generalized}
\eeq
This expression is not very convenient to handle directly because the variables $\s_i$ have complicated interactions under this law: $\s_i^T$ is indeed a function of all variables $\s_j$ on the vertices $j$ at distance smaller than $T$ from $i$. A way to circumvent this difficulty and to turn the interactions of the model into local ones has been proposed in~\cite{Torino1,Torino2}, and we shall follow the same approach here.

Let us first define $t_i(\us)$ as the time of activation of site $i$ in the dynamical process generated by the initial configuration $\us$, i.e. $t_i(\us)=\min \{ t : \s_i^t=1 \}$, with conventionally $t_i(\us)=\infty$ if this time is strictly greater than the time horizon $T$. These variables obey the following equations:
\beq
t_i(\us) = f(\s_i,\{t_j(\us)\}_{j \in \di};l_i) \ \ \ \forall \, i \in V \ ,
\eeq
where the function $f$ is defined as
\beq
f(\s,t_1,\dots,t_n;l) = \begin{cases} 0 & \text{if} \ \s=1 \\
1+\lmin(t_1,\dots,t_n) & \text{if} \ \s=0 \ \text{and} \ 1+\lmin(t_1,\dots,t_n) \le T \\
\infty & \text{otherwise}
\end{cases} \ .
\eeq
Here $\lmin(t_1,\dots,t_n)$ is the $l$-th smallest $t_i$, i.e ordering the arguments as $t_1 \le t_2 \le \dots \le t_n$ one has $\lmin(t_1,\dots,t_n)=t_l$. This translates the dynamic rules (\ref{eq_dynrules}) in terms of the activation times, a site $i$ activating at the time following the first time where at least $l_i$ of its neighbours are active. In the following $f(0,t_1,\dots,t_n;l)$ will be abbreviated in $f(t_1,\dots,t_n;l)$. Reciprocally one can show that if a set of $\{t_i\}_{i\in V}$ verifies the condition that for all $i$ either $t_i=0$ or $t_i=f(\{t_j\}_{j \in \di};l_i)$, then they correspond to the activation times for the dynamics started from the initial condition $\us$ such that $\s_i=1$ if and only if $t_i=0$. These two descriptions in terms of $(\s_1,\dots,\s_N)$ and $(t_1,\dots,t_N)$ are thus equivalent, yet the great advantage of the latter is that the conditions to enforce among the $t_i$ are local along the graph, and that they contain in an obvious way the information on $\s_i^T$ that was lacking to deal with (\ref{eq_eta_us_generalized}).

\begin{figure}
\includegraphics[width=4.5cm]{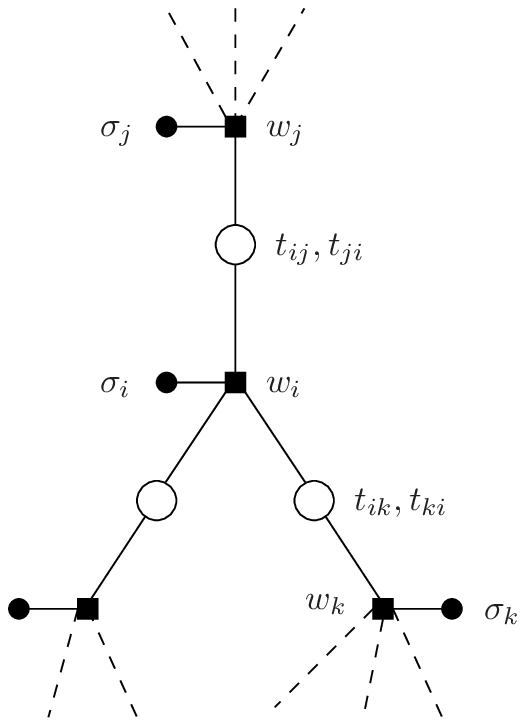}
\caption{A portion of the factor graph corresponding to the measure (\ref{eq_eta_ust}).}
\label{fig_factor}
\end{figure}

Finally a last twist on Eq.~(\ref{eq_eta_us_generalized}) will be to ``duplicate'' the activation time $t_i$ on all edges connecting $i$ to one of its neighbour $j$, introducing redundant variables $t_{ij}$ to be finally constrained to be all equal to $t_i$. Let us denote $\ut$ the collective configurations of all these $2M$ variables $t_{ij},t_{ji}$ on each edge $\la i,j\ra$ of the graph, that take values in $\{0,1,\dots,T,\infty\}$, and consider the following probability measure on $(\us,\ut)$:
\beq
\eta(\us,\ut) = \frac{1}{Z(G,\{l_i\},T,\{\mu_i,\epsilon_i\})} \prod_{i=1}^N w_i(\s_i,\{t_{ij},t_{ji}\}_{j \in \di}) \ ,
\label{eq_eta_ust}
\eeq
where the functions $w_i$ are defined by
\beq
w_i(\s_i,\{t_{ij},t_{ji}\}_{j \in \di}) = 
e^{\mu_i \s_i} e^{-\epsilon_i \ind(f(\s_i,\{t_{ki}\}_{k \in \di};l_i)=\infty)} 
\prod_{j\in\di} \ind(t_{ij}=f(\s_i,\{t_{ki}\}_{k \in \di};l_i)) \ .
\eeq
The above observations imply that the marginal of $\us$ under $\eta(\us,\ut)$ is precisely the desired one from Eq.~(\ref{eq_eta_us_generalized}), and that in the support of the law the $\ut$ are strictly constrained to be the activation times for the dynamics starting from $\us$. This correspondance being one-to-one the partition function is the same in (\ref{eq_eta_us_generalized}) and (\ref{eq_eta_ust}). A portion of the factor graph~\cite{factorgraph} associated to the probability law (\ref{eq_eta_ust}) is sketched in Fig.~\ref{fig_factor}, with black squares representing the function nodes (interactions) $w_i$, black circles the variables $\s_i$, and white circles the variables $t_{ij},t_{ji}$. One notes that if the original graph $G$ is a tree (resp. is locally a tree) then the corresponding factor graph is a tree (resp. is locally a tree). This fact was the motivation for the ``duplication'' of the $t_i$ variables on the surrounding edge, without it short loops of interactions would still be present in the factor graph.

\subsection{Replica Symmetric (RS) formalism}

Let us now explain how the probability law (\ref{eq_eta_ust}) and its associated normalization $Z$ can be handled within the cavity formalism, first at the simplest, so called Replica Symmetric (RS), level.

\subsubsection{Single sample equations}

If the graph $G$ were a finite tree the factor graph associated to (\ref{eq_eta_ust}) would be a tree, hence $Z$ and the marginals of $\eta$ could be computed exactly via the recursive equations that we are about to write down. If the graph is only locally tree-like these equations are only approximate, they correspond to the (loopy) Belief Propagation equations, valid under some assumptions of long-range correlation decay under the measure $\eta$. This recursive computation of $Z$ amounts to introduce on each directed edge $i\to j$ of the graph a ``message'' $\eta_{i \to j}(t_{ij},t_{ji})$, which is a normalized probability distribution over a pair of activation times. These messages obey recursion relations of the form $\eta_{i \to j} = \hg(\{ \eta_{k \to i} \}_{k \in \dimj};l_i,\epsilon_i,\mu_i)$, where the mapping $\eta=\hg(\eta_1,\dots,\eta_k;l,\epsilon,\mu)$ is given by
\bem
\eta(t,t') =
\frac{1}{\hziter(\eta_1,\dots,\eta_k;l,\epsilon,\mu)}
\left[ \delta_{t,0} e^\mu 
\prod_{i=1}^k\left( \sum_{t''} \eta_i(t'',0)\right) \right.  \\ \left.
+ e^{- \epsilon \delta_{t,\infty}} \sum_{t_1,\dots,t_k} \eta_1(t_1,t) \dots \eta_k(t_k,t) \ind(t=f(t_1,\dots,t_k,t';l)) \right] \ .
\label{eq_recurs_eta}
\end{multline}
Here and in the following unprecised summations over a time index go along $\{0,\dots,T,\infty\}$. The function $\hziter(\eta_1,\dots,\eta_k;l,\epsilon,\mu)$ is defined by normalization, in such a way that $\sum_{t,t'}\eta(t,t')=1$.

The knowledge of the messages $\eta_{i \to j}$ on all edges of the graph allows to compute the free-entropy density, according to the Bethe formula:
\beq
\phi = \frac{1}{N}\ln Z = \frac{1}{N} \sum_{i=1}^N \ln \hzsite(\{\eta_{j \to i} \}_{j \in \di};l_i,\epsilon_i,\mu_i) 
-\frac{1}{N} \sum_{\la i,j \ra \in E}  \ln \hzedge (\eta_{i \to j},\eta_{j \to i}) \ ,
\label{eq_lnZ_eta}
\eeq
where the second sum runs over the (undirected) edges of the graph, and the local partition functions are
\bea
\hzsite(\eta_1,\dots,\eta_{k+1};l,\epsilon,\mu) 
&=& e^\mu \prod_{i=1}^{k+1} \left(\sum_{t'} \eta_i(t',0)\right) + 
\sum_{t=1}^T \sum_{t_1,\dots,t_{k+1}} \eta_1(t_1,t) \dots \eta_{k+1}(t_{k+1},t)
\ind(t=1+\lmin(t_1,\dots,t_{k+1})) \nonumber \\
&&+ e^{-\epsilon} \sum_{t_1,\dots,t_{k+1}} \eta_1(t_1,\infty) \dots \eta_{k+1}(t_{k+1},\infty)
\ind(\lmin(t_1,\dots,t_{k+1}) \ge T)
\label{eq_hzsite} \\
\hzedge (\eta_1,\eta_2) &=& \sum_{t,t'} \eta_1(t,t') \eta_2(t',t) \ .
\label{eq_hzedge}
\eea
The marginals of the law (\ref{eq_eta_ust}) can also be deduced from the messages, for instance the probability distribution of the activation time $t_i$ for the vertex $i$ reads $\eta(t_i)=\hetasite(\{\eta_{j \to i} \}_{j \in \di};l_i,\epsilon_i,\mu_i)(t_i)$, where
\bea
\hetasite (\eta_1,\dots,\eta_{k+1};l,\epsilon,\mu) (t) &=& 
\frac{1}{\hzsite(\eta_1,\dots,\eta_{k+1};l,\epsilon,\mu)}
 \left\{
\delta_{t,0} e^{\mu} \prod_{i=1}^{k+1} \left(\sum_{t'} \eta_i(t',0) \right)
\right. \nonumber \\ &+& \left. 
(1-\delta_{t,0}-\delta_{t,\infty}) \sum_{t_1,\dots,t_{k+1}} \eta_1(t_1,t) \dots \eta_{k+1}(t_{k+1},t)
\ind(t=1+\lmin(t_1,\dots,t_{k+1})) \right. \nonumber \\ 
&+ &\left. \delta_{t,\infty} e^{-\epsilon} \sum_{t_1,\dots,t_{k+1}} \eta_1(t_1,\infty) \dots \eta_{k+1}(t_{k+1},\infty)
\ind(\lmin(t_1,\dots,t_{k+1}) \ge T) \right\} \ .
\label{eq_hetasite}
\eea
The probability that the vertex $i$ is active in the initial condition is then deduced as $\eta(\s_i=1)=\eta(t_i=0)$. As explained above in Eq.~(\ref{eq_Legendre_both}), one can deduce from the above results the entropy density $s(\theta,\theta')$ for initial configurations with a fraction $\theta$ of active sites leading to a fraction $\theta'$ of active sites after $T$ steps, taking $\mu_i=\mu$ and $\epsilon_i = \epsilon$ for all sites, with
\beq
s(\theta,\theta') = \phi(\mu,\epsilon) - \mu \theta + \epsilon (1-\theta') \ ,
\qquad \theta = \frac{1}{N} \sum_{i=1}^N \eta(t_i=0) \ , \qquad
\theta' = \frac{1}{N} \sum_{i=1}^N \eta(t_i \le T) \ .
\eeq
Note that the derivatives of $\phi$ with respect to $\mu$ and $\epsilon$ can be taken only on the explicit dependence in (\ref{eq_lnZ_eta}), the recursion equations on the messages $\eta_{i \to j}$ being precisely the stationarity condition of $\phi$ with respect to the $\eta$'s.

\subsubsection{A more compact parametrization of the messages}

Each probability distribution $\eta(t,t')$ is a priori described by $(T+2)^2-1$ independent real numbers (the times run over $T+2$ values, including $\infty$, and there is a global normalization constraint). We shall see however that a much more compact parametrization is possible, which will be very useful for the further analytical treatment of the model. From now on we shall assume that $\mu_i=\mu$ and $\epsilon_i =\epsilon$ for all vertices. To unveil these simplifications let us first rewrite Eq.~(\ref{eq_recurs_eta}) more explicitly:
\bea
\eta(0,t') &=& \frac{1}{\hziter} e^\mu \prod_{i=1}^k (\eta_i(0,0)+\eta_i(1,0)+\dots+\eta_i(T,0)+\eta_i(\infty,0)) \\
\eta(t,t') &=& \frac{1}{\hziter} \sum_{t_1,\dots,t_k} \eta_1(t_1,t) \dots \eta_k(t_k,t)  \  \ind(t=1+\lmin(t_1,\dots,t_k,t')) \qquad \text{for} \ t\in\{1,\dots,T\}  \\
\eta(\infty,t') &=& \frac{1}{\hziter} e^{- \epsilon} \sum_{t_1,\dots,t_k} \eta_1(t_1,\infty) \dots \eta_k(t_k,\infty) \  \ind(\lmin(t_1,\dots,t_k,t') \ge T) 
\eea
where in all the three cases $t'$ can take any value in $\{0,1,\dots,T,\infty\}$. Now the condition ``$\lmin(t_1,\dots,t_k,t')=t-1$'' is easily seen to be equivalent to ``at least $l$ of the time arguments are $\le t-1$ and at most $l-1$ of them are $\le t-2$''. Similarly the condition ``$\lmin(t_1,\dots,t_k,t') \ge T$'' is equivalent to ``at most $l-1$ times are $\le T-1$''. This observation allows to rewrite the above equations under the following form:
\bea
\eta(0,t') &=& \frac{1}{\hziter} e^\mu \prod_{i=1}^k (\eta_i(0,0)+\eta_i(1,0)+\dots+\eta_i(T,0)+\eta_i(\infty,0)) \label{eq_eta_0tp} \\
\eta(t,t') &=&  \frac{1}{\hziter} \sum_{\substack{ I,J,K \\ |I| + \ind(t'\le t-2) \le l-1 \\ |I| + |J| + \ind(t'\le t-1) \ge l } } 
\prod_{i\in I} \left(\sum_{t''=0}^{t-2} \eta_i(t'',t) \right) 
\prod_{i\in J} \eta_i(t-1,t)
\prod_{i \in K} \left( \sum_{t''\ge t} \eta_i(t'',t)\right) \label{eq_eta_ttp} 
\\
\eta(\infty,t') &=& \frac{1}{\hziter} e^{- \epsilon} 
\sum_{\substack{ I,J \\ |I| + \ind(t'\le T-1) \le l-1}} 
\prod_{i\in I} \left(\sum_{t''=0}^{T-1} \eta_i(t'',\infty) \right) 
\prod_{i \in J} \left(\sum_{t''\ge T} \eta_i(t'',\infty)\right)
\label{eq_eta_inftytp}
\eea
where the summation in the second (resp. third) line is over the partitions $I,J,K$ (resp. $I,J$) of $\{1,\dots,k\}$. These expressions reveal a first simplification, as already noticed in~\cite{Torino1,Torino2}: among the $(T+2)^2$ elements of $\eta(t,t')$ only $3T+2$ are distinct. Indeed $\eta(0,t')$ is independent of $t'$, for a given value of $t \in \{1,\dots,T\}$ $\eta(t,t')$ takes at most three distinct values, whether $t'\ge t$, $t'=t-1$, or $t' \le t-2$ and finally $\eta(\infty,t')$ takes two values whether $t'\le T-1$ or $t' \ge T$. There is however a further simplification to perform: in the right hand sides of the above equations the $\eta_i$'s always appear under the form of particular linear combinations. In particular the elements under the diagonal of the matrices $\eta_i$, i.e. $\eta_i(t,t')$ with $t\ge t'$, always intervene under the form $\sum_{t \ge t'}\eta(t,t')$. This allows to reduce further the number of relevant linear combinations of elements of the $\eta$'s. A convenient parametrization of the messages $\eta$ is thus provided by the numbers $a_t$ for $t \in \{0,1,\dots,T\}$ and $b_t$ for $t \in \{1,\dots,T\}$, defined
by:
\beq
e^{\mu a_t} = \frac{\eta(0,0)}{\sum_{t'} \eta(t',t)}  \ , \qquad
e^{\mu b_t} = \frac{\eta(0,0)}{\sum_{t'=0}^t \eta(t',t)} = \frac{\eta(0,0)}{\sum_{t'=0}^t \eta(t',t'')} \ \forall t'' \ge t \ . 
\label{eq_def_at}
\eeq
One can consistently extend these definitions with $b_0=0$, and it will be useful to adopt the convention $e^{-\mu b_{-1}}=0$ in order to simplify some expressions. Let us denote $h=(a_0,a_1,\dots,a_T,b_{T-1},\dots,b_1)$ the vector of $2T$ reals encoding in this way a matrix $\eta$; $h$ will be called a cavity field in the following (note that we excluded $b_T$ which disappears from the final expressions). The recursion relations (\ref{eq_eta_0tp}-\ref{eq_eta_inftytp}) should now be transformed into a relation between cavity fields, i.e. $h=g(h_1,\dots,h_k)$, with $h_i=(a_0^{(i)},a_1^{(i)},\dots,a_T^{(i)},b_{T-1}^{(i)},\dots,b_1^{(i)})$. Inserting the definitions (\ref{eq_def_at}) into the equations (\ref{eq_eta_0tp}-\ref{eq_eta_inftytp}) leads to the explicit form for $g$,

\bea
e^{- \mu a_t}
&=& 1 + e^{-\mu} \sum_{t'=1}^T \sum_{\substack{ I,J,K \\ |I| + \ind( t' \ge t+2) \le l-1 \\ |I| + |J| + \ind(t' \ge t+1) \ge l } } 
\P_{t'}(h_1,\dots,h_k;I,J,K) +
e^{-\mu-\epsilon}
\sum_{\substack{ I,J,K \\ |I|+|J| + \ind(t\le T-1) \le l-1}} 
\P_T(h_1,\dots,h_k;I,J,K) \nonumber \\
e^{- \mu b_t} &=& 1 + e^{-\mu} \sum_{t'=1}^t 
\sum_{\substack{I,J,K \\  |I| \le l-1 \\ |I|+|J| \ge l}} 
\P_{t'}(h_1,\dots,h_k;I,J,K) 
\label{eq_g}
\eea
where we defined
\beq
\P_t(h_1,\dots,h_k;I,J,K) = 
e^{\mu \underset{i=1}{\overset{k}{\sum}} a_0^{(i)}}  
\prod_{i \in I} e^{-\mu b_{t-2}^{(i)}}
\prod_{i \in J} ( e^{-\mu b_{t-1}^{(i)}} -  e^{-\mu b_{t-2}^{(i)}} )
\prod_{i \in K} ( e^{-\mu a_t^{(i)}} -  e^{-\mu b_{t-1}^{(i)}} ) \ .
\eeq
It can be checked that for $T=1$ and $\epsilon=+\infty$ these equations correspond, as they should, to the one of the Biroli-M\'ezard model (see Eqs.~(108,109) of~\cite{bm}). One can also express the partial partition functions $\hzsite$ and $\hzedge$ in terms of these fields. It will be more convenient to factor out a common part in the site and edge contributions to the free-entropy. Denoting $r(\eta)=\sum_t \eta(t,0)$, we define $\zedge$ as:
\bea
\zedge(h_1,h_2) &=& \frac{\hzedge (\eta_1,\eta_2)}
{r(\eta_1)r(\eta_2)}  \label{eq_zedge} \\
&=&  e^{\mu (a_0^{(1)}+a_0^{(2)})} \left\{
e^{-\mu (a_T^{(1)}+a_T^{(2)}) } +
\sum_{t=0}^{T-1} \left[ 
\left( e^{-\mu a_t^{(1)}}-e^{-\mu a_{t+1}^{(1)}}\right)
e^{-\mu b_t^{(2)}} 
+ e^{-\mu b_t^{(1)}} 
\left( e^{-\mu a_t^{(2)}}-e^{-\mu a_{t+1}^{(2)}}\right)
\right] 
\right\}\ ,
\nonumber
\eea
where the explicit expression is obtained from Eq.~(\ref{eq_hzedge}).
Similarly, exploiting Eq.~(\ref{eq_hzsite}), we get for the site term (factoring also a contribution from the chemical potential):
\bea
\zsite(h_1,\dots,h_{k+1};l,\epsilon;\mu) &=& 
\frac{e^{-\mu}\hzsite(\eta_1,\dots,\eta_{k+1};l,\epsilon;\mu)}
{r(\eta_1) \dots r(\eta_{k+1})  } \label{eq_zsite}  \\ &=& 
1 + e^{-\mu}\sum_{t=1}^T 
\sum_{\substack{I,J,K \\ |I| \le l-1 \\ |I|+|J| \ge l}} 
\P_t(h_1,\dots,h_{k+1};I,J,K)
+ e^{-\mu- \epsilon} \sum_{\substack{I,J,K \\ |I|+|J|\le l-1}} \P_T(h_1,\dots,h_{k+1};I,J,K)
\nonumber
\eea
where as above in the summations $I,J,K$ denotes a partition of $\{1,\dots,k+1\}$.

To summarize the results of this reparametrization, on a given graph one has cavity fields $h_{i\to j}$ on each directed edge, obeying the Belief Propagation equations $h_{i\to j}=g(\{h_{k \to i}\}_{k \in \dimj})$, with the $g$ defined in Eq.~(\ref{eq_g}), and the Bethe free-entropy density is computed from these cavity fields according to
\beq
\phi = \mu+ \frac{1}{N} \sum_i \ln \zsite(\{ h_{j \to i} \}_{j \in \di};l_i,\epsilon,\mu) 
- \frac{1}{N} \sum_{\la i,j \ra \in E} \ln \zedge(h_{i \to j},h_{j \to i}) \ , 
\label{eq_phi_h_BP}
\eeq
with $\zsite$ and $\zedge$ defined in Eqs.~(\ref{eq_zsite}) and (\ref{eq_zedge}) respectively. Note that the factors $r$ introduced in the definitions of $\zsite$ and $\zedge$ compensate because in the expression of the Bethe free-energy of Eq.~(\ref{eq_lnZ_eta}) the messages on each directed edge appear exactly once in the site term and once in the edge term. The marginals of the law $\eta(\us,\ut)$ can also be computed from the cavity fields $h$, in particular from the expression (\ref{eq_hetasite}) one obtains the marginal of one activation time from the incident cavity fields as
\bea
\etasite(h_1,\dots,h_{k+1};l,\epsilon;\mu)(t) = \frac{1}{\zsite(h_1,\dots,h_{k+1};l,\epsilon;\mu)} &&\left\{
\delta_{t,0} \,  
+ (1-\delta_{t,0}-\delta_{t,\infty}) e^{-\mu}
\sum_{\substack{I,J,K \\ |I| \le l-1 \\ |I|+|J| \ge l}} 
\P_t(h_1,\dots,h_{k+1};I,J,K)
\right. \nonumber \\ 
&&+ \left. \delta_{t,\infty} \, e^{-\mu-\epsilon}
\sum_{\substack{I,J,K \\ |I|+|J|\le l-1}} \P_T(h_1,\dots,h_{k+1};I,J,K)
\right\} \ .
\label{eq_etasite}
\eea

\subsubsection{Random (regular) graphs}
\label{sec_cavity_RS_rg}

The replica symmetric cavity method, for generic models defined on sparse random graphs, postulates the asymptotic validity of the above computations, exact on finite trees, thanks to the local convergence of random graphs to trees and an assumption of correlation decay at large distance. The order parameter is then a probability distribution over cavity fields, the randomness arising from the fluctuations of the degrees of the vertices in the graph and/or the randomness in the local interactions.

In the case of random regular graphs with no disorder in the coupling the situation is even simpler, as one can look for a ``factorized'' solution with all cavity fields equal. In particular for the model at hand on a $k+1$ random regular graph, with the same threshold of activation $l$ for all vertices, the RS prediction for the typical free-entropy density in the thermodynamic limit defined in Eq.~(\ref{eq_def_phi}) reads
\beq
\phi(k,l,T,\mu,\epsilon) = \mu + \ln \left( \zsite(h,\dots,h)  \right) 
- \frac{k+1}{2} \ln\left( \zedge(h,h) \right) \ ,
\eeq
which is easily obtained from (\ref{eq_phi_h_BP}) noting that $2M=(k+1)N$ in a regular graph. The field $h$ is the fixed-point solution of the cavity recursion (\ref{eq_g}),
\beq
h=g(h,\dots,h) \ .
\label{eq_RS}
\eeq
The marginal law for the activation time is obtained from (\ref{eq_etasite}) by setting all the fields to $h$, which allows finally to compute the entropy density from the Legendre inverse transform discussed in (\ref{eq_Legendre_both}).

We shall discuss the results obtained from this RS prediction in the next Section, more explicit formulas for the RS equation in this case, along with some technical details on their resolution being displayed in the Appendix~\ref{sec_app}. One can however anticipate that in some regime of parameters the RS hypothesis will be violated. This is for instance known for $T=1$, $\epsilon=+\infty$, which corresponds to the Biroli-M\'ezard model; it was indeed shown in~\cite{bm} that for large negative values of $\mu$ the predictions of the RS ansatz are unphysical, the frustration arising from the contradictory constraints of putting as few active vertices in the initial condition as possible while imposing that all vertices become active at a latter time induces long-range correlations between variables that are incompatible with the RS ansatz. This limit $\mu \to -\infty$ being the interesting case for the computations of the minimal density of contagious sets, we shall now see how to include the effects of replica symmetry breaking in this model.

\subsection{One step of Replica Symmetry Breaking (1RSB) formalism}

The long-range correlation decay assumption underlying the RS cavity method breaks down for models with too much frustration. In this case one has to picture the configuration space as fractured into pure states, or clusters, that we shall index here by $\gamma$, such that the correlation decay assumption only holds for the Gibbs-Boltzmann probability law restricted to one pure-state. The partition function restricted to a given pure-state is denoted $Z_\gamma$, in such a way that $Z=\sum_\gamma Z_\gamma$. The replica symmetry breaking version of the cavity method then postulates some properties of this decomposition into pure states, which are compatible with the local convergence of the graph under study to a tree. In the first non-trivial version of the RSB formalism, so called one-step RSB (1RSB), one assumes the existence of a complexity function, also called configurational entropy in the context of glasses, $\Sigma(\phi)$, such that the number of pure states with an internal free-entropy density $\phi_\gamma = \frac{1}{N} \ln Z_\gamma$ close to some value $\phi$ is, at the leading exponential order, $e^{N \Sigma(\phi)}$. The computation of $\Sigma(\phi)$ is performed via the 1RSB potential with a parameter $m$ (known as the Parisi breaking parameter), related to $\Sigma$ through a Legendre transform structure~\cite{Mo95}:
\beq
\Phi(m) = \frac{1}{N} \ln \sum_\gamma Z_\gamma^m = 
\sup_\phi \, [ \Sigma(\phi) + m \, \phi ] \ .
\label{eq_def_Phi}
\eeq
The function $\Sigma(\phi)$ can be reconstructed in a parametric way varying $m$, with
\beq
\Sigma(\phi_{\rm int}(m)) = \Phi(m) - m \phi_{\rm int}(m) \ , \qquad
\phi_{\rm int}(m) = \Phi'(m) \ ,
\label{eq_invLegendre_1RSB}
\eeq
$\phi_{\rm int}(m)$ denoting the internal free-entropy density of the clusters selected by the corresponding value of $m$. The value $m=1$ plays a special role in this approach, as it corresponds a priori to the original computation of the free-entropy density of the model. However a so-called condensation (or Kauzmann) transition can occur, signaled by the vanishing of the complexity $\Sigma$ associated to $m=1$. In this case the Gibbs-Boltzmann measure is dominated by a sub-exponential number of pure-states, corresponding to a parameter $m_{\rm s}<1$ with $\Sigma(m_{\rm s})=0$. In the following paragraphs we shall derive the 1RSB equations and the expression of the 1RSB potential for the model under study, before discussing the concrete results for random regular graphs in the next Section.

\subsubsection{Single sample equations}

Let us first discuss the 1RSB formalism with the basic messages represented in terms of the matrices $\eta(t,t')$. In the RS description one had a message $\eta_{i \to j}$ on each directed edge of the graph, solution of the recurrence equations $\eta_{i \to j} = \hg(\{ \eta_{k \to i} \}_{k \in \dimj};l_i,\epsilon,\mu)$, see Eq.~(\ref{eq_recurs_eta}). At the 1RSB level one introduces instead a distribution $\hP_{i \to j}(\eta)$ on each directed edge, the randomness being over the choice of the pure-state $\gamma$ with a weight proportional to $Z_\gamma^m$. These distributions are thus found to obey the recurrence equations $\hP_{i \to j}=\hG[\{ \hP_{k \to i} \}_{k \in \dimj}]$, where $\hP=\hG(\hP_1,\dots,\hP_k)$ means
\beq
\hP(\eta) = \frac{1}{\hZiter(\hP_1,\dots,\hP_k)} 
\int \dd \hP_1(\eta_1) \dots \dd \hP_k(\eta_k) \ 
\delta(\eta - \hg(\eta_1,\dots,\eta_k)) \ \hziter(\eta_1,\dots,\eta_k)^m \ ,
\eeq
with $\hg$ and $\hziter$ defined in Eq.~(\ref{eq_recurs_eta}), and $\hZiter$ normalizes the distribution $\hP$. The 1RSB potential $\Phi(m)$ defined above is then computed from the solution of these equations, according to
\beq
\Phi(m)=
\frac{1}{N} \sum_{i=1}^N \ln \hZsite(\{\hP_{j \to i} \}_{j \in \di};l_i,\epsilon_i,\mu_i) 
- \frac{1}{N} \sum_{\la i,j \ra \in E}  \ln \hZedge (\hP_{i \to j},\hP_{j \to i}) \ ,
\label{eq_Phi_single}
\eeq
where 
\bea
\hZsite(\hP_1,\dots,\hP_{k+1}) &=& \int \dd \hP_1(\eta_1) \dots \hP_{k+1}(\eta_{k+1}) \ \hzsite(\eta_1,\dots,\eta_{k+1})^m \ , \\
\hZedge(\hP_1,\hP_2) &=& \int \dd \hP_1(\eta_1) \hP_2(\eta_2) \ \hzedge(\eta_1,\eta_2)^m 
\eea
are weighted averages, over the pure-states distribution, of the site and edge contributions to the free-entropy defined in (\ref{eq_hzsite},\ref{eq_hzedge}). Similarly the marginal distribution of an activation time can be computed as a weighted average of the RS expression in the various pure-states, i.e.
\beq
\eta(t) = \frac{1}{\hZsite(\hP_1,\dots,\hP_{k+1})} 
\int \dd \hP_1(\eta_1) \dots \hP_{k+1}(\eta_{k+1}) \ 
\hetasite(\eta_1,\dots,\eta_{k+1})(t) \ 
\hzsite(\eta_1,\dots,\eta_{k+1})^m \ .
\eeq
Note that the derivative $\Phi'(m)$, which plays an important role to compute the complexity from Eq.~(\ref{eq_invLegendre_1RSB}), can be taken in (\ref{eq_Phi_single}) on the explicit dependence on $m$ only, the recursion relations on the $\hP_{i\to j}$ being the stationarity conditions of (\ref{eq_Phi_single}) with respect to the $\hP$'s.

As we have seen in the discussion of the RS cavity method the matrices $\eta$ can be parametrized in a more economic way by the fields $h$ (vectors of $2T$ real numbers). The expressions of the 1RSB quantities can also be rewritten using this parametrization. After a few lines of algebra one finds that the potential $\Phi(m)$ reads
\beq
\Phi(m) = \mu m + \sum_{i=1}^N \ln \Zsite(\{P_{j \to i} \}_{j \in \di};l_i,\epsilon,\mu) 
- \sum_{\la i,j \ra \in E}  \ln \Zedge (P_{i \to j},P_{j \to i}) \ ,
\eeq
with
\bea
\Zsite(P_1,\dots,P_{k+1}) &=& \int \dd P_1(h_1) \dots P_{k+1}(h_{k+1}) \ \zsite(h_1,\dots,h_{k+1})^m \ , \\
\Zedge(P_1,P_2) &=& \int \dd P_1(h_1) P_2(h_2) \ \zedge(h_1,h_2)^m \ ,
\eea
the weighted averages of the quantities defined in (\ref{eq_zedge},\ref{eq_zsite}). The field distributions $P_{i \to j}(h)$ are solutions of the recurrence equations $P_{i \to j}=G(\{P_{k \to i} \}_{k \in \dimj})$, where the mapping $P=G(P_1,\dots,P_k)$ is given explicitly by
\beq
P(h) = \frac{1}{\Ziter(P_1,\dots,P_k)} 
\int \dd P_1(h_1) \dots \dd P_k(h_k) \ 
\delta(h - g(h_1,\dots,h_k)) \ \ziter(h_1,\dots,h_k)^m \ .
\label{eq_1RSB_Ph}
\eeq
$\Ziter$ is a normalizing factor ensuring that the left hand side is a probability distribution, $g$ is the function defined in Eq.~(\ref{eq_g}), and the reweighting factor reads
\beq
\ziter(h_1,\dots,h_k) = \frac{e^{-\mu} \hziter(\eta_1,\dots,\eta_k)r(\hg(\eta_1,\dots,\eta_k)) }{r(\eta_1) \dots r(\eta_k)} = e^{-\mu a_0(h_1,\dots,h_k)} \ ,
\label{eq_ziter}
\eeq
the last equality following from Eqs.~(\ref{eq_eta_0tp},\ref{eq_def_at}).

\subsubsection{Random regular graphs}

For the reasons explained in the context of the RS ansatz in Sec.~\ref{sec_cavity_RS_rg} one can look for a simple factorized solution of the 1RSB equations in the case of a $k+1$ regular random graph with all thresholds of activation equal to $l$. In this case one has to find a distribution $P(h)$ solution of
\beq
P(h) = \frac{1}{{\Ziter}} \int \dd P(h_1) \dots \dd P(h_k) \ \delta(h-g(h_1,\dots,h_k)) \ \ziter(h_1,\dots,h_k)^m \ ,
\label{eq_1RSB}
\eeq
where $m \in [0,1]$ is the Parisi breaking parameter and the functions $g$ and $\ziter$ are the ones defined in Eqs.~(\ref{eq_g},\ref{eq_ziter}). The 1RSB potential is then computed as
\bea
\Phi(m) = \mu m &+& \ln \left( \int \dd P(h_1) \dots \dd P(h_{k+1})\ \zsite(h_1,\dots,h_{k+1})^m  \right) \nonumber \\
&-& \frac{k+1}{2} \ln\left(\int \dd P(h_1)\dd P(h_2) \ \zedge(h_1,h_2)^m \right) \ ,
\label{eq_phi1RSB_m}
\eea
with the functions $\zsite,\zedge$ defined in Eqs.~(\ref{eq_zedge},\ref{eq_zsite}). As already mentioned above $\Phi'(m)$ can be computed by taking into account only the explicit dependence on $m$ of (\ref{eq_phi1RSB_m}).

\subsection{``Energetic'' 1RSB formalism}
\label{sec_1RSB_y}

Even within the simplified case of the factorized ansatz for regular graphs the 1RSB equations are relatively complicated, as they involve the resolution of a distributional equation on $P(h)$. However we are ultimately interested in a particular limit for the computation of the minimal density of contagious sets, namely the case where $\epsilon=+\infty$ (to take into account only the fully activating configurations), and in the limit $\mu \to -\infty$ (to select the initial configurations with the minimal number of active sites). It turns out that a simplified version of the 1RSB formalism can be devised in this case, corresponding to the ``energetic'' version of the 1RSB cavity method, first developed in~\cite{cavity_T0,MeZe}, see in particular Sec. 5 of~\cite{bm} for such a treatment of the related Biroli-M\'ezard model. This simplified treatment amounts to take simultaneously the limit $m\to 0$ and $\mu \to - \infty$, with a fixed finite value of a new parameter $y=-\mu m$. To explain the meaning of this limit let us rewrite more explicitly the expression of the 1RSB potential of Eq.~(\ref{eq_def_Phi}) in the case $\epsilon=+\infty$, introducing the complexity $\Sigma(s,\theta)$ counting the (exponential) number of clusters containing a number of order $e^{N s}$ of activating initial configurations with a fraction $\theta$ of active sites, hence with a free-entropy density $\phi=\mu \theta +s$:
\beq
\Phi(m) = \sup_{\theta,s}[\Sigma(s,\theta)+m (\mu \theta +s) ] \ .
\eeq
In the limit $m\to 0$, $\mu \to - \infty$ with $y=-\mu m$ this function becomes
\beq
\Phi_{\rm e}(y) = \sup_\theta [\Sigma_{\rm e}(\theta) - y \theta ] \ , \qquad
\Sigma_{\rm e}(\theta) = \sup_s \Sigma(s,\theta) \ .
\eeq
The ``energetic'' complexity $\Sigma_{\rm e}(\theta)$ can thus be computed via an inverse Legendre transform of the potential $\Phi_{\rm e}(y)$,
\beq
\Sigma_{\rm e}(\theta(y)) = \Phi_{\rm e}(y) + y \theta(y) \ , \qquad \theta(y) = - \Phi'_{\rm e}(y) \ .
\label{eq_Legendre_y}
\eeq
As we shall see the ``energetic'' 1RSB cavity equations leading to the computations of $\Phi_{\rm e}(y)$ are much simpler than the initial 1RSB ones at finite values of $\mu$ and $m$. The price to pay for this simplification is the loss of information on the entropy of the clusters when going from $\Sigma(s,\theta)$ to $\Sigma_{\rm e}(\theta)$. However this is not a problem for the determination of $\tmin$: its estimate at the 1RSB level, to be denoted $\tmino$, is the smallest value of $\theta$ with $\Sigma_{\rm e}(\theta) \ge 0$. Indeed the least dense activating configurations have to be in some pure states, whatever their entropy.

\subsubsection{Simplification of the cavity field recursion (Warning Propagation equations)}

We want to simplify the equations (\ref{eq_g}) giving $h=g(h_1,\dots,h_k)$ with $\epsilon=+\infty$ and in the limit $\mu \to -\infty$. First let us make some remarks, valid when $\epsilon=+\infty$ for any value of $\mu$. From the definition (\ref{eq_def_at}) of the fields $b_t$, or from their expressions in (\ref{eq_g}), it is obvious that
\beq
e^{-\mu b_T} \ge e^{-\mu b_{T-1}} \ge \dots \ge e^{-\mu b_1} \ge e^{-\mu b_0} = 1 \ . 
\eeq
One can also notice that for $\epsilon=+\infty$ one has, for any $\mu$, the equality $a_T=b_T$: this appears both from the definition (\ref{eq_def_at}) of the fields, as $\eta(\infty,t)=0$ when $\epsilon=+\infty$, and from the recursion relations (\ref{eq_g}), the last term in the first line of (\ref{eq_g}) disappearing when $\epsilon=+\infty$. To continue the above chain of inequalities let us first compute from (\ref{eq_g})
\beq
e^{-\mu a_{T-1}} - e^{-\mu a_T} = e^{-\mu+\mu \overset{k}{\underset{i=1}{\sum}} a_0^{(i)} } \sum_{\underset{|I|=l-1}{I,J}}
\prod_{i \in I} e^{- \mu b_{T-1}^{(i)}}
\prod_{i \in J} \left(e^{-\mu b_T^{(i)}} - e^{- \mu b_{T-1}^{(i)}}
\right) \ ,
\label{eq_diff_aT}
\eeq
where $I,J$ forms a partition of $\{1,\dots,k\}$. This shows that $e^{-\mu a_{T-1}} \ge e^{-\mu a_T} = e^{-\mu b_T}$, because in the right-hand side $e^{-\mu b_T^{(i)}} \ge e^{- \mu b_{T-1}^{(i)}}$. These inequalities can then be continued by recurrence, as for $t \in \{0,\dots,T-2\}$ one obtains from (\ref{eq_g})
\beq
e^{-\mu a_t} - e^{-\mu a_{t+1}} = e^{-\mu +\mu \overset{k}{\underset{i=1}{\sum}} a_0^{(i)}} \sum_{\underset{|I|=l-1}{I,J}}
\prod_{i \in I} e^{-\mu b_{t}^{(i)}}
\left( \prod_{i \in J} (e^{- \mu a_{t+1}^{(i)}} - e^{-\mu b_t^{(i)}}) 
-\prod_{i \in J} (e^{- \mu a_{t+2}^{(i)}} - e^{-\mu b_t^{(i)}}) 
\right)
\ ,
\label{eq_diff_at}
\eeq
hence
\beq
e^{-\mu a_0} \ge a^{-\mu a_1} \ge \dots \ge e^{-\mu a_{T-1}} \ge e^{-\mu a_T} =  e^{-\mu b_T} \ge e^{-\mu b_{T-1}} \ge \dots \ge e^{-\mu b_1} \ge e^{-\mu b_0} = 1 \ , 
\eeq
and for any $\mu \le 0$:
\beq
a_0 \ge a_1 \ge \dots \ge a_{T-1} \ge a_T = b_T \ge b_{T-1} \ge \dots \ge b_1 
\ge b_0 = 0 \ .
\label{eq_ineq_ab}
\eeq

Let us now take the limit $\mu \to -\infty$ in the equations (\ref{eq_g}), assuming that $a_t$ and $b_t$ have finite limits. Treating (\ref{eq_g}) at the leading exponential order one obtains
\bea
a_t &=& \max\left(0, \max_{t'\in[1,T]} \max_{\substack{ I,J,K \\ |I| + \ind( t' \ge t+2) \le l-1 \\ |I| + |J| + \ind(t' \ge t+1) \ge l } } \S_{t'}(h_1,\dots,h_k;I,J,K) \right) \ ,  \label{eq_at_WP}\\
b_t &=& \max\left(0, \max_{t'\in[1,t]} 
\max_{\substack{I,J,K \\  |I| \le l-1 \\ |I|+|J| \ge l}}
\S_{t'}(h_1,\dots,h_k;I,J,K) \right) \ , \label{eq_bt_WP}
\eea
where
\beq
\S_t(h_1,\dots,h_k;I,J,K) = 1 - \sum_{i \in I} (a_0^{(i)} - b_{t-2}^{(i)}) 
- \sum_{i \in J}(a_0^{(i)} - b_{t-1}^{(i)})
-\sum_{i \in K}(a_0^{(i)} - a_t^{(i)}) \ .
\eeq
Now from the inequalities (\ref{eq_ineq_ab}) it appears that $\S_t \le 1$, hence that the $a$'s and $b$'s belong to the interval $[0,1]$. It is however natural to assume that they are integers, as in the limit $\mu \to -\infty$ they can be interpreted as differences between number of particles in constrained groundstate configurations (see~\cite{bm,cavity_T0} for more details). Within this ansatz the $a$'s and $b$'s can only be equal to $0$ or $1$; using in addition the inequalities (\ref{eq_ineq_ab}) one realizes that the fields $h$ can only take $2T+1$ possible values, that we shall call $A_t$ for $t \in \{0,1,\dots,T-1\}$ and $B_t$ for $t\in\{0,1,\dots,T\}$. These are defined as follows; $A_t$ denotes the case where $a_0=\dots=a_t=1$, all the other $a$'s and $b$'s vanishing. For $t\in\{2,\dots,T\}$, $B_t$ means that $b_1=\dots=b_{t-1}=0$, all the other $a$'s and $b$'s being equal to 1. Finally $B_1$ corresponds to the case where all $a$'s and $b$'s are equal to 1, and $B_0$ to the case where they all vanish. Note that one can consistently extend these definitions to $A_T=B_T$, as by definition $a_T=b_T$.

It remains to determine the value of $h=g(h_1,\dots,h_k)$ in this $\mu \to -\infty$ limit, when all the fields $h_1,\dots,h_k$ belong to the set $\{A_0,A_1,\dots,A_{T-1},A_T=B_T,B_{T-1},\dots,B_1,B_0 \}$ of ``hard fields'', or Warning Propagation messages. Some algebra, sketched in Appendix \ref{app_eqs_WP}, leads to:
\beq
g(B_{t_1},\dots,B_{t_n},A_{t_{n+1}},\dots,A_{t_k}) = \begin{cases}
B_{1+\lmin(t_1,\dots,t_n)} & \text{if} \ n \ge l \ \text{and} \ \min(t_{n+1},\dots,t_k)  \ge 1+ \lmin(t_1,\dots,t_n) \\
A_{\min(t_{n+1},\dots,t_k) -1 } & 
\text{if} \ n \ge l-1 \ \text{and} \\
&1+ \lmunmin(t_1,\dots,t_n) \le \min(t_{n+1},\dots,t_k) \le \lmin(t_1,\dots,t_n)
\\
B_0 & \text{otherwise}
\end{cases}
\label{eq_g_ABt}
\eeq
where $t_1,\dots,t_n \in \{0,\dots,T-1\}$ and $t_{n+1},\dots,t_k \in \{0,\dots,T\}$. We assumed conventionally that $\lmin(t_1,\dots,t_{l-1})=\infty$.

The equation (\ref{eq_g_ABt}) can be given a very intuitive interpretation. The messages $h \in \{A_0,\dots,A_{T-1},B_0,\dots,B_T\}$ can be interpreted as ``warnings'' sent from one vertex of the graph to one of its neighbours, with the following meanings. A vertex $i$ sends a message $h_{i \to j}=B_t$ to one of its neighbour $j$ to say: ``if $j$ is kept inactive at all times the configuration of $i$ and of its sub-tree (the one rooted at $i$ and excluding $j$) leads to complete activation of the sub-tree within the time horizon $T$, and $i$ activates itself at time $t$''. In particular $h_{i \to j}=B_0$ means that $i$ is activated in the initial configuration. On the contrary $i$ sends the message $h_{i \to j}=A_t$ to $j$ to express: ``the complete activation of $i$ and its sub-tree requires that $j$ becomes activated at time $t$''. The rules of Eq.~(\ref{eq_g_ABt}) for the combination of these messages are then obtained by finding the configuration compatible with them, containing the minimal number of active sites in the initial configuration (because of the $\mu \to -\infty$ limit): 
\begin{itemize}
\item if strictly less than $l-1$ incoming messages are of the type $B_{t_i}$, with $t_i \in \{0,\dots,T-1\}$, the central site $i$ will never have more than $l$ active neighbours (even with the participation of the receiving site $j$) if it is initially inactive, hence the only way for $i$ to be active at time $T$ is to be active in the initial configuration, which implies $h_{i \to j}=B_0$.

\item if at least $l$ of the incoming messages are of the type $B_{t_i}$, with $t_i \in \{0,\dots,T-1\}$, say $(B_{t_1},\dots,B_{t_n})$, the central site $i$ will become active at time $t=1+\lmin(t_1,\dots,t_n)$, without the ``help'' of the activation of the site $j$ receiver of the message. This situation thus leads to a message of type $B_t$, at the condition that all other incoming messages of type $\{A_0,\dots,A_T\}$ do not require the activation of the central site $i$ at a time strictly earlier than $t=1+\lmin(t_1,\dots,t_n)$.

\item the participation of the activation of the receiving site $j$ is required at some time $t$ when the above condition is not fulfilled, i.e. when the incoming messages $(A_{t_{n+1}},\dots,A_{t_k})$ require the activation of the central site at some time $t_{\rm act}=\min (t_{n+1}, \dots ,t_k) < 1+ \lmin(t_1,\dots,t_n)$. This mechanism is possible if at time $t_{\rm act} -1$ already $l-1$ of the neighbours sending messages of type $B$ are active, i.e. it requires $\lmunmin(t_1,\dots,t_n)\le t_{\rm act} -1$. The ``help'' needed from the receiving site is that it is active at some time before $t_{\rm act} -1$; in the limit $\mu \to -\infty$ the least dense configurations, and thus the least stringent constraint on the time of activation is privileged, hence the message sent in this case is $h_{i\to j}= A_{t_{\rm act} -1}$.

\item all cases not fulfilling one of the conditions above require that $i$ is active in the initial configuration to be active at time $T$, hence the message sent is $h_{i\to j}= B_0$.
\end{itemize}

\subsubsection{Energetic 1RSB single sample equations}

Within this ansatz the 1RSB distributions $P(h)$ greatly simplify, as they are supported on the discrete set $h \in \{A_0,A_1,\dots,A_{T-1},A_T=B_T,B_{T-1},\dots,B_1,B_0 \}$. We shall denote $p_t$ the weight in $P(h)$ of the event $h=A_t$, and similarly $q_t$ for $h=B_t$ (with again the convention $p_T=q_T$ to simplify notations), i.e.
\beq
P(h) = \sum_{t=0}^{T-1} p_t \ \delta(h - A_t  ) 
+ \sum_{t=0}^T q_t \ \delta(h - B_t  ) \ .
\label{eq_def_pqt}
\eeq

The 1RSB recursion relation (\ref{eq_1RSB_Ph}) now reduces to a recursion between these finite-dimensional vectors of probabilities; inserting the definition (\ref{eq_def_pqt}) in the right hand side of (\ref{eq_1RSB_Ph}) and exploiting the combination rule (\ref{eq_g_ABt}) between hard fields, one obtains the following limit for the recursion relation $P=G[P_1,\dots,P_k]$:
\bea
p_t &=&  \frac{1}{Z[P_1,\dots,P_k]} e^y \tp_t \ , \quad
\tp_t = \sum_{\substack{I,J,K \\ |I| = l-1 \\ |J| \ge 1 }}
\prod_{i \in I}\left(\sum_{t'=0}^t q_{t'}^{(i)}\right) 
\prod_{i \in J} p_{t+1}^{(i)}
\prod_{i \in K}
\left(\sum_{t'=t+1}^T q_{t'}^{(i)} + \sum_{t'=t+2}^{T-1} p_{t'}^{(i)}
\right) \ \text{for} \ t \in \{0, \dots , T-1\} \nonumber \\
q_t &=& \frac{1}{Z[P_1,\dots,P_k]} e^y \tq_t \ , \quad
\tq_t = \sum_{\substack{I,J,K \\ |I| \le l-1 \\ |I| +|J| \ge l }}
\prod_{i \in I}\left(\sum_{t'=0}^{t-2} q_{t'}^{(i)}\right)   
\prod_{i \in J} q_{t-1}^{(i)}
\prod_{i \in K}\left(\sum_{t'=t}^T q_{t'}^{(i)} + \sum_{t'=t}^{T-1} p_{t'}^{(i)} \right)\ \text{for} \ t \in \{1, \dots , T\}
\nonumber \\
q_0 &=& \frac{1}{Z[P_1,\dots,P_k]} \left[1-\sum_{t=0}^{T-1} \tp_t - \sum_{t=1}^T \tq_t \right] \ , \qquad \qquad Z[P_1,\dots,P_k] = 1 + (e^y -1) \left[\sum_{t=0}^{T-1} \tp_t + \sum_{t=1}^T \tq_t \right]
\label{eq_G_1RSBy}
\eea
the reweighting term of Eq.~(\ref{eq_ziter}) becoming indeed 
$\ziter(h_1,\dots,h_k)^m = e^{y a_0(h_1,\dots,h_k)}$, hence the factor $e^y$ 
multiplying the probabilities of all warnings except $B_0$; this is indeed the only case where an active site has to be inserted in the initial configuration.

To compute the 1RSB potential we have to study the limit of the contribution of site and edge terms in the limit $\mu\to-\infty$, $m\to 0$. We have from Eq.~(\ref{eq_zsite})
\beq
\zsite(h_1,\dots,h_{k+1})^m \to \exp\left[y
\max\left(0, \max_{t\in[1,T]} 
\max_{\substack{I,J,K \\  |I| \le l-1 \\ |I|+|J| \ge l}}
\S_t(h_1,\dots,h_{k+1};I,J,K) \right)
\right] \ ,
\eeq
which can be simplified following the same reasoning than the one which led to (\ref{eq_g_ABt}). This yields
\beq
\Zsite(P_1,\dots,P_{k+1}) \to
1+(e^y-1) \sum_{t=1}^T \sum_{\substack{I,J,K \\ |I| \le l-1 \\ |I| +|J| \ge l }}
\prod_{i \in I}\left(\sum_{t'=0}^{t-2} q_{t'}^{(i)}\right)   
\prod_{i \in J} q_{t-1}^{(i)}
\prod_{i \in K}\left(\sum_{t'=t}^T q_{t'}^{(i)} + \sum_{t'=t}^{T-1} p_{t'}^{(i)} \right) \ ,
\label{eq_Zsite_y}
\eeq
where $I,J,K$ is a partition of $\{1,\dots,k+1\}$. This expression can be interpreted intuitively in terms of the warnings defined above; the factor multiplying $(e^y-1)$ is indeed the probability of complete activation, at time $t \in \{1,\dots,T\}$, for an initially empty site receiving messages $(h_1,\dots,h_{k+1})$ from its neighbours, with their respective distributions $P_1,\dots,P_{k+1}$. As a matter of fact, for its activation to occur at time $t$ at least $l$ neighbours must have activated without any help from the central site at time $t-1$, no more than $l-1$ must be active at time $t-2$ (otherwise the activation time would be strictly less than $t$), and the neighbours sending messages of type $A_{t'}$ should not require activation at a time $t'<t$.

For the edge term we obtain from Eq.~(\ref{eq_zedge})
\beq
\zedge(h_1,h_2)^m \to  \exp\left[ - y \min_{t\in[0,T]} \min(
(a_0^{(1)}-b_t^{(1)})+ (a_0^{(2)}-a_t^{(2)}), (a_0^{(1)}-a_t^{(1)})+ (a_0^{(2)}-b_t^{(2)}))
 \right] \ ,
\eeq
hence
\beq
\Zedge(P_1,P_2) \to e^{-y}+(1-e^{-y}) \left[ 
\left(\sum_{t=0}^T q_t^{(1)} \right)
\left(\sum_{t=0}^T q_t^{(2)} \right)
+ \sum_{t=0}^{T-1} p_t^{(1)} \sum_{t'=0}^t q_{t'}^{(2)}
+ \sum_{t=0}^{T-1} p_t^{(2)} \sum_{t'=0}^t q_{t'}^{(1)}
\right]  \ .
\label{eq_Zedge_y}
\eeq
One can interpret the factor multiplying $(1-e^{-y})$ as the probability of complete activation when two messages $(h_1,h_2)$ drawn with the probabilities $P_1,P_2$ are sent in the two opposite directions of an edge.

Let us summarize the main findings of this subsection. In the limit $\mu \to -\infty$, $m\to 0$ with $y = -\mu m$ the 1RSB formalism simplifies in the following way. The cavity field distributions $P_{i \to j}(h)$ have now a discrete support with $2T$ possible values, each of them is thus described by a (normalized) vector of $2T$ probabilities denoted $\{p_t,q_t\}$. These vectors are solutions of recurrence equations of the form $P_{i\to j}=G(\{P_{k\to i} \}_{k\in\dimj})$, the mapping $G$ being defined in Eq.~(\ref{eq_G_1RSBy}). The energetic limit of the 1RSB potential is then computed as
\beq
\Phi_{\rm e}(y) = - y + 
\frac{1}{N} \sum_{i=1}^N \ln\left( \Zsite(\{ P_{j\to i} \}_{j \in \di} )\right)
-\frac{1}{N}\sum_{\la i,j\ra} \ln\left( \Zedge(P_{i \to j},P_{j\to i}) \right)
\ ,
\label{eq_Phie_single_sample}
\eeq
with the expression of $\Zsite$ and $\Zedge$ given in Eqs.~(\ref{eq_Zsite_y},\ref{eq_Zedge_y}). This expression of $\Phi_{\rm e}$ is variational, its derivative with respect to $y$ (which is needed in the computation of the inverse Legendre transform in (\ref{eq_Legendre_y})) can be taken on the explicit dependence only.

\subsubsection{Random regular graphs}

For the reasons already exposed in the context of the RS and of the full 1RSB cavity formalism a factorized solution of the energetic 1RSB equations can be searched for when dealing with random $k+1$ regular graphs with a constant threshold of activation $l$. One has thus a single vector of probabilities $P=(\{p_t,q_t\})$, fixed-point solution of Eq.~(\ref{eq_G_1RSBy}), from which the energetic 1RSB potential is obtained as
\beq
\Phi_{\rm e}(y) = - y + \ln\left( \Zsite(P,\dots,P )\right)
-\frac{k+1}{2} \ln\left( \Zedge(P,P) \right) \ ,
\label{eq_Phie_reg}
\eeq
with $\Zsite$ and $\Zedge$ defined in Eqs.~(\ref{eq_Zsite_y},\ref{eq_Zedge_y}).

\section{Results of the cavity method for random regular graphs}
\label{sec_results_rrg}

We shall present now the results of the resolution of the cavity equations for random regular graphs of degree $k+1$, with an activation threshold equal to $l$ for all vertices. In all this discussion it will be understood that $\epsilon=+\infty$, i.e. we only consider initial configurations that activate the whole graph in $T$ steps. We will first present in Sec.~\ref{sec_res_finiteT} the results for finite values of $T$, which are qualitatively the same for all values of $k,l$ and $T$; the behaviour of the replica symmetric cavity method are first displayed, then we turn to the effects of replica symmetry breaking, in particular in the ``energetic'' limit to compute the minimal density of initially active sites in activating configurations. In a second part (Sec.~\ref{sec_res_largeT}) we shall discuss the limit $T\to\infty$, in which some further analytical computations can be performed. In this case several qualitatively distinct phenomena emerge, depending on the values of $k$ and $l$. 

\subsection{Finite $T$ results}
\label{sec_res_finiteT}
\subsubsection{Replica symmetric formalism}

\begin{figure}
\begin{center}
\includegraphics[width=7cm]{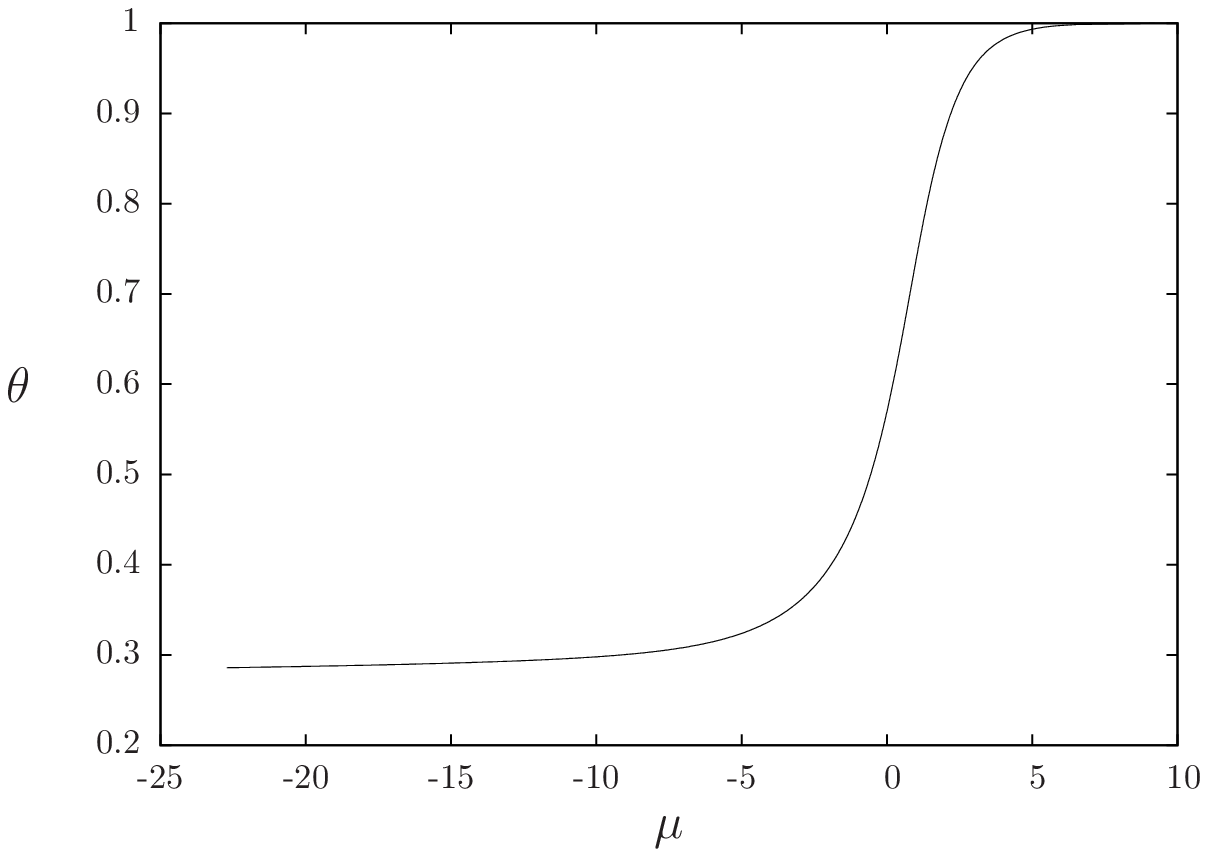}
\hspace{1cm}
\includegraphics[width=7cm]{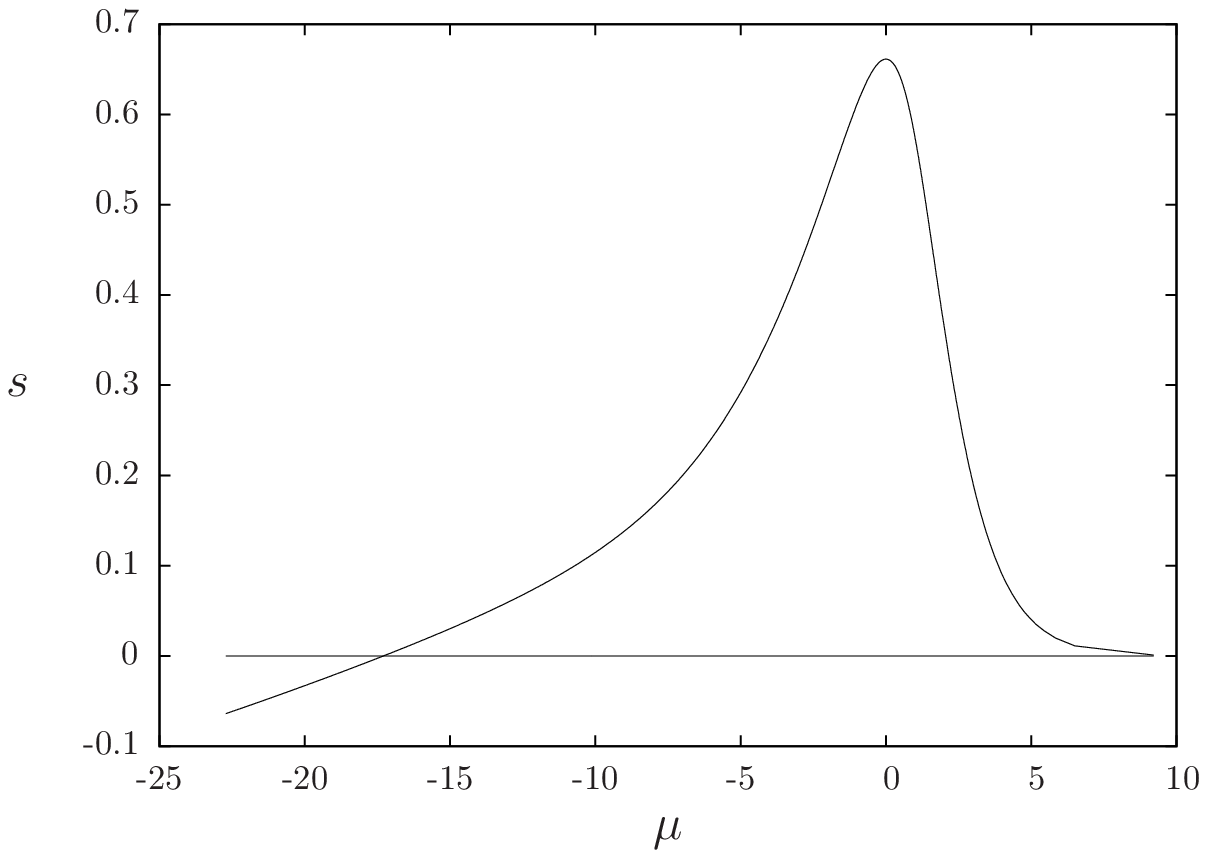}
\end{center}
\caption{The density of initially active sites $\theta$ (left panel) and the entropy $s$ (right panel) as a function of the chemical potential $\mu$, computed from the replica symmetric cavity equations, for $k=l=2$ and $T=3$.}
\label{fig_RS_k2l2_T3}
\end{figure}

\begin{figure}
\begin{center}
\includegraphics[width=7cm]{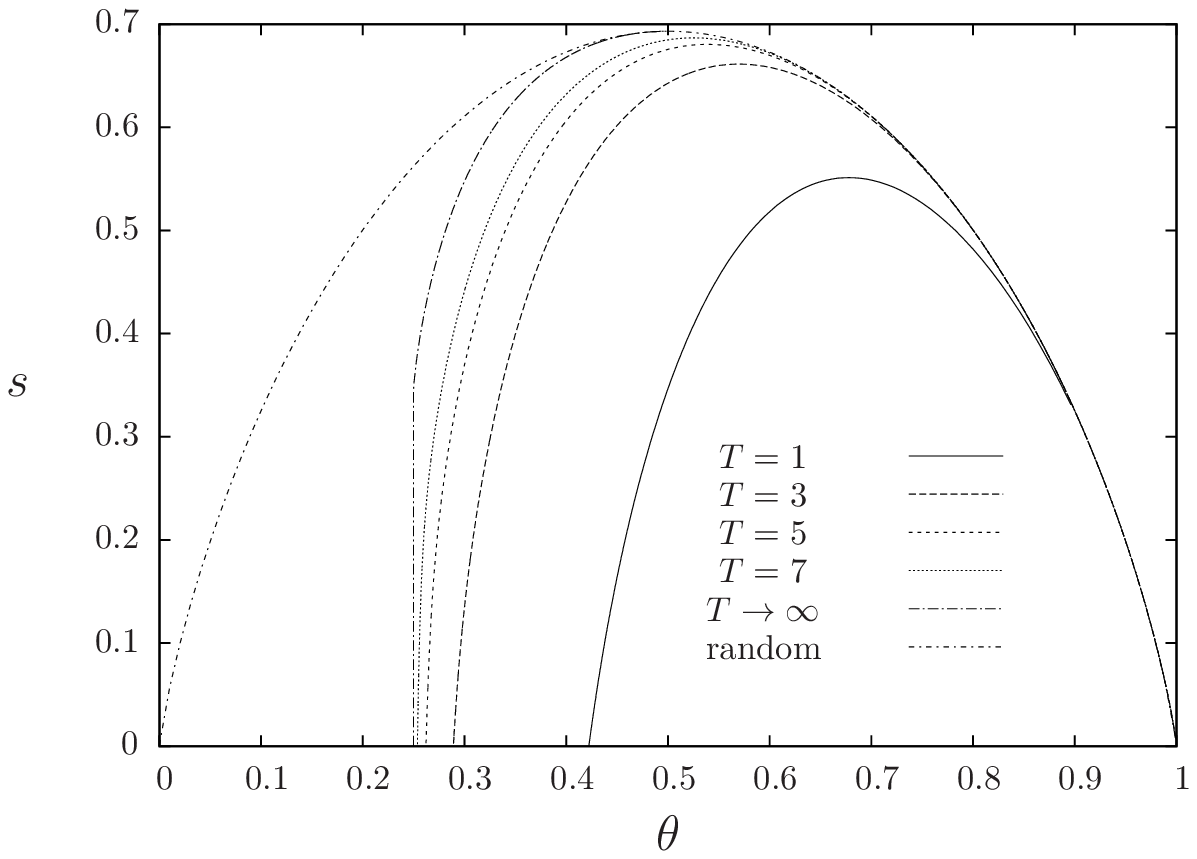}
\hspace{1cm}
\includegraphics[width=7cm]{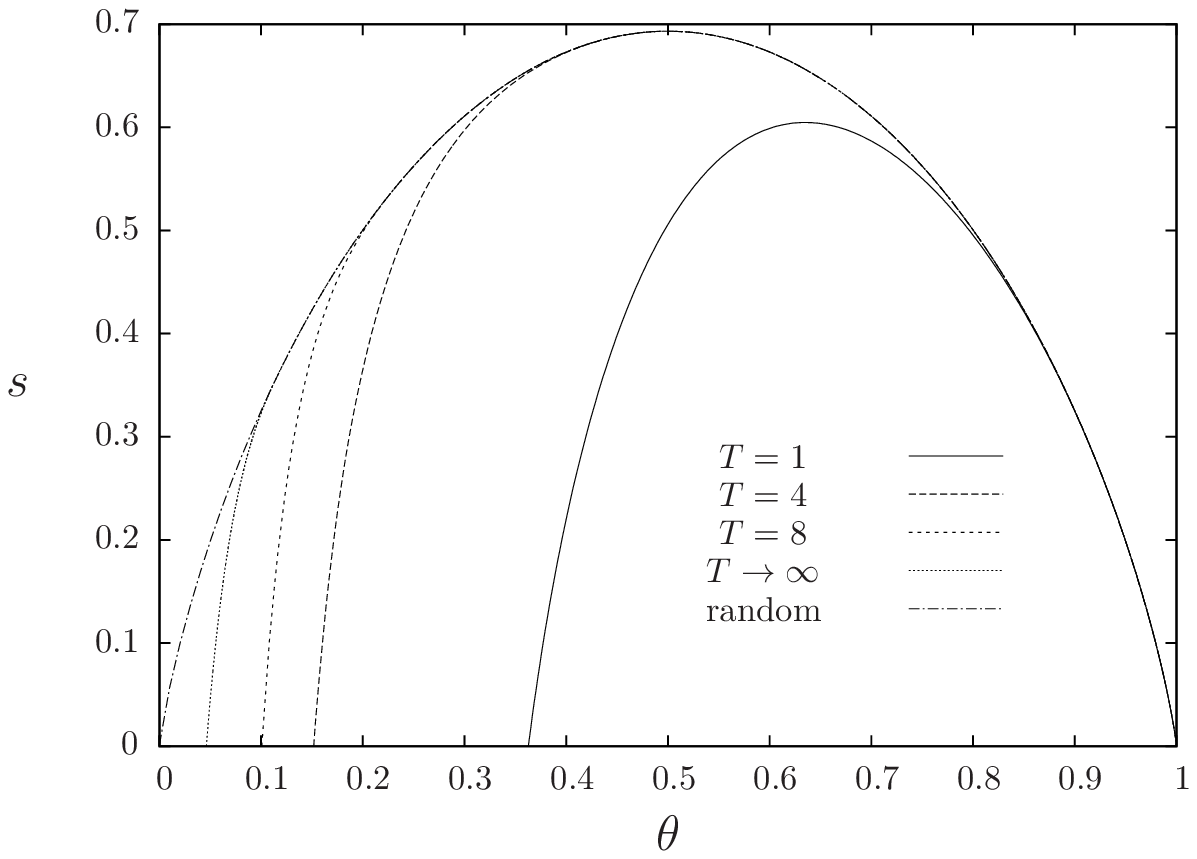}
\end{center}
\caption{The RS entropy $s(\theta)$ of configurations with a fraction $\theta$ of initially active sites able to activate completely the graph within time $T$, for $k=l=2$ (left panel) and $k=3$, $l=2$ (right panel). The curve labelled ``random'' is the binary entropy function $-\theta \ln \theta - (1-\theta) \ln(1-\theta)$ that counts all configurations with such an initial density. The curves in the limit $T\to\infty$ are computed analytically, from Eq.~(\ref{eq_s_klequal_Tinfty}) for the left panel and (\ref{eq_lmk_Tinfty_RS}) for the right panel, see Sec.~\ref{sec_res_largeT} for a further discussion of this limit.}
\label{fig_RS_k2l2_softheta}
\end{figure}

The technical details of the resolution of the RS equation $h=g(h,\dots,h)$, where $g$ is given in Eq.~(\ref{eq_g}), and of the computation of the free-entropy density, are deferred to the Appendix \ref{sec_app}. From a numerical point of view it is an easy task, as it corresponds essentially to the resolution of a set of $2T$ equations on $2T$ unknowns. Let us discuss the numerical results obtained in this way. On the left panel of Fig.~\ref{fig_RS_k2l2_T3} we display the curve $\theta(\mu)$ of the average fraction of initially active sites as a function of the chemical potential $\mu$; the curve is for $k=l=2$ and $T=3$, the qualitative features are independent of these precise values. This function is increasing as it should, and reaches a finite limit when $\mu \to -\infty$, that would be the candidate value for $\tmin$ if the RS computation was correct in this limit. This however cannot be true, as revealed from the computation of the entropy, displayed in the right panel of Fig.~\ref{fig_RS_k2l2_T3}: for $\mu < \mu_{s=0}$ the RS entropy becomes negative, which is a certain indication of the inadequacy of the RS theory in this regime. In Fig.~\ref{fig_RS_k2l2_softheta} we display the results for the entropy $s(\theta)$ of the number of configurations with a fraction $\theta$ of initially active sites, for the regime of positive entropies where the RS prediction cannot be ruled out at once (for the cases $k=l=2$ and $k=3$, $l=2$). For increasing values of $T$ these curves converge to a limit, this will be further discussed in Sec.~\ref{sec_kequall_Tinfty}. The numerical values of the chemical potential and of the fraction of active sites at the point of entropy cancellation, which would be the best guess from the RS computation of the value of $\tmin$, denoted respectively $\mu_{s=0}$ and $\tminz$, can be found for various values of $T$ in the Tables \ref{tab_k2l2}, \ref{tab_k3l3} and \ref{tab_k3l2} for the cases $k=l=2$, $k=l=3$ and $k=3$, $l=2$ respectively. For $T=1$ they reproduce, as they should, the results of the Biroli-M\'ezard model given in~\cite{bm}.

\begin{table}
\begin{tabular}{|c||c|c||c|c|c|c||c|c|}
\hline
 & \multicolumn{2}{c||}{RS} & \multicolumn{4}{c||}{1RSB} & \multicolumn{2}{c|}{energetic 1RSB} \\
\hline
$T$  & $\mu_{s=0}$ & $\tminz$ & $\mud$ & $\td$ & $\muc$ & $\tc$ & $\ys$ & $\tmino$  \\
\hline
1 &  -7.403996 & 0.422251 & -6.49 & 0.4292 & -6.69 & 0.4275 & 5.563433 & 0.424257\\
\hline
2 & -11.374979 & 0.325742 & -9.89 & 0.3291 & -11.23 & 0.3260 & 10.826348 & 0.325882 \\
\hline
3 & -17.292682 & 0.289093 & -13.7 & 0.2922 &-17.28 & 0.2890 &  17.232166 & 0.289097 \\
\hline
4 & -24.936318 & 0.271564 & -20.9 & 0.2731 &-24.93 & 0.2715 &  24.933659 & 0.271564 \\
\hline
5 & -34.966263 & 0.262167 & -31.3 & 0.2628 &-34.63 & 0.2622 &  34.966225 & 0.262167\\
\hline
6 & -49.901175 & 0.256844 & & & & &  49.901175 & 0.256844 \\
\hline
7 & -74.984724 & 0.253779 & & & & &  74.984724 & 0.253779 \\
\hline
8 & -120.79085 & 0.252036 & & & & &  120.79085 & 0.252036 \\
\hline
10 & -378.44778 & 0.250553 & & & & &  378.44778 & 0.250553 \\
\hline
15 & $-1.069 \ 10^4$ & 0.250018 & & & & & $1.069 \ 10^4$ & 0.250018 \\
\hline
20 & $-3.4 \ 10^5$ & 0.250000 & & & & & $3.4 \ 10^5$ & 0.250000  \\
\hline
$\infty$ & $-\infty$ & $\frac{1}{4}$ & & & & &  $+\infty$ & $\frac{1}{4}$ \\ 
\hline
\end{tabular}
\caption{Numerical results from the cavity computations at finite $T$ for $k=l=2$; the results in the limit $T\to\infty$ are explained in Sec.~\ref{sec_res_largeT}.}
\label{tab_k2l2}
\end{table}

\subsubsection{1RSB results}

As we have seen above the hypothesis underlying the RS computation must go wrong when $\mu$ is decreased towards $-\infty$, as the entropy computed within the RS scheme becomes negative for $\mu < \mu_{s=0}$; a 1RSB computation is thus required to investigate the limit $\mu \to -\infty$ and hence the properties of the least dense activating initial conditions, in particular their density $\tmin$.

We have thus solved numerically the 1RSB equations (\ref{eq_1RSB}) using population dynamics methods~\cite{cavity}, i.e. representing $P(h)$ as a weighted sample of fields $h_i$. This method has become fairly standard and we shall not give more details on the procedure, see for instance~\cite{cavity,MeMo_book} for detailed presentations. In the particularly important $m=1$ case we used a version of this procedure, inspired by the tree reconstruction problem, that allows to get rid of the reweighting terms in (\ref{eq_1RSB}) and is thus much more precise and efficient numerically, see~\cite{MeMo06,MRS08} for more technical details.

\begin{figure}
\includegraphics[width=8.5cm]{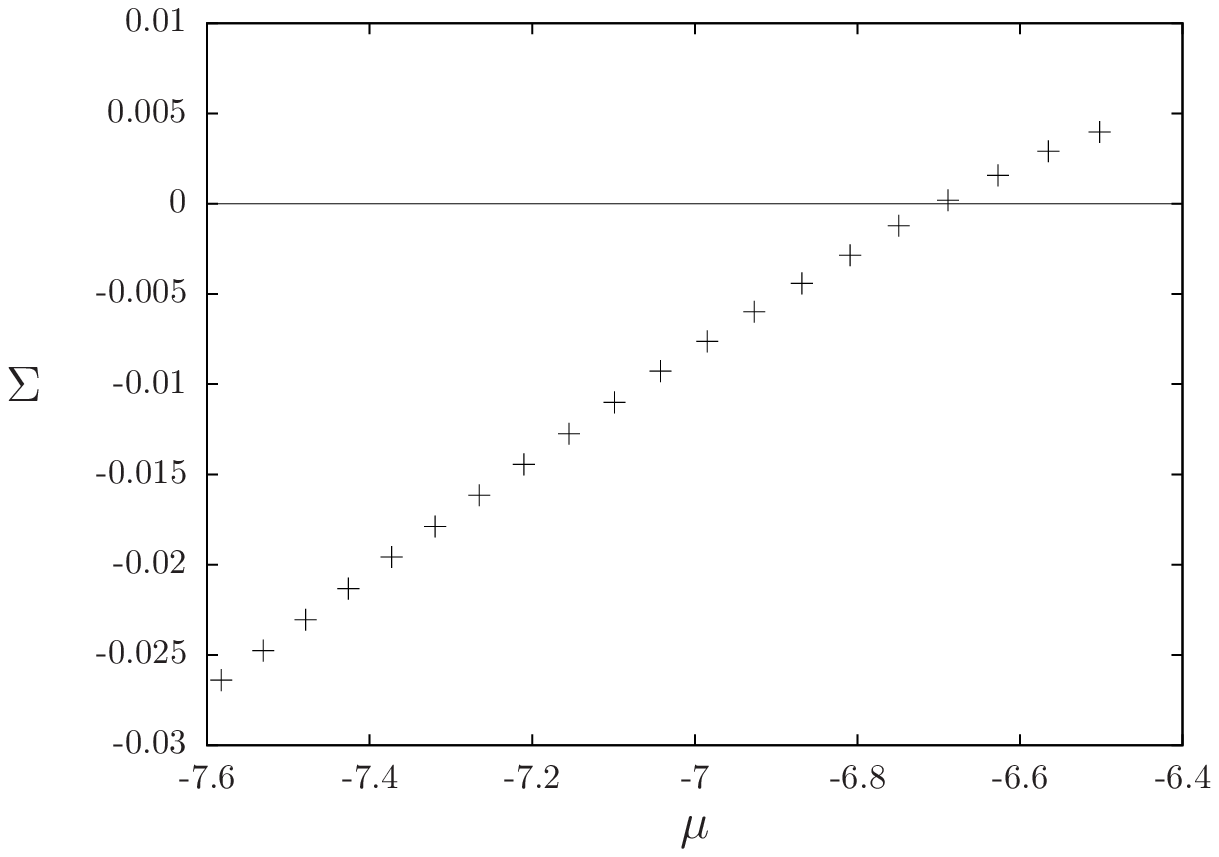}
\caption{The complexity at $m=1$ as a function of the chemical potential $\mu$, for $k=l=2$ and $T=1$. The function is defined for $\mu<\mud \approx -6.49$, the complexity being positive for $\mu > \muc \approx -6.69$.}
\label{fig_k2l2T1m1}
\end{figure}

The results of these investigations follow the usual pattern encountered in constraint satisfaction problems~\cite{KrMoRiSeZd}: for large enough values of $\mu$ (i.e. for dense enough initial configurations) there is no non-trivial solution of the 1RSB equation with $m=1$; decreasing $\mu$ a non-trivial solution appears discontinuously at a threshold $\mud$ (the ``dynamic'' transition). Its complexity (or configurational entropy) $\Sigma$ is positive in an interval $\mu \in [\muc,\mud]$, which thus corresponds to a ``dynamic 1RSB phase'' with an exponential number of clusters contributing to the Gibbs measure, see Fig.~\ref{fig_k2l2T1m1} for an illustration in the case $T=1$. The numerical values of $\mud$ and $\muc$ (as well as the associated densities of initially active sites $\td$ and $\tc$), can be found for several values of $T$ in the Tables \ref{tab_k2l2}, \ref{tab_k3l3} and \ref{tab_k3l2}. For the values of $\mu$ in the interval $[\muc,\mud]$ the thermodynamic predictions of the RS computations are correct. Note that in all the cases we investigated ($k=2,3$, $2\le l \le k$ and $T\le 5$) we always found a discontinuous transition with $\muc < \mud$; we cannot rule out the possibility that for other values of the parameters the replica symmetry breaking transition is continuous with $\muc=\mud$ (as happens for instance in the independent set problem at low degrees~\cite{is_japan}).

\begin{figure}
\includegraphics[width=8cm]{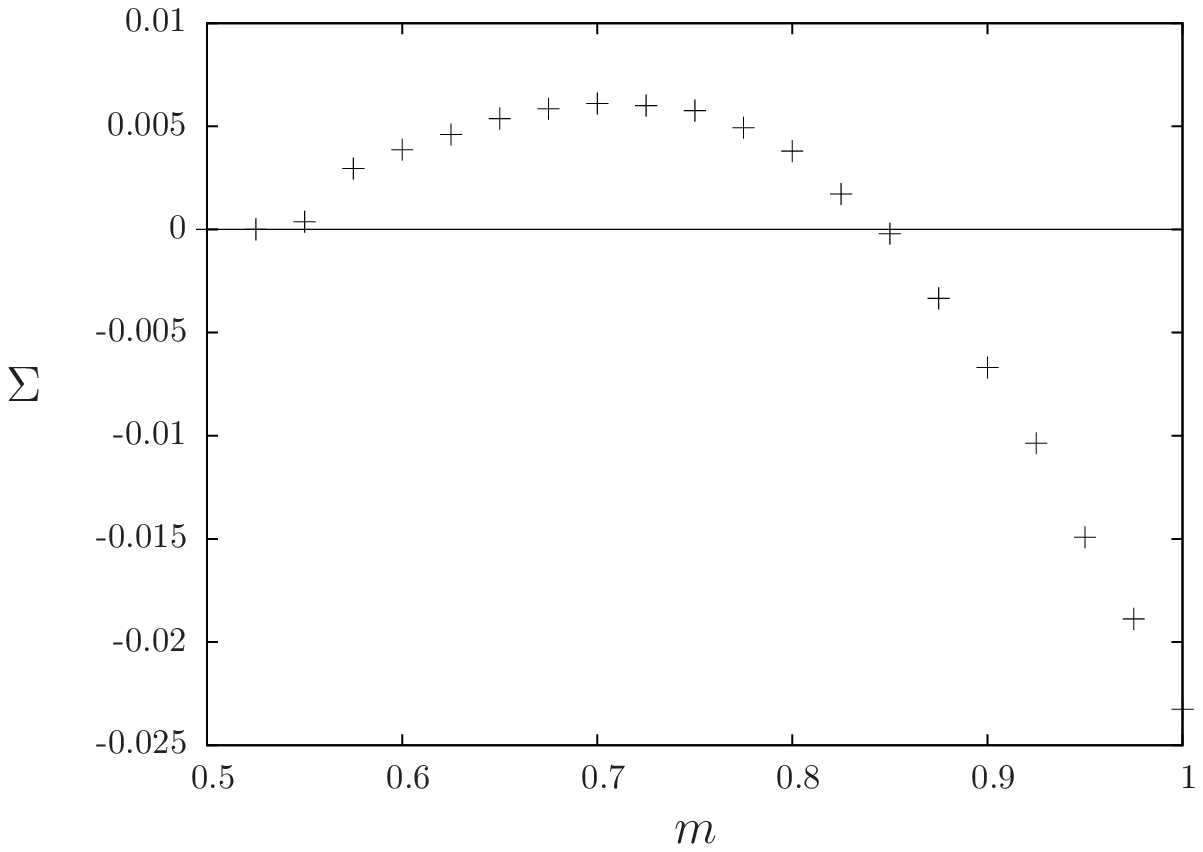}\hspace{1cm}
\includegraphics[width=8cm]{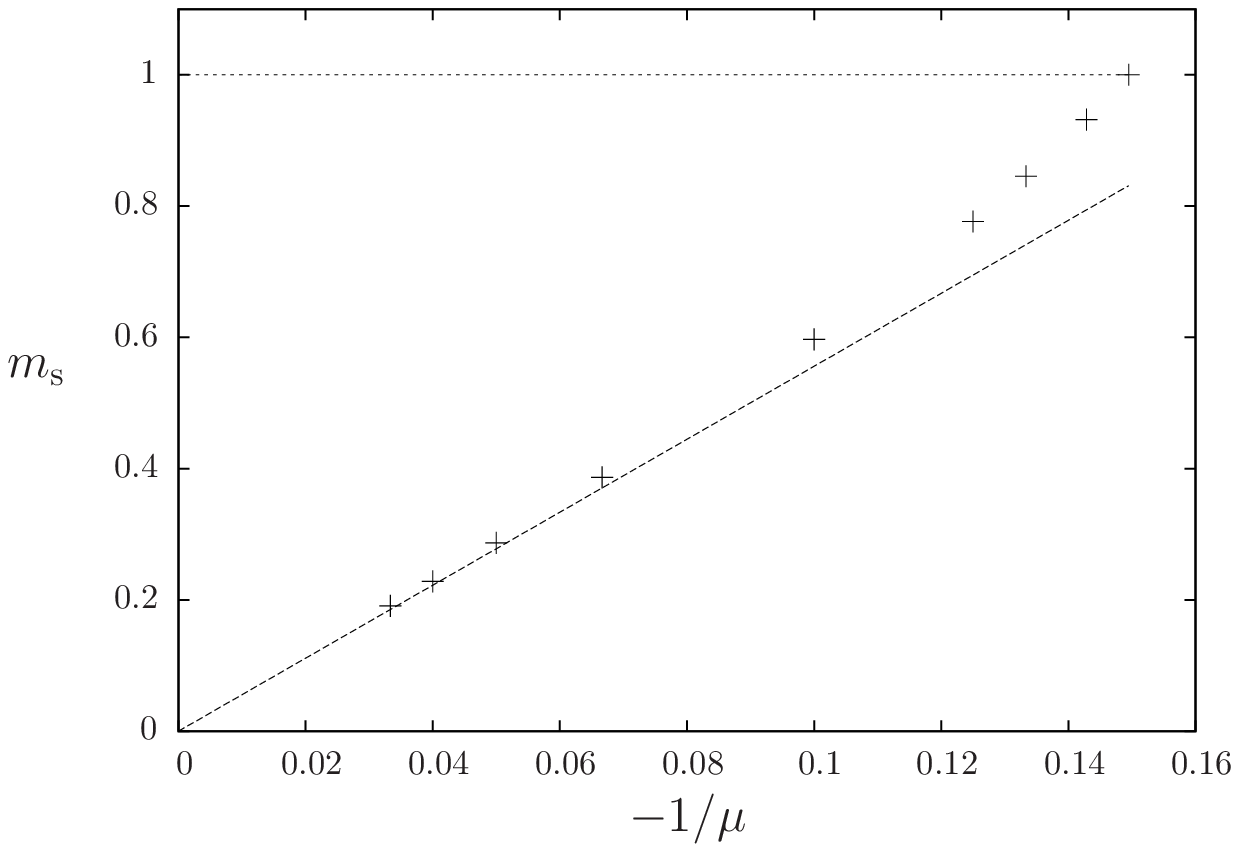}
\caption{Study of the condensed phase for $k=l=2$ and $T=1$.
 Left panel: complexity as a function of $m$ for $\mu=-7.5<\muc$, the complexity vanishes for $\ms \approx 0.84$. Right panel: Parisi parameter $\ms$ as a function of $-1/\mu$, departing from 1 for $\mu<\muc$; the dashed line corresponds to the linear behaviour $-\mu \ms=5.56$ that fits the $\mu \to -\infty$ limit.}
\label{fig_k2l2T1condensed}
\end{figure}

Lowering further the chemical potential, i.e. in the regime $\mu<\muc$, the complexity at $m=1$ becomes negative. This is thus a true replica symmetry breaking phase with only a sub-exponential number of clusters contributing to the Gibbs measure; $\muc$ corresponds to the ``condensation'' transition. In this phase the thermodynamic properties of the model differ from the RS prediction and are given by the properties of the clusters selected by the static value of the Parisi parameter, $\ms(\mu)$, for which the complexity vanishes. This value can be determined by computing the complexity as a function of $m$, for a fixed value of $\mu$, see left panel of Fig.~\ref{fig_k2l2T1condensed} for an example. 

To compute the minimal density $\tmin(T)$ one has to take the limit $\mu\to-\infty$; we have introduced above in Sec.~\ref{sec_1RSB_y} a simplifying ansatz in this limit, assuming in particular a finite value of $-\mu m$. To check the consistency of this ansatz we solved the complete 1RSB equations for $T=1$ and several values of $\mu$ large and negative. The Parisi parameter $\ms$ is plotted as a function of $-1/\mu$ in the right panel of Fig.~\ref{fig_k2l2T1condensed}; in the limit $\mu \to -\infty$ one obtains indeed a linear behaviour, corresponding to a finite limit of $-\mu \ms$.

\subsubsection{Energetic 1RSB results}

We turn now to the results obtained via the energetic 1RSB cavity method, i.e. taking simultaneously the limits $\mu \to -\infty$ and $m \to 0$ with a finite value for $y=-\mu m$. The equations to solve in this case amounts to find the fixed point of Eq.~(\ref{eq_G_1RSBy}), from which one obtains the 1RSB potential (\ref{eq_Phie_reg}) and the energetic complexity $\Sigma_{\rm e}(\theta)$ from the Legendre transform structure explained in (\ref{eq_Legendre_y}), as a parametric plot varying the parameter $y$. The computational complexity of this problem is drastically reduced compared to the complete 1RSB equations: as in the RS case one has a set of (roughly) $2T$ equations on $2T$ real unknowns, instead of an equation on a probability distribution of fields. More technical details on the procedure to solve these equations can be found in Appendix \ref{sec_app}.

Fig.~\ref{fig_1RSBy_k2l2_Sigma} displays the energetic complexity $\Sigma_{\rm e}(\theta)$ for a few values of $T$, in the cases $k=l=2$ and $k=3$, $l=2$. The expert reader will notice that we restricted the range of $y$ used in this plot to the so-called physical branch, in such a way that $\Sigma_{\rm e}$ is a concave function of $\theta$. The most important characteristics of these curves are the values of $\tmino$ where the complexity vanishes, and the corresponding values $\ys$ of the parameter $y$; these are reported for several values of $T$ in the last columns of the Tables \ref{tab_k2l2}, \ref{tab_k3l3} and \ref{tab_k3l2}. Indeed $\tmino$ is the 1RSB prediction for $\tmin$, as it corresponds to the smallest density of active sites in initial configurations belonging to clusters with a non-negative complexity. For $T=1$ these values can be successfully cross-checked with the results of the Biroli-M\'ezard model~\cite{bm}, and the parameter $\ys$ agrees with the fit of $-\mu \ms(\mu)$ in the limit $\mu \to -\infty$ obtained from the full 1RSB equations (cf. right panel of Fig.~\ref{fig_k2l2T1condensed}).

\begin{figure}
\includegraphics[width=8cm]{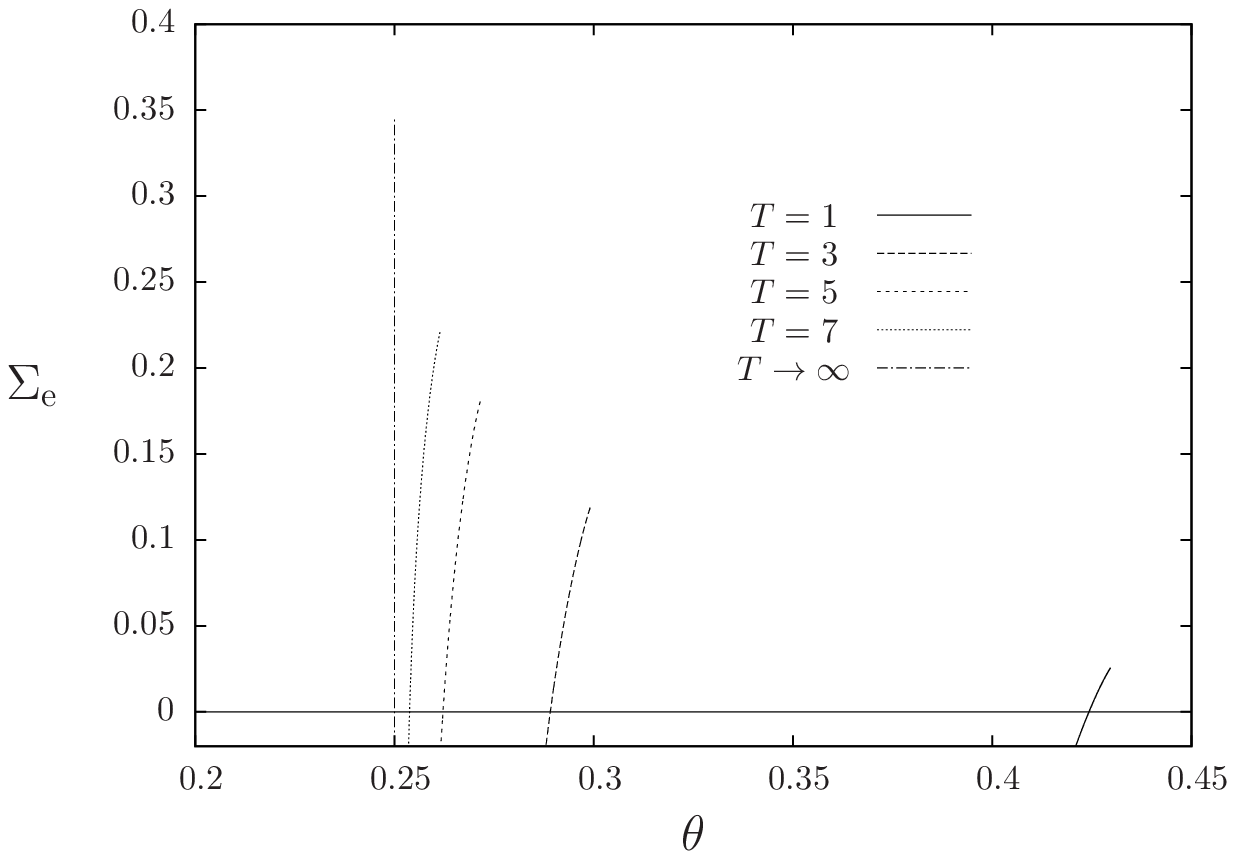}
\hspace{1cm}
\includegraphics[width=8cm]{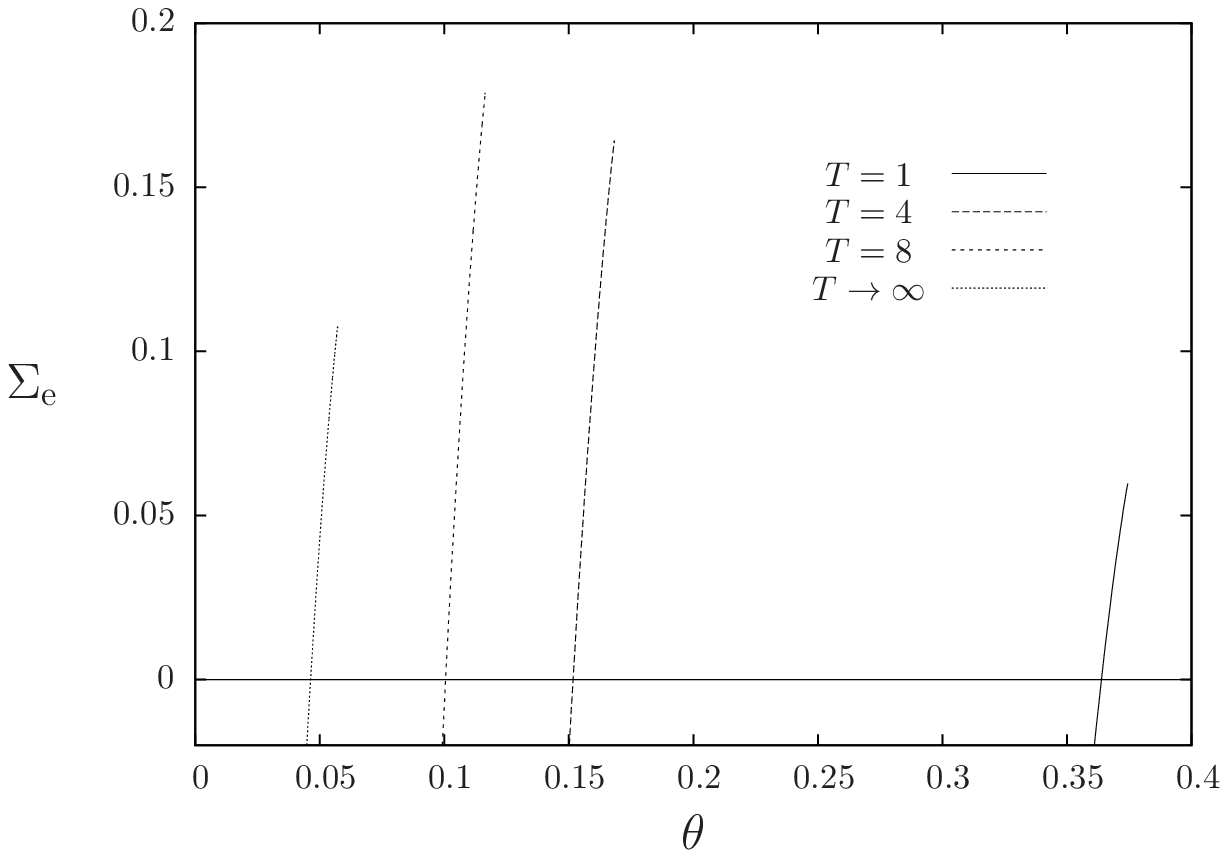}
\caption{The complexity $\Sigma_{\rm e}(\theta)$ obtained from the energetic 1RSB cavity formalism, for $k=l=2$ (left panel) and $k=3$, $l=2$ (right panel); see Sec.~\ref{sec_res_largeT} for explanations on the $T\to\infty$ result.}
\label{fig_1RSBy_k2l2_Sigma}
\end{figure}

\begin{table}
\begin{tabular}{|c||c|c||c|c|c|c||c|c|}
\hline
 & \multicolumn{2}{c||}{RS} & \multicolumn{4}{c||}{1RSB} & \multicolumn{2}{c|}{energetic 1RSB} \\
\hline
$T$  & $\mu_{s=0}$ & $\tminz$ & $\mud$ & $\td$ & $\muc$ & $\tc$ & $\ys$ & $\tmino$  \\
\hline
1 & -6.113951 & 0.479455 & -5.35 & 0.4906 & -5.39 & 0.4900 & 4.644980 & 0.482712 \\
\hline
2 & -8.175902 & 0.397326 & -7.38 & 0.4027 & -7.95 & 0.3988 & 7.485437 & 0.397922 \\
\hline
3 & -10.381917 & 0.366187 & -8.63 & 0.3725 & -10.33 & 0.3663 & 10.077681 & 0.366291 \\
\hline
4 & -13.140888 & 0.351221 & -9.59 & 0.3583 & -13.11 & 0.3513 & 13.037666 & 0.351234 \\
\hline
5 & -17.249334 & 0.343205 & -10.3 & 0.3507 & -17.36 & 0.3432 & 17.232334 & 0.343206 \\
\hline
6 & -24.322138 & 0.338721 & & & & & 24.321721 & 0.338721  \\
\hline
7 & -35.739653 & 0.336191 & & & & & 35.739653 & 0.336191 \\
\hline
8 & -54.198587 & 0.334760 & & & & & 54.198587 & 0.334760 \\
\hline
$\infty$ & $-\infty$ & $\frac{1}{3}$ & & & & & $+\infty$ & $\frac{1}{3}$ \\
\hline
\end{tabular}
\caption{Numerical results from the cavity computations at finite $T$ for $k=l=3$; the results in the limit $T\to\infty$ are explained in Sec.~\ref{sec_res_largeT}.}.
\label{tab_k3l3}
\end{table}

\begin{table}
\begin{tabular}{|c||c|c||c|c|c|c||c|c|}
\hline
 & \multicolumn{2}{c||}{RS} & \multicolumn{4}{c||}{1RSB} & \multicolumn{2}{c|}{energetic 1RSB} \\
\hline
$T$  & $\mu_{s=0}$ & $\tminz$ & $\mud$ & $\td$ & $\muc$ & $\tc$ & $\ys$ & $\tmino$  \\
\hline
1 & -7.730059 & 0.362794 & -7.06 & 0.3681 & -7.38 & 0.3654 & 6.778540 & 0.363813 \\
\hline
2 & -10.21534 & 0.236821 & -9.16 & 0.2416 & -10.12 & 0.2372 & 9.873120 & 0.237009 \\
\hline
3 & -11.90150 & 0.182272 & -10.38 & 0.1875 & -11.85 & 0.1824 & 11.72892 & 0.182338 \\
\hline
4 & -13.03158 & 0.151659 & -11.45 & 0.1563 & -13.00 & 0.1517 & 12.92114 & 0.151693\\
\hline
5 & -13.80059 & 0.132014 & -12.47 & 0.1354 &-13.78 & 0.1321 & 13.71834 & 0.132036\\
\hline
6 & -14.33193 & 0.118324 & & & & &         14.26439 & 0.118341\\
\hline
7 & -14.70251 & 0.108237 & & & & &         14.64332 & 0.108251 \\
\hline
8 & -14.96150 & 0.100498 & & & & &         14.90729 & 0.100510 \\
\hline
10 &-15.26375 & 0.089415& & & & &          15.21429 & 0.089425\\
\hline
15 & -15.42086 & 0.074242 & & & & &       15.37163 & 0.074251 \\ 
\hline
20 & -15.27922 & 0.066569 & & & & &       15.22489 & 0.066579 \\ 
\hline
30 & -14.85174 & 0.058995 & & & & &       14.78367 & 0.059008 \\ 
\hline
$\infty$ & -12.72072 & 0.046283 & & & & &  12.54796 & 0.046328 \\ 
\hline
\end{tabular}
\caption{Numerical results from the cavity computations at finite $T$ for $k=3$, $l=2$; the results in the limit $T\to\infty$ are explained in Sec.~\ref{sec_res_largeT}.}
\label{tab_k3l2}
\end{table}

\subsection{The large $T$ limit}
\label{sec_res_largeT}

The limit case $T\to\infty$ is particularly interesting as it corresponds to the original influence maximization problem with no constraint on the time taken to activate the whole graph. This limit can be performed analytically for the RS and energetic 1RSB formalism; the technical details of these computations can be found in Appendix \ref{sec_app_Tinfty}, we present here the results of these analytical simplifications. It turns out that the case $k=l$ is qualitatively different from the case $k>l$, we shall thus divide this section according to this distinction. 

\subsubsection{The case $k=l$}
\label{sec_kequall_Tinfty}

Let us first recall that when $k=l$ the dynamics from a random initial configuration of density $\theta$ has a continuous transition at $\tr(k,k)=\frac{k-1}{k}$ (see Sec.~\ref{sec_reminder_random}); we also saw in Sec.~\ref{sec_connections} that minimal contagious sets (with no constraint on the activation time) correspond to minimal decycling sets, which led to the bound $\tmin(k,k) \ge \frac{k-1}{2 k} =\frac{\tr(k,k)}{2}$. In the rest of this subsection we shall for simplicity abbreviate $\tr(k,k)$ by $\tr$.

As suggested by the left panel of Fig.~\ref{fig_RS_k2l2_softheta} in the case $k=l=2$, the RS entropy $s(\theta)$ converges to a limit curve when $T\to\infty$. This limit curve can actually be computed analytically for all $k$; we defer the details of the computation to App.~\ref{sec_app_Tinfty_klequal} and only state here the properties of this limit curve. For $\theta \ge \tr$ it coincides with the binary entropy function $-\theta \ln \theta - (1-\theta) \ln (1-\theta)$; this is a posteriori obvious. Indeed by definition of $\tr$ typical configurations in this density range do activate the whole graph, hence the number of activating initial configurations coincide (at the leading exponential order) with the total number of configurations of this density. A non-trivial portion of the limit curve arises in the density range $[\tr/2,\tr]$, where it is given by
\beq
s(\theta) = - \frac{k}{2} \left(2 \theta - \tr \right) \ln (2 \theta - \tr) + k \theta \ln \theta
+(1-\theta) \ln(k-1) - \frac{k+1}{2} \ln\left(\frac{k-1}{k} \right) \ .
\label{eq_s_klequal_Tinfty}
\eeq
This function has the same value and the same first derivative than the binary entropy function in $\tr$, while at the lower limit $\tr/2$ of its range of definition it has an infinite derivative with a finite value
\beq
s(\tr/2) = \ln k - \frac{k-1}{2k} \ln (k-1) - \frac{k-1}{2} \ln 2 \ .
\label{eq_s_trovertwo}
\eeq
The parametric plot of $s(\theta)$ also contains a vertical segment for $\theta =  \tr/2$, from $-\infty$ to the value given in (\ref{eq_s_trovertwo}).

The complexity $\Sigma_{\rm e}(\theta)$ of the energetic 1RSB formalism also converges to a limit curve when $T\to\infty$, as shown in Fig.~\ref{fig_1RSBy_k2l2_Sigma} and obtained analytically in App.~\ref{sec_app_Tinfty_klequal}. This limit curve has the same vertical segment in $\tr/2$ from $-\infty$ to the value (\ref{eq_s_trovertwo}); the non-trivial part of the curve is given in a parametrized form as follows:
\bea
\Sigma_{\rm e}(\tlambda)&=& \ln \Zsite(\tlambda) - \frac{k+1}{2} \ln \Zedge(\tlambda) - y(\tlambda) (1-\theta(\tlambda)) \ , \label{eq_1RSBy_klequal_Tinfty} \\
\theta(\tlambda) &=& 1 
- \frac{e^{y(\tlambda)}}{e^{y(\tlambda)}-1} \frac{\Zsite(\tlambda) -1}{\Zsite(\tlambda) }
- \frac{k+1}{2} \frac{1}{e^{y(\tlambda)}-1} \frac{1-\Zedge(\tlambda)}{\Zedge(\tlambda) } \ ,
\eea
where $\tlambda$ is the positive parameter along the curve, the Parisi parameter
\beq
y(\tlambda)= \ln \left(\frac{(1+\tlambda)^k - k \, \tlambda^{k-1} - \tlambda^k}{(k-1)\, \tlambda^k} \right) \ , 
\eeq
is the slope of the tangent to the curve $\Sigma_{\rm e}(\theta)$, and
\bea
\Zsite(\tlambda) &=& 1+ \frac{(k+1+\tlambda)((1+\tlambda)^{k-1}-k\, \tlambda^{k-1})}{(k-1)(1+\tlambda)^k} \ ,\\
\Zedge(\tlambda) &=&\frac{\tlambda}{1+\tlambda} \left(1+\frac{(1+\tlambda)^{k-1}-\tlambda^{k-1}}{(1+\tlambda)^k - k \, \tlambda^{k-1} - \tlambda^k} \right) \ .
\eea
When $\tlambda \to 0^+$ this part of the curve connects with the vertical segment in $\tr/2$. The large values of $\tlambda$ yield a non-concave branch of $\Sigma_{\rm e}$ that has to be discarded.

\begin{figure}
\begin{center}
\includegraphics[width=7cm]{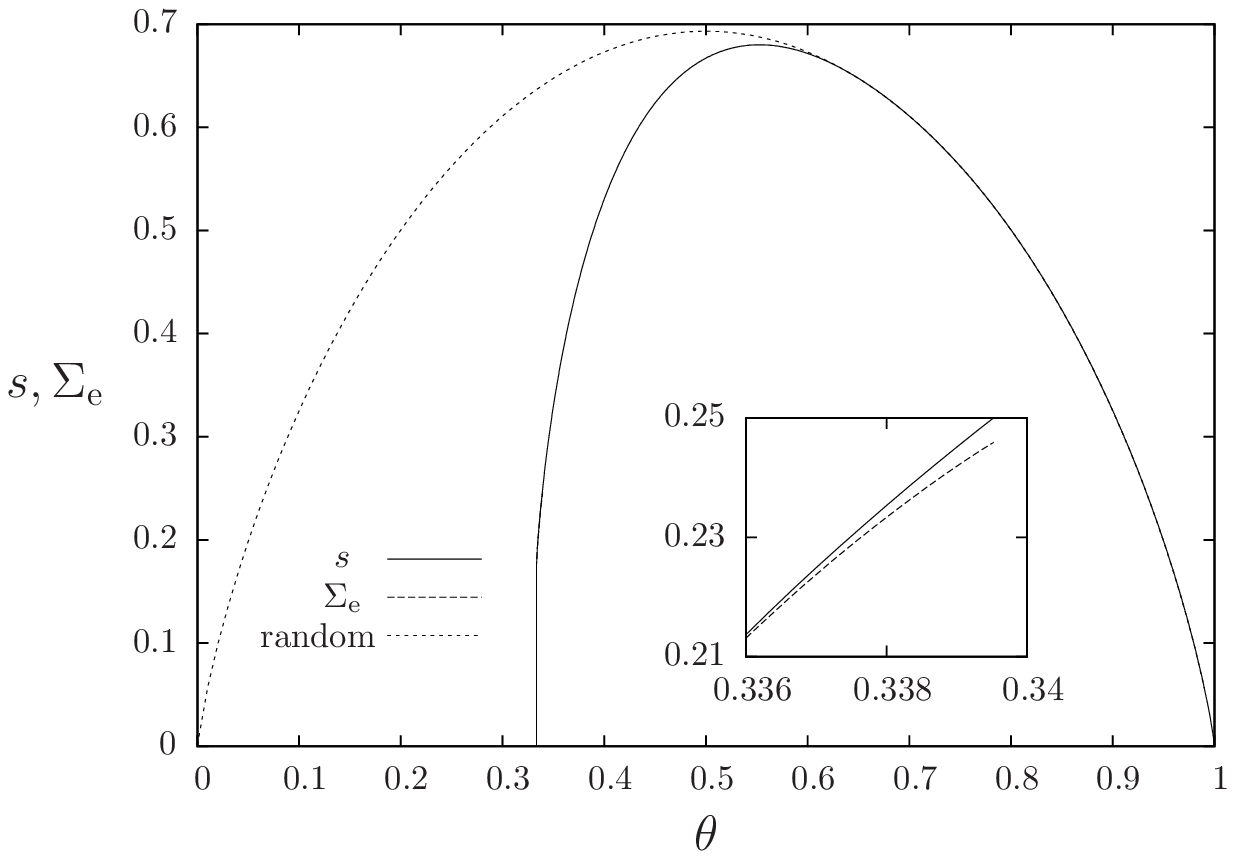}
\hspace{1cm}
\includegraphics[width=7cm]{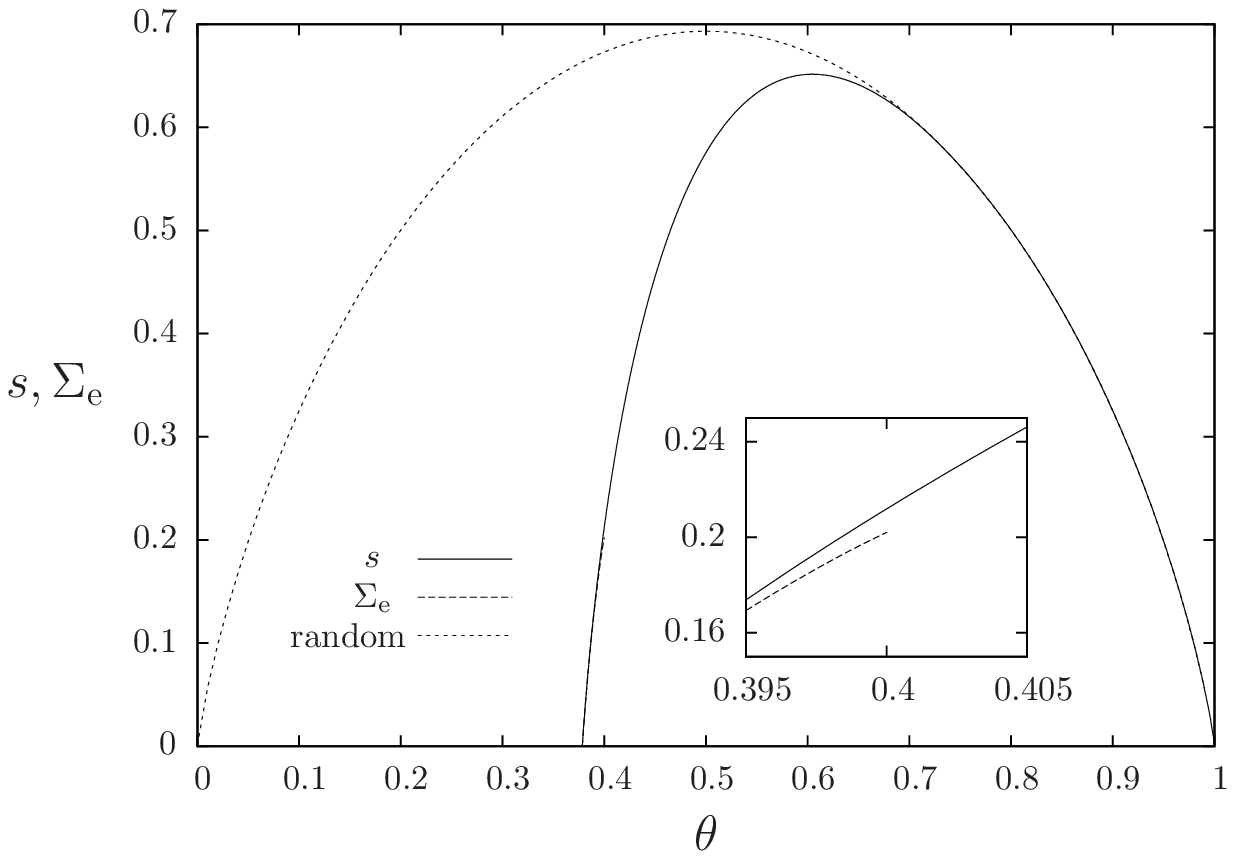}
\end{center}
\caption{The RS entropy $s(\theta)$ and energetic 1RSB complexity $\Sigma_{\rm e}(\theta)$ in the $T\to \infty$ limit, for $k=l=3$ (left panel) and $k=l=4$ (right panel). The binary entropy function is also plotted for comparison (the RS entropy coincides with it for $\theta \ge \tr$). The physical part of the complexity extends on a small range of $\theta$, on which it is only slightly smaller than the RS entropy, the inset allows to see this small difference at the end of the domain of definition of $\Sigma_{\rm e}$.} 
\label{fig_Tinfty_klequal}
\end{figure}

Depending on the value of $k$ qualitatively different behaviours emerge from the analysis of the RS entropy and 1RSB energetic complexity:
\begin{itemize}

\item
For $k=l=2$ the entropy of the endpoint in $\tr/2$ given in (\ref{eq_s_trovertwo}) is strictly positive (it is equal to $(\ln 2)/2$); moreover the energetic complexity curve converges, in the $T\to\infty$ limit, to a vertical segment (the non-trivial part parametrized by $\tlambda$ is convex and has thus to be discarded). This leads to the conclusion that $\tmin=\tr/2=1/4$ in this case, saturating the lowerbound of (\ref{eq_bound_kk}), and recovering the rigorous result of~\cite{decycling} on the decycling number of 3-regular graphs. This is a reassuring evidence in favour of the validity of the approach, in particular on the interversion of the $T\to\infty$ and $N\to\infty$ limit. It would be an even more challenging computation to determine the limit of $\td$ and $\tc$ as $T$ diverges; we are however tempted to conjecture that they both go to $1/4$ and that the effects of replica symmetry breaking are irrelevant in this limit. A numerical argument in favour of this conjecture will be presented in Sec.~\ref{sec_single_sample}, where it is shown that a simple greedy algorithm is able to find contagious sets of these densities. Assuming this is true, the expression (\ref{eq_s_klequal_Tinfty}) would give for $k=2$ the typical (quenched) entropy of the decycling sets of 3-regular random graphs in their non-trivial regime of densities $[1/4,1/2]$. Note that the coincidence of the RS entropy and 1RSB energetic complexity at $\tmin$ is reminiscent of the phenomenology discussed for the matching problem in~\cite{ZdMe_matchings}, which might suggest that the minimal density activating configurations are at a large Hamming distance in configuration space one from the other.

\item
For $k=l=3$ the expression (\ref{eq_s_trovertwo}) of the entropy in $\tr/2$ is still positive (equal to $\ln 3 - (4/3) \ln 2$), hence the endpoint of the non-trivial part of both the RS entropy and the 1RSB complexity curves occurs in $\tminz=\tmino=\tr/2=1/3$, saturating again the bound (\ref{eq_bound_kk}). This leads to the conclusion that $\tmin=1/3$ in this case, as was also conjectured in~\cite{decycling}. However, at variance with the previous case, the energetic complexity curve has a non-trivial part for $\theta > \tmin$, as shown in the left panel of Fig.~\ref{fig_Tinfty_klequal}. We thus expect that the limits of $\td$ and $\tc$ when $T\to\infty$ are strictly greater than $1/3$, hence that simple algorithms would have difficulties to find the minimal contagious sets (see Sec.~\ref{sec_single_sample} for a numerical check of this statement), and that the RS entropy (\ref{eq_s_klequal_Tinfty}) is incorrect for some regime of densities close to $1/3$.

\item 
Finally when $k=l\ge 4$ the entropy in (\ref{eq_s_trovertwo}) is negative, the cancellation of $s$ occurs at a value $\tminz$ strictly between $\tr/2$ and $\tr$, see the right panel of Fig.~\ref{fig_Tinfty_klequal}. The energetic complexity vanishes on its non-trivial part parametrized by $\tlambda$, at a value $\tmino$ slightly larger than $\tminz$, see Table~\ref{table_res_Tinfty} for some numerical values. Whether $\tmino$ should be taken as a conjectured exact value for $\tmin$ or simply as a lowerbound is dubious and might depend on the value of $k$. Indeed one should test the stability of the 1RSB ansatz against further levels of replica symmetry breaking. This computation is in principle doable along the lines of~\cite{stab1,stab2,bm}, but has not been performed yet. It is however relatively easy to set up an asymptotic expansion at large $k$ of the thresholds $\tminz$ and $\tmino$ from the expressions (\ref{eq_s_klequal_Tinfty},\ref{eq_1RSBy_klequal_Tinfty}). One finds that the first terms of the expansion are equal at the RS and 1RSB level, it is thus natural to conjecture that they are indeed the correct expansion of $\tmin$, namely
\beq
\tmin(k,k) = 1- \frac{2 \ln k}{k} - \frac{2}{k} + O\left(\frac{1}{k \ln k} \right) \ .
\eeq
This conjecture is in agreement with the rigorous lowerbound proven in~\cite{lb_klequal_rig},
\beq
\tmin(k,k) \ge 1- \frac{2 \ln k}{k} - \frac{4-2\ln 2}{k} + o\left(\frac{1}{k} \right) \ .
\eeq
It can also be compared with the asymptotic expansion in the case $l=k+1$~\cite{is_Frieze} where the inactive vertices have to form an independent set of the graph:
\beq
\tmin(k,k+1) = 1- \frac{2 \ln k}{k} + \frac{2 \ln\ln k}{k} + \frac{2 \ln 2 -2}{k} + o\left(\frac{1}{k} \right) \ .
\eeq
The third term of this expansion is of a larger order; indeed the condition imposed on the graph induced by the inactive vertices is much more stringent when $l=k+1$ (it has to be made of isolated vertices) with respect to the case $l=k$ (it only has to be acyclic).

\end{itemize}

Let us mention at this point that $\tmin(T)$, the minimal density of initial configuration percolating within $T$ steps of the dynamics, reaches its asymptotic value $\tmin$ as $T\to\infty$ with different finite $T$ corrections in the various cases listed above. The analysis of App.~\ref{sec_app_Tinfty_klequal} shows that for $k=l=2$ (resp. $k=l=3$) these corrections are of order $2^{-T}$ (resp. $3^{-T}$), which is in agreement with a numerical fit of the data in Table \ref{tab_k2l2} (resp. Table \ref{tab_k3l3}). On the contrary for $k=l\ge 4$ these corrections are only polynomially small in $T$. 

\begin{figure}
\includegraphics[width=8cm]{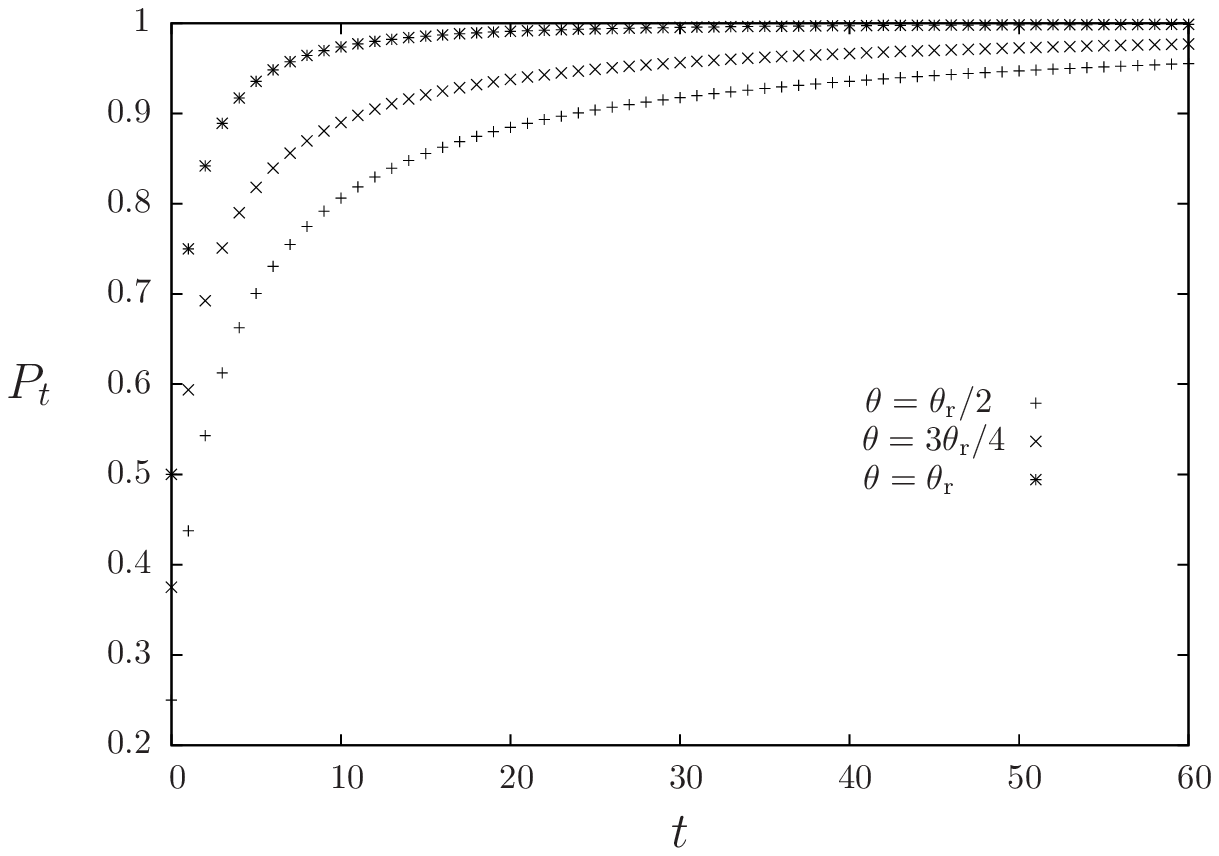}
\caption{The integrated distribution of activation times (\ref{eq_Pt_klequal_Tinfty}) for percolating initial conditions of density $\theta \in [\tr/2,\tr]$. The curves are presented in the case $k=l=2$.}
\label{fig_k2l2_tact_analytic}
\end{figure}

Finally, we could also compute analytically the distribution of activation times, within the RS formalism, for the initial configurations with a non-trivial density $\theta$ of active vertices in the interval $[\tr/2,\tr]$. Their cumulative distribution function $P_t=\eta(t_i\le t)$ obtained from the marginals of the law (\ref{eq_eta_ust}) reads in the $T\to\infty$ limit with $t$ kept fixed:
\beq
P_{t+1}=\theta + \frac{(2\theta-\tr)(1-\tr)}{\tr} w_t^{k+1} + (1-\tr)(k+1) w_t^k \left(\frac{\theta}{\tr}-\frac{2\theta-\tr}{\tr} w_t \right) \ ,
\label{eq_Pt_klequal_Tinfty}
\eeq
where $w_t$ is a series defined recursively by
\beq
w_0 = \tr \ , \qquad w_{t+1}=\tr + (1-\tr) w_t^k \ .
\label{eq_wt}
\eeq
Examples of this cumulative distribution are displayed in Fig.~\ref{fig_k2l2_tact_analytic}. As explained above the predictions of the RS cavity method are not expected to be correct for $\theta < \tc$; in the particular case $k=l=2$ we however expect this result to be true down to $\theta=\tmin=1/4$. Note that $P_t$ goes to $1$ when $t\to\infty$, in other words in the limit $T\to\infty$ the support of the distribution of activation times does not scale with $T$ and remains of order 1. One can also check that when $\theta=\tr$, the prediction $P_t$ of (\ref{eq_Pt_klequal_Tinfty}) coincides, as it should, with the distribution of activation times for random initial conditions of density $\tr$ given in Eq.~(\ref{eq_random_x}); to see this one can notice that $w_t$ is equal to the series $\tx_t$ defined in Eq.~(\ref{eq_random_tx}) for the study of random initial conditions, when $k=l$ and $\theta=\tr$. At the lower limit of the interval of density, $\theta=\tr/2$, one obtains instead a simple expression,
\beq
P_{t+1} = \frac{\tr}{2} + (k+1) \frac{1-\tr}{2} w_t^k \ .
\label{eq_Pt_trsudue}
\eeq
A straightforward analysis of (\ref{eq_Pt_klequal_Tinfty},\ref{eq_wt}) reveals that for all $\theta < \tr$ the cumulative distribution $P_t$ reaches 1 with corrections of order $1/t$, in other words the probability $P_t-P_{t-1}$ that a vertex activates precisely at time $t$ has a power-law tail with exponent $-2$. On the contrary the random initial conditions of density $\tr$ have $1-P_t$ of order $1/t^2$, hence the exponent of the tail is $-3$; random initial conditions with $\theta>\tr$ have instead an exponentially decaying tail for their distribution of activation times.

\subsubsection{The case $k>l$}
\label{sec_lmk_Tinfty}

We shall now turn to a description of the limit as $T\to\infty$ of the RS and energetic 1RSB results when $k>l$, with again the technical details relegated in the Appendix \ref{sec_app_Tinfty_lmk}.
The RS entropy $s(\theta)$ coincides with the binary entropy function for $\theta \ge \tr$, for exactly the same reasons as explained above in the case $k=l$ (here and in the rest of this subsection we denote $\tr$ the threshold $\tr(k,l)$). The non-trivial part of $s(\theta)$ and $\Sigma_{\rm e}(\theta)$ are obtained in a parametric way, with unfortunately rather long expressions that we shall now progressively describe. We keep implicit below the dependency of all quantities on $k$ and $l$ when there is no risk of confusion.

This parametrization is given in terms of a real $\lambda$ in the range $]0,\lambda_{\rm r}]$, where this upper limit is expressed in terms of the threshold $\tr$ for activation from a random initial condition as $\lambda_{\rm r}=(1-\tr)\theta_{\rm r}^{k-1}$. We need first to introduce some auxiliary functions $\hu(\lambda)$, $\hv(\lambda)$, $u_*(\lambda)$ and $v_*(\lambda)$. The first two are given explicitly as
\beq
\hu(\lambda) = \left(\frac{1-\tr}{\lambda}\right)^{\frac{1}{k-1}} \ , \qquad
\hv(\lambda) = \txr \left(\frac{1-\tr}{\lambda}\right)^{\frac{1}{k-1}} \ ,
\label{eq_uvhat}
\eeq
where we recall that $\txr$ is the fixed-point of Eq.~(\ref{eq_random_tx}) at the bifurcation $\tr$, see also (\ref{eq_tandx_r}). The last one, $v_*(\lambda)$, is defined as the smallest positive solution of
\beq
v=1+\lambda \sum_{p=l}^k \binom{k}{p} \left(\lambda l \binom{k}{l} \right)^{-\frac{k-p}{k-l}} v^{\frac{p(k-1)-k(l-1)}{k-l}}  \ , \qquad 
\label{eq_vstar}
\eeq
then $u_*(\lambda)$ can be deduced as the solution of
\beq
1 = \lambda l \binom{k}{l} v_*(\lambda)^{l-1} (u_*(\lambda)-v_*(\lambda))^{k-l} 
\qquad \text{with} \ \ u_*(\lambda)\ge v_*(\lambda) \ .
\label{eq_ustar}
\eeq
One can check that $u_*(\lambda) \ge \hu(\lambda) \ge \hv(\lambda) \ge v_*(\lambda)$ on the interval $\lambda \in ]0,\lambda_{\rm r}]$, and that $u_*=\hu=1/\tr$ and $v_*=\hv=\txr/\tr$ in $\lambda=\lambda_{\rm r}$. We then define two functions $\Fsite(\lambda)$ and $\Fedge(\lambda)$ through
\bea
\Fsite(\lambda) &=& \frac{\lambda}{u_*}\left[
\hu^{k+1} + (k+1) \sum_{p=l}^k \binom{k}{p} \left[
\frac{l-1}{k-l} I_{p-1} - I_p \right]\right]\label{eq_Fsite_lmk} \\
\Fedge(\lambda) &=& \frac{1}{u_*}\left[(\hu-\hv)^2 + 2 u_* v_* - v_*^2 + 2 \lambda l \binom{k}{l}  I_{l-1} \right] \label{eq_Fedge_lmk}
\eea
where for clarity we kept implicit the $\lambda$ dependency of $\hu,\hv,u_*$ and $v_*$, and we introduced
\bea
I_p &=& \left(\lambda l \binom{k}{l} \right)^{-\frac{k-p}{k-l}}
\int_{v_*}^{\hv} \dd v \ v^{\frac{p(k-1)-k(l-1)}{k-l}}  \ , \\
&=& \left(\lambda l \binom{k}{l} \right)^{-\frac{k-p}{k-l}} 
\times
\begin{cases}
\ln \left(\frac{\hv}{v_*} \right) \qquad \text{if} \ p=l-1 \ \text{and} \ k=2l-1 \ , & \\
\frac{k-l}{(p+1)(k-1)-(k+1)(l-1)} \left( \hv^{\frac{(p+1)(k-1)-(k+1)(l-1)}{k-l}} - v_*^{\frac{(p+1)(k-1)-(k+1)(l-1)}{k-l}} \right) & \text{otherwise} \ .
\end{cases} \nonumber
\eea
We can finally give the parametric form of the RS entropy $s(\theta)$:
\bea
s(\lambda) &=& \ln(1+\Fsite(\lambda)) - \frac{k+1}{2} \ln \left(\frac{\Fedge(\lambda)}{u_*(\lambda)} \right) +\mu(\lambda) (1-\theta(\lambda))  \ , \nonumber \\
\theta(\lambda) &=& \frac{1}{1+\Fsite(\lambda)} \ , \nonumber \\
\mu(\lambda) &=& - \ln(\lambda \, u_*(\lambda)^k)\ ,
\label{eq_lmk_Tinfty_RS}
\eea
where $\mu(\lambda)$ is the opposite of the derivative of $s(\theta)$ in the point $\theta(\lambda)$. Thanks to the values $\hu,\hv,u_*$ and $v_*$ assume in $\lambda_{\rm r}$ this curve joins the binary entropy function in $\tr$ with a continuous slope.

Similarly the 1RSB entropic complexity $\Sigma_{\rm e}(\theta)$ is obtained parametrically as
\bea
\Sigma_{\rm e}(\lambda) &=& \ln \left(1+\left(1-\frac{1}{\lambda \, u_*(\lambda)^{k-1}} \right) \Fsite(\lambda) \right) 
-\frac{k+1}{2} \ln \left(\frac{1+ (\lambda \, u_*(\lambda)^{k-1} - 1)\, \Fedge(\lambda)}{\lambda \, u_*(\lambda)^k - u_*(\lambda) +1} \right) - y(\lambda) (1-\theta(\lambda))
\ , \nonumber \\
\theta(\lambda) &=& 
\frac{1-\frac{1}{\lambda \, u_*(\lambda)^k} \Fsite(\lambda)}{1 + \left(1 - \frac{1}{\lambda \, u_*(\lambda)^{k-1}} \right) \Fsite(\lambda)} 
- \frac{k+1}{2} \frac{1-\frac{1}{u_*(\lambda)} \Fedge(\lambda)}{1+(\lambda\, u_*(\lambda)^{k-1}-1) \Fedge(\lambda)}
\ , \nonumber \\
y(\lambda) &=& \ln(\lambda \, u_*(\lambda)^k - u_*(\lambda) +1 ) \ ,
\label{eq_lmk_Tinfty_1RSB}
\eea
with $y(\lambda)$ giving the slope of the tangent of $\Sigma_{\rm e}(\theta)$ in the point $\theta(\lambda)$.

\begin{figure}
\begin{center}
\includegraphics[width=8cm]{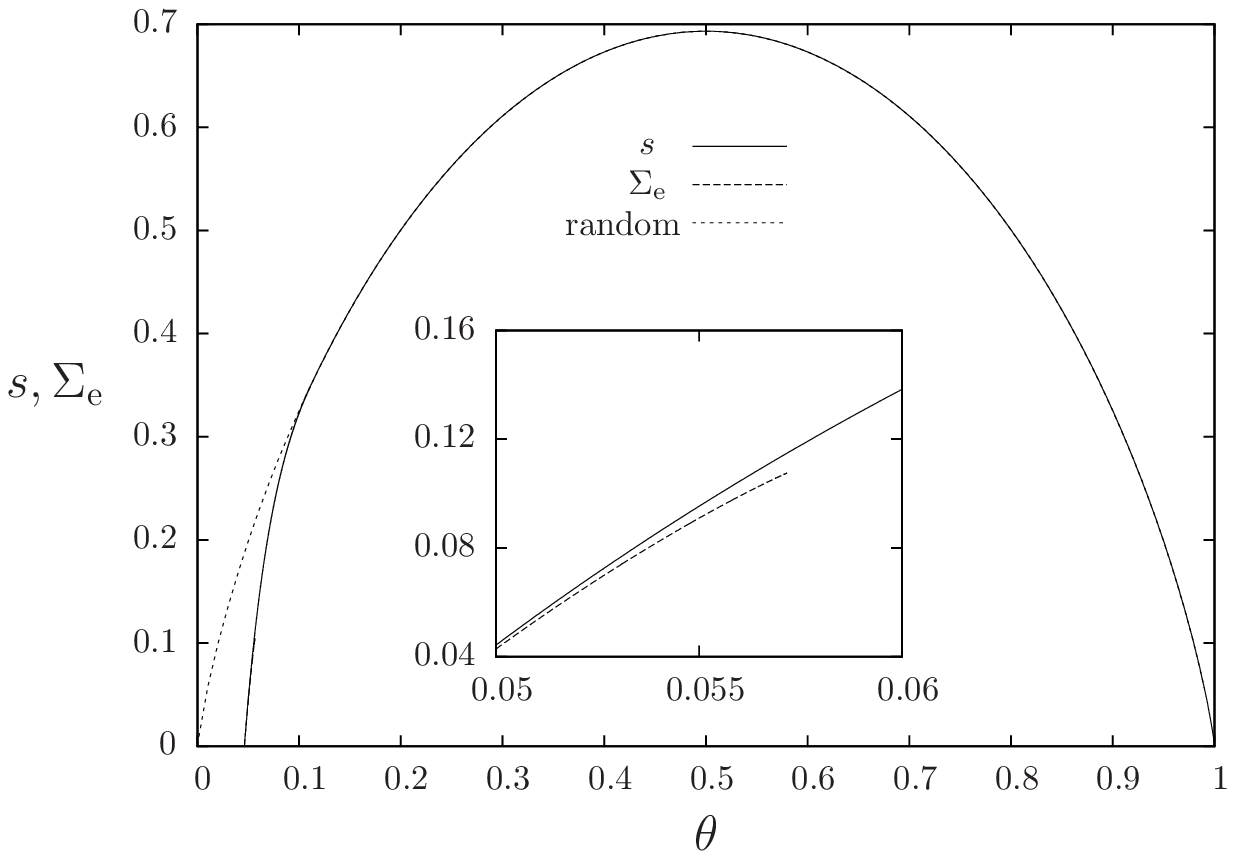}
\end{center}
\caption{The RS entropy $s(\theta)$ and energetic 1RSB complexity $\Sigma_{\rm e}(\theta)$ in the $T\to \infty$ limit, for $k=3$, $l=2$, from the analytical formulas given in (\ref{eq_lmk_Tinfty_RS},\ref{eq_lmk_Tinfty_1RSB}).}
\label{fig_Tinf_k3l2}
\end{figure}

An example of the limit for the RS entropy can be found in the right panel of Fig.~\ref{fig_RS_k2l2_softheta} for $k=3$, $l=2$, along with some finite $T$ curves, and a similar plot for the energetic complexity is displayed in the right panel of Fig.~\ref{fig_1RSBy_k2l2_Sigma}. The entropy and energetic complexity for this case in the limit are compared in Fig.~\ref{fig_Tinf_k3l2}. The values $\tminz$ and $\tmino$ where $s(\theta)$ and $\Sigma_{\rm e}(\theta)$ vanish are easily determined numerically from the above representation, and are collected in Table~\ref{table_res_Tinfty} for various values of $k$ and $l$. For most of the cases one finds $\tmino$ to be slightly larger than $\tminz$; as explained above the exactness of this 1RSB prediction has still to be assessed from a computation of the stability with respect to further replica symmetry breaking.

\begin{table}
\begin{tabular}{|c|c|c||c|c||c|c|}
\hline
$k$ & $l$ & $\tr$ & $\mu_{s=0}$ & $\tminz$ & $y_{\rm s}$ & $\tmino$ \\
\hline
$2$ & $2$ & $\frac{1}{2}$ & $-\infty$ & $\frac{1}{4}$ & $\infty$ &$\frac{1}{4}$ \\ 
\hline
\hline
$3$ & $2$ & $0.111111$  & -12.720727 & 0.046283 & 12.547960 & 0.046328 \\ 
\hline
$3$ & $3$ & $\frac{2}{3}$ &  $-\infty$ & $\frac{1}{3}$ & $\infty$ & $\frac{1}{3}$  \\ 
\hline
\hline
4 & 2 & 0.050781 & -9.633812 & 0.013108 & 9.125975 &  0.013258\\
\hline
4 & 3 & 0.275158  &  $-\infty$  & $\frac{1}{6}$ & $\infty$ & $\frac{1}{6}$  \\ 
\hline
$4$ & $4$ & $\frac{3}{4}$  & -14.904539 & 0.378463 &  14.883293 & 0.378465  \\ 
\hline
\hline
5 & 2 & 0.029096  & -9.499859 & 0.005715   & 8.891066 & 0.005820  \\
\hline
5 & 3 & 0.165116  & -12.395257 & 0.076228  & 12.333754 & 0.076247  \\
\hline
5 & 4 & 0.397212  & $-\infty$ & $\frac{1}{4}$  & $\infty$ & $\frac{1}{4}$  \\
\hline
$5$ & $5$ & $\frac{4}{5}$ & -9.786306 & 0.422619    & 9.647302 & 0.422695  \\ 
\hline
\hline
6 & 2 & 0.018854  & -9.675930 & 0.003098   & 9.026488 & 0.003166  \\
\hline
6 & 3 & 0.112870  & -10.396651 & 0.042825   & 10.234248 & 0.042894  \\
\hline
6 & 4 & 0.269022  & -16.484079 & 0.150054   & 16.480311 & 0.150055  \\
\hline
6 & 5 & 0.486312  & -40.532392 & 0.300090  & 40.532392 & 0.300090  \\
\hline
$6$ & $6$ & $\frac{5}{6}$ & -8.403727 & 0.460014  & 8.191036 & 0.460228  \\ 
\hline
\end{tabular}
\caption{The predictions of the RS and energetic 1RSB cavity method in the $T\to \infty$ limit.}
\label{table_res_Tinfty}
\end{table}

There are however two special cases which stand on a different footing, namely $(k,l)=(4,3)$ and $(k,l)=(5,4)$. Indeed in these two cases one has the same phenomenology than for $k=l=3$, namely a coincidence of $\tminz$ and $\tmino$ due to a vertical segment in the curves $s(\theta)$ and $\Sigma_{\rm e}(\theta)$ extending to positive values. This phenomenon can be understood by studying the limit $\lambda \to 0$ of the above representation of these curves. After some algebra one finds indeed that for $k<2l-1$,
\beq
\underset{\lambda \to 0}{\lim} \ \theta(\lambda) = \frac{2l-k-1}{2l} \ , \qquad 
\underset{\lambda \to 0}{\lim} \ s(\lambda)
=\underset{\lambda \to 0}{\lim}\ \Sigma_{\rm e}(\lambda) = 
\frac{k+1}{2l} \ln \left( \frac{l^l}{(l-1)^{l-1}} \binom{k}{l} \right)
- \frac{k-1}{2} \ln \left( \frac{2l}{2l-k-1}\right) \ ,
\eeq 
the limiting value for $\theta$ being valid both for the RS (\ref{eq_lmk_Tinfty_RS}) and 1RSB (\ref{eq_lmk_Tinfty_1RSB}) expressions. It turns out that for $k=4$, $l=3$ and $k=5,l=4$, the latter expression for the entropy $s$ and complexity $\Sigma_{\rm e}$ is strictly positive, hence the simple predictions $1/6$ and $1/4$ for $\tmin$ in these two cases respectively, that saturate the lowerbound of (\ref{eq_bound_kl}). We did not find any other values of $k,l$ that produce the same phenomenon.

\begin{figure}
\includegraphics[width=8cm]{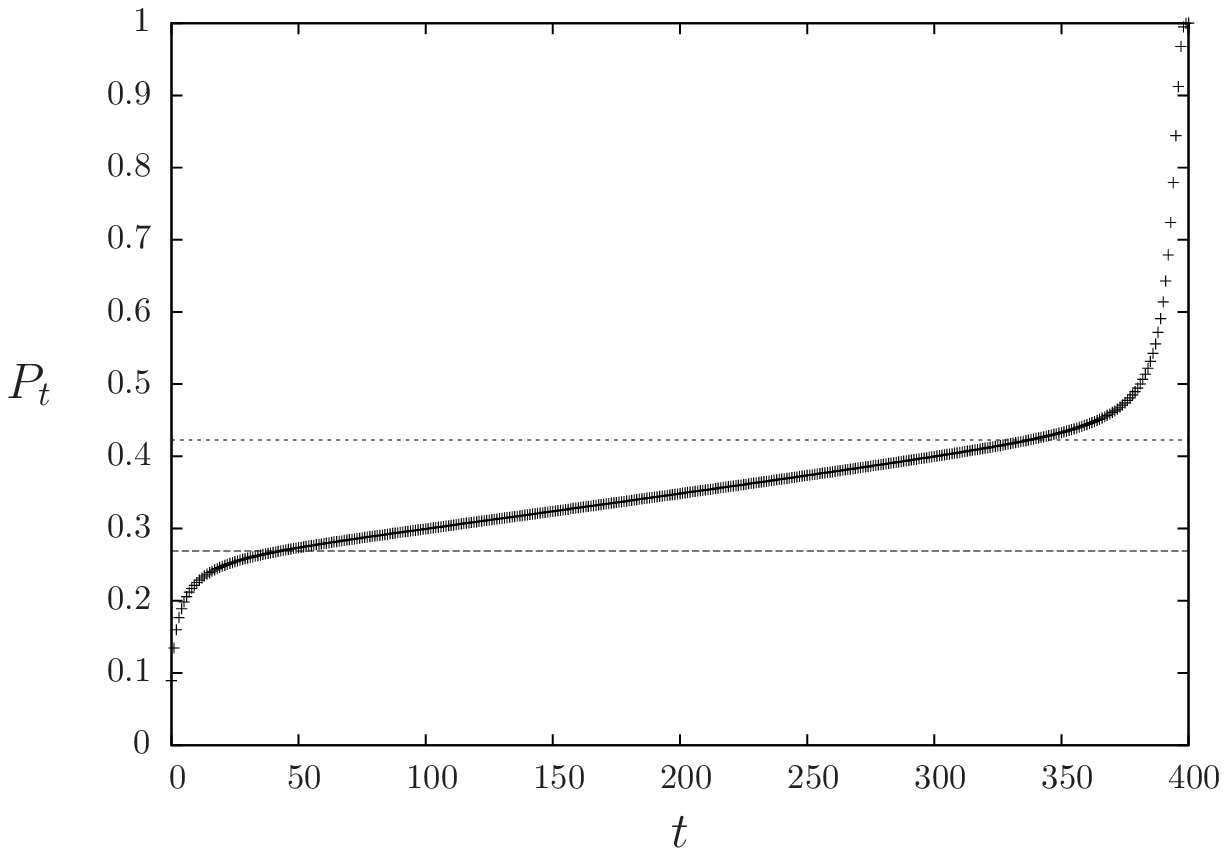}
\caption{An example of the cumulative distribution of activation times for $k=3$, $l=2$, obtained with the numerical resolution of the RS equations for a large but finite value of $T=400$, with a parameter $\lambda=0.005$, corresponding to an initial density of active sites of $0.089$. The two horizontal lines corresponds to $P(s=0^+)$ and $P(s=1^-)$ from Eq.~(\ref{eq_Pslimits}), delimiting the fraction of vertices that activate within a finite time after the beginning of the process (resp. before its end).}
\label{fig_k3l2_tact}
\end{figure}

Finally the distribution of activation times in the RS formalism exhibits a very different pattern with respect to the case $k=l$ (see Fig.~\ref{fig_k3l2_tact} for an illustration). As a matter of fact, in the limit $T\to\infty$ the activation times $t$ of the vertices have to be divided in three categories, each of them comprising a finite fraction of the $N$ vertices: (i) $t=O(1)$ (ii) $t=O(T)$ (iii) $t=T-O(1)$. The category (ii) of vertices can be described by a scaling function for the cumulative distribution, $P(s)=P_{t=sT}$, with $s\in]0,1[$ a reduced time. One has $P(s=0^+)>0$ and $1-P(s=1^-)>0$, these two numbers representing the fractions of vertices of type (i) and (iii) respectively. They can be computed following the techniques of the Appendix \ref{sec_app_Tinfty_lmk}, yielding for initial configurations with a fraction $\theta(\lambda) < \tr$ of active vertices:
\bea
P(s=0^+)&=& \theta + \theta \frac{\lambda}{u_*} \sum_{p=l}^{k+1} \binom{k+1}{p} v_*^p\, (u_*-v_*)^{k+1-p} \ , \nonumber \\
1-P(s=1^-) &=& \theta \frac{\lambda}{u_*} \left(\frac{1-\tr}{\lambda} \right)^{\frac{k+1}{k-1}}\left[1-\sum_{p=l}^{k+1} \binom{k+1}{p} \txr^p \, (1-\txr)^{k+1-p} \right] \ .
\label{eq_Pslimits}
\eea

\section{Algorithmic results}
\label{sec_single_sample}

We shall present in this Section the results of numerical experiments performed on finite size random regular graphs, for which we have constructed explicitly some activating initial configurations. We have used two strategies to do so, one based on a simple greedy heuristic, the other inspired by the results of the cavity method. Both of them build iteratively a percolating initial configuration, starting from the configuration with all vertices inactive, and adding one active vertex at a time (another route would be to start from the all active configuration and sequentially reduce the number of active vertices, but we did not investigate this alternative strategy). We shall denote $\tau$ the number of addition steps performed by the algorithm, and $\us(\tau)$ the initial configuration considered at this point (that contains by definition $\tau$ active vertices). The configuration denoted $\us^T(\tau)$ (resp. $\us^{\rm f}(\tau)$) is thus the configuration obtained after $T$ (resp. an infinite) number of steps of the dynamics defined in (\ref{eq_dynrules}) from the initial configuration $\us(\tau)$; we will denote $|\us^T(\tau)|$ the number of active vertices in this configuration. The algorithm stops when this number reaches $N$, as $\us(\tau)$ is then the first percolating initial configurations encountered. The difference in the two algorithms to be presented below lies in the rule used to choose which additional active vertex to add in the initial configuration in a step $\tau \to \tau +1$.

\subsection{A greedy algorithm}
\label{sec_single_sample_greedy}

Let us first consider the case of a finite time horizon $T$, i.e. the problem of finding an initial configuration $\us$ with $\us^T$ the fully active configuration and $\us$ containing the smallest possible number of active vertices. The simplest strategy is to choose at each time step $\tau\to\tau+1$ the inactive vertex of $\us(\tau)$ whose activation leads to the largest possible value of $|\us^T(\tau+1)|$, and stop at the first time $\tau$ such that $\us^T(\tau)$ is the fully active configuration. This can be immediately generalized to the case $T=\infty$ by including at each time step the vertex whose activation increases most $|\us^{\rm f}(\tau+1)|$; this version of the greedy procedure was actually a tool in the rigorous bounds on $\tmin$ for graphs with good expansion properties of~\cite{bounds_Amin}. If several vertices lead to the same increase the ties can be broken arbitrarily. The time complexity of the greedy algorithm is a priori cubic in the number $N$ of vertices: a linear number of steps $\tau\to\tau +1$ have to be performed before finding a percolating initial configuration. For each of these steps a number of order $N$ of candidate new configurations $\us(\tau+1)$ have to be considered, the computation of $\us^T(\tau+1)$ requiring itself a linear number of operations for each configuration. It is however easy to reduce significantly this complexity when $T=\infty$. As explained at the end of Sec.~\ref{sec_def_dyn} , in this case the final configuration of the dynamical process can be obtained sequentially, regardless of the order of the activations. By monotonicity the configuration $\us^{\rm f}(\tau+1)$ can be computed by adding one active vertex to $\us^{\rm f}(\tau)$ (instead of $\us(\tau)$) and determining the number (of order 1) of additional activations that can be triggered by this addition. This reduces the total complexity to a quadratic scaling with $N$.

\begin{figure}
\includegraphics[width=8cm]{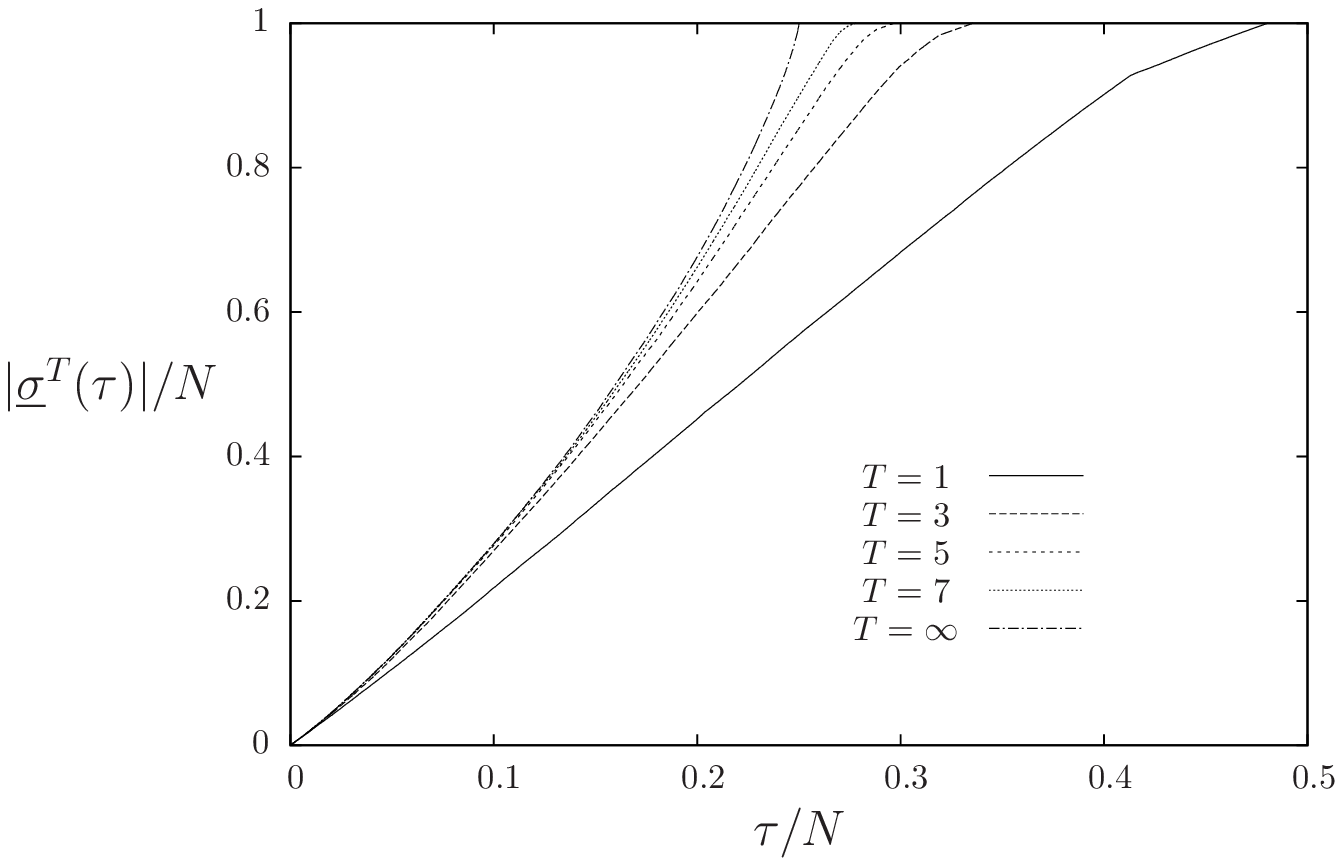}
\hspace{1cm}
\includegraphics[width=8cm]{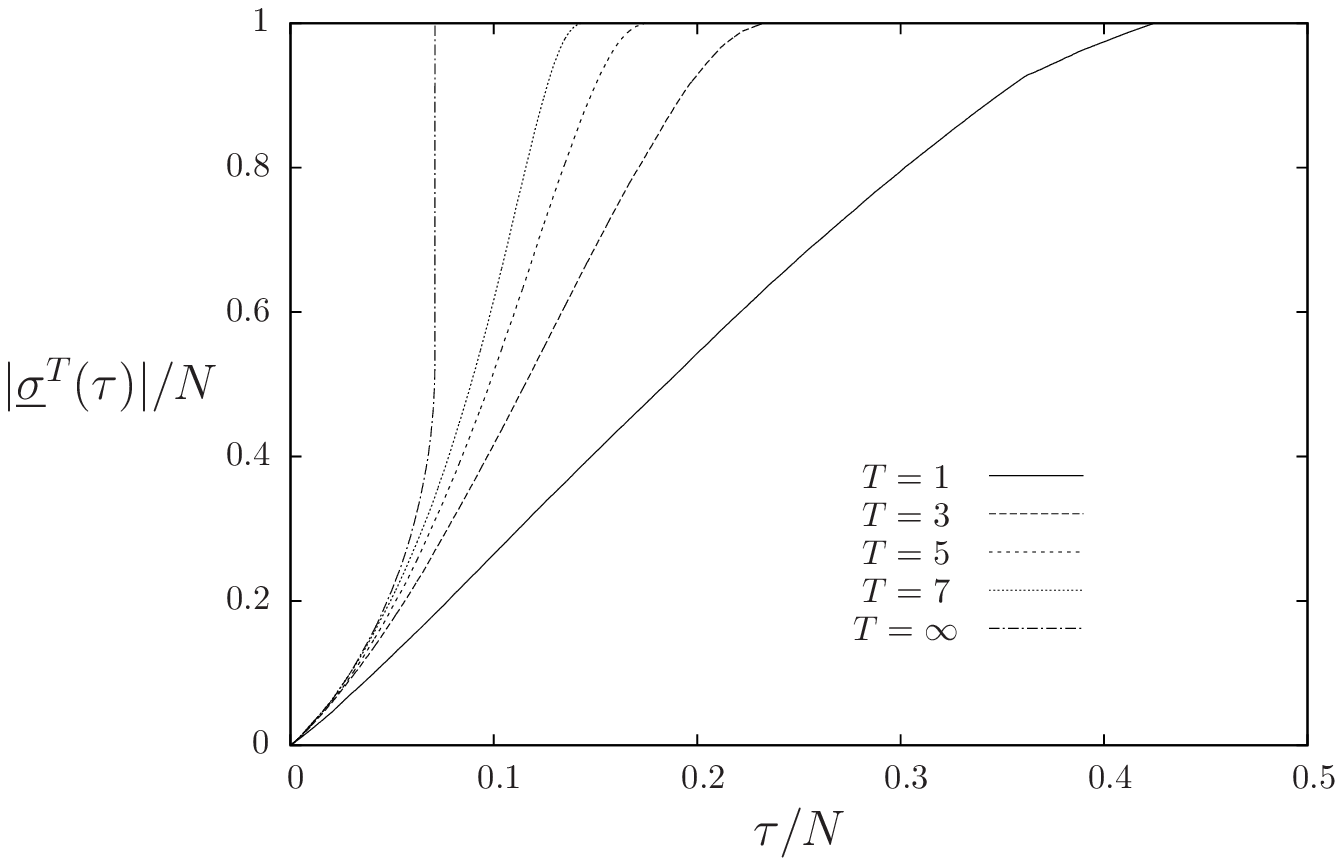}
\caption{The density of active vertices in the configuration $\us^T(\tau)$ after $\tau$ steps of the greedy algorithm, for $k=l=2$ (left panel) and $k=3$, $l=2$ (right panel). Each curve corresponds to a single run of the algorithm on a graph of $N=10^4$ vertices.}
\label{fig_greedy}
\end{figure}

In Fig.~\ref{fig_greedy} we plot the fraction of active vertices in the configuration $\us^T(\tau)$ as a function of the density $\tau/N$ of the active vertices in the initial configuration obtained after $\tau$ steps of this greedy procedure; when the curve reaches 1 we have thus obtained an initial configuration that percolates within $T$ steps (note that the part of the curve for smaller $\tau$ corresponds to the alternative optimization problem labelled (i) in the introduction). The density of the contagious sets reached in this way are summarized in Table~\ref{tab_single_sample_finiteT}; as expected these densities are strictly greater than the prediction $\tmino$ of the 1RSB cavity method, and also than the ones reached by more involved message-passing algorithms (see the discussion in next subsection).

\begin{table}
\begin{tabular}{|c|c|c|c|c||c|c|c|c|}
\hline
 & \multicolumn{4}{c||}{$k=l=2$} & \multicolumn{4}{c|}{$k=3$, $l=2$} \\
\hline
$T$ & $\tmino$ & $\theta_{\rm sp}$ & $\theta_{\rm maxsum}$~\cite{private} & $\theta_{\rm greedy}$ & $\tmino$ & $\theta_{\rm sp}$ & $\theta_{\rm maxsum}$~\cite{private} & $\theta_{\rm greedy}$ \\
\hline
1 & 0.424257 & 0.426 & 0.427 & 0.482 & 0.363813 & 0.366 & 0.370 & 0.426 \\
\hline
2 & 0.325882 & 0.328 & 0.330 &0.376 & 0.237009 & 0.240 & 0.243 & 0.291 \\
\hline
3 & 0.289097 & 0.291& 0.293 &0.335 & 0.182338 & 0.185 & 0.190 & 0.233 \\
\hline
4 & 0.271564 & 0.273 & 0.275 &0.311 & 0.151693 & 0.156 & 0.164 & 0.197 \\
\hline
5 & 0.262167 & 0.263 & 0.266 &0.296 & 0.132036 & 0.142 & 0.146 & 0.174 \\
\hline
7 & 0.253779 &           & 0.257 &0.278  & 0.108251 & 0.127 & 0.125 & 0.144 \\
\hline
10 & 0.250553 &         & 0.251 &0.265  & 0.089425 &           & 0.108 &0.119 \\
\hline
\end{tabular}
\caption{The density of (finite time) contagious sets reached by the greedy and message-passing algorithms, compared to the predictions of the cavity method for their minimal size. The data for the algorithmic results correspond to averages over ten graphs of size $N=10^4$.}
\label{tab_single_sample_finiteT}
\end{table}

One can clearly see a qualitative difference between the cases $k=l$ and $k>l$ in the two panels of Fig.~\ref{fig_greedy}: in the latter case as $T$ gets larger the last active vertices added in the initial configuration before finding a percolating one provoke a very steep increase in the final size of the activated set.
As said above the greedy procedure can easily be generalized to $T=\infty$; the density of the smallest contagious sets constructed in this way are presented in Table~\ref{tab_single_sample_Tinfty} for various values of $k$ and $l$. As these results demonstrate the greedy algorithm is able, in all cases we investigated, to find contagious sets with a density strictly smaller than $\tr$, the density above which typical uncorrelated configurations are percolating. However in general the density reached by this simple procedure is strictly greater than the prediction $\tmino$ of the cavity method for their minimal size; this is in agreement with the interpretation of the replica symmetry breaking creating metastable states that trap simple local search procedures and prevent them from reaching global optima of the cost function landscape in which the search moves. The only exception is the case $k=l=2$, for which the minimal density $1/4$ (corresponding to the decycling number of 3-regular random graphs~\cite{decycling}) is actually reached by the greedy procedure; this result is in line with the analysis of Sec.~\ref{sec_kequall_Tinfty}, which revealed a disappearance of the RSB phase in the large $T$ limit for this peculiar case.

\begin{table}
\begin{tabular}{|c|c|c|c|c|}
\hline
$k$ & $l$ & $\tr$ & $\tmino$ & $\theta_{\rm greedy}$ \\
\hline
2 & 2 & $\frac{1}{2}$ & $\frac{1}{4}$ & 0.250 \\
\hline
3 & 2 & 0.111111 & 0.046328 & 0.070 \\
\hline
3 & 3 & $\frac{2}{3}$ & $\frac{1}{3}$ & 0.387 \\
\hline
4 & 4 & $\frac{3}{4}$ & 0.378465 & 0.482 \\
\hline
5 & 5 & $\frac{4}{5}$ & 0.422695 & 0.551 \\
\hline
\end{tabular}
\caption{The density of (infinite time) contagious sets reached by the greedy algorithm, compared to the predictions of the cavity method. The algorithm was run on ten graphs of size $N=10^4$, the last column is the average over these repetitions. Experiments with graphs of different sizes revealed a very clear $1/N$ dependency of the finite-size corrections of $\theta_{\rm greedy}$ in the cases with $k=l$. We could not get such a clear dependency when $k>l$, slower finite-size corrections might be at play in these cases.}
\label{tab_single_sample_Tinfty}
\end{table}

\begin{figure}
\includegraphics[width=8cm]{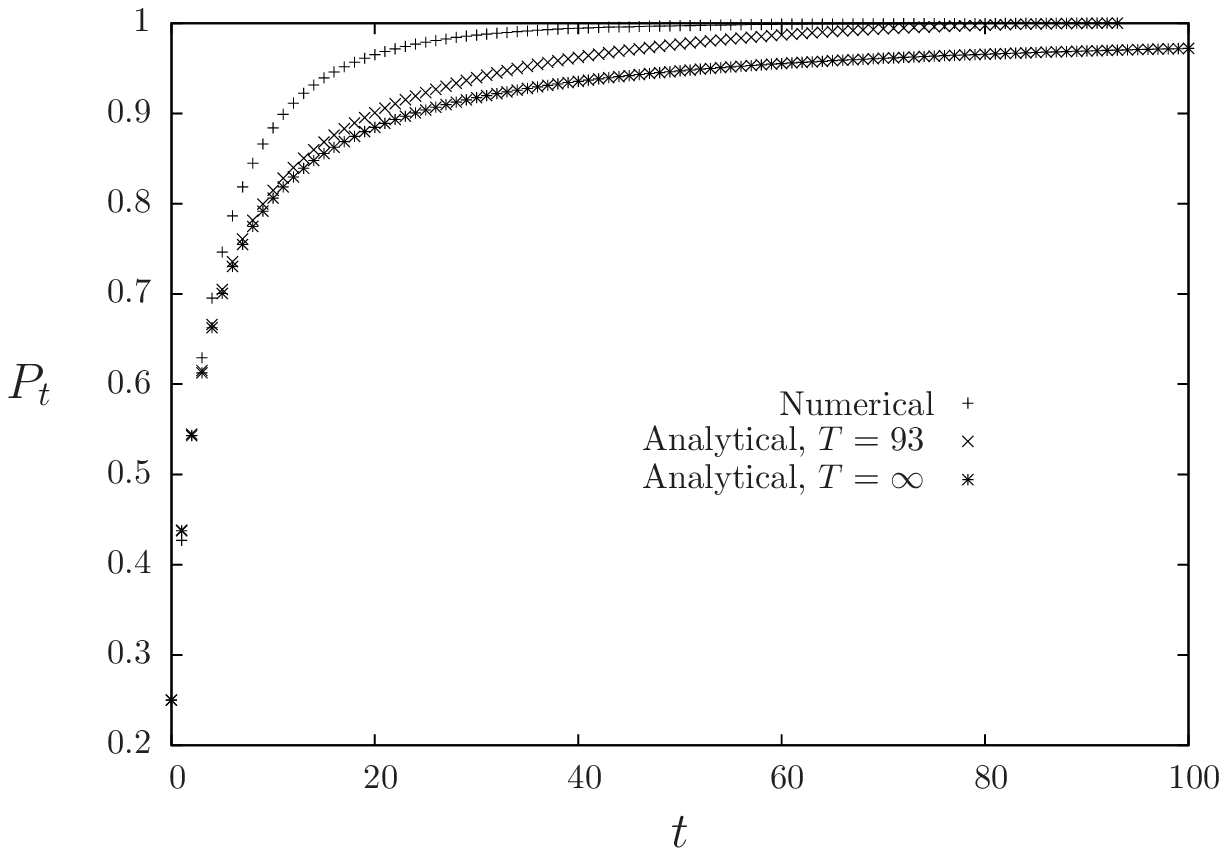}
\hspace{1cm}
\includegraphics[width=8cm]{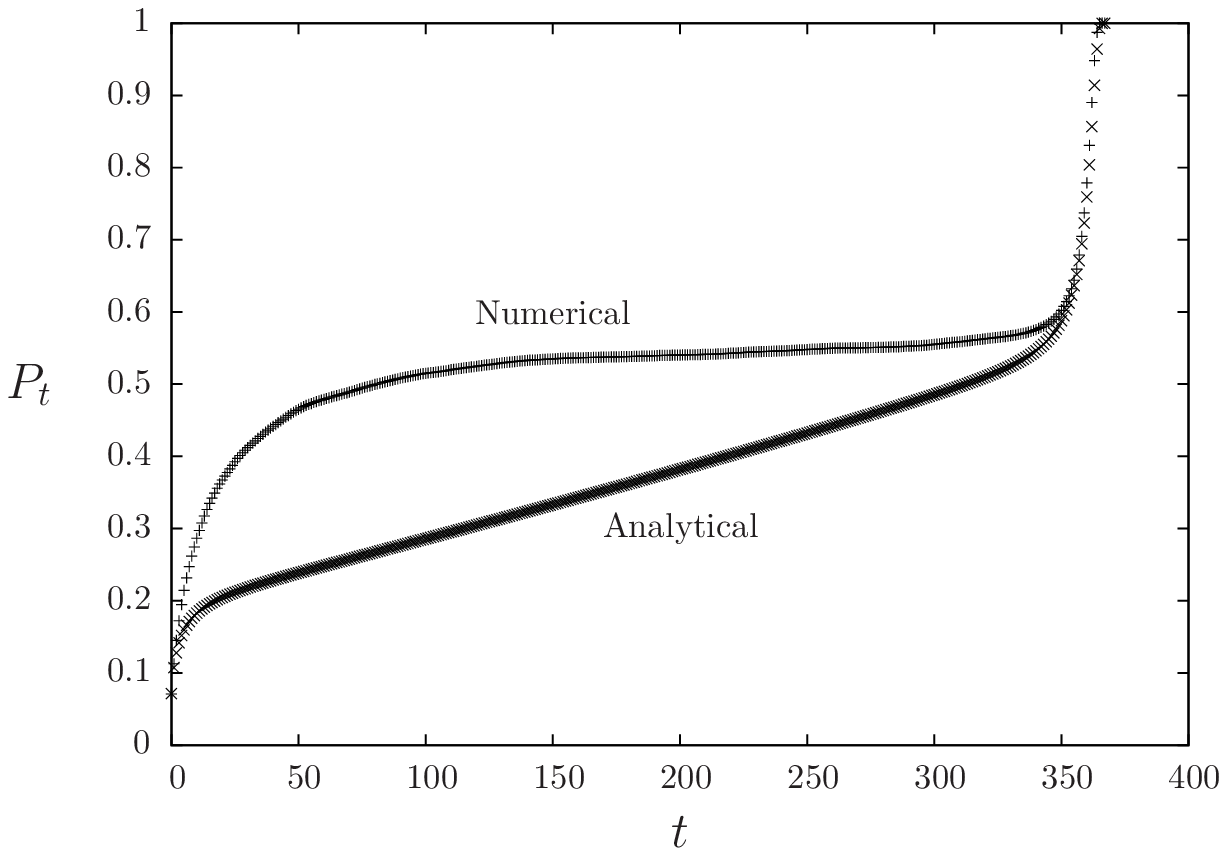}
\caption{The ``numerical'' curves represent the distribution of activation times for the least dense activating initial configurations found by the greedy algorithm for $T=\infty$, for $k=l=2$ (left panel) and $k=3$, $l=2$ (right panel). In both cases the graph studied contained $N=8 \cdot 10^4$ vertices, in the left panel the complete activation is reached in 93 steps, in the right one it takes 367 steps. For comparison in the left panel the analytical prediction is plotted both for $T=\infty$ (see Eq.~(\ref{eq_Pt_trsudue})) and for $T=93$, in the right panel the analytical curve corresponds to $T=367$.}
\label{fig_tact_compa}
\end{figure}

Further information on the minimal contagious sets produced by the greedy algorithm with $T=\infty$ can be obtained from the distribution of the activation times of the vertices they induce, which are plotted in Fig.~\ref{fig_tact_compa}. Of course as the graphs under study are finite the support of these distributions is bounded; in all cases we investigated we found that the time to reach total activation from these initial configurations scales logarithmically with the number of vertices of the graph (see also Fig.~\ref{fig_tact_compa} for a comparison between two different sizes of the graph). The qualitative difference between the cases $k=l$ and $k>l$ expected from the discussion of the $T\to\infty$ limit of Sec.~\ref{sec_res_largeT} is indeed apparent on these curves; in the latter case a finite fraction of the vertices are activated at the very end of the dynamical process. However the activation time distributions induced by the configurations produced by the greedy algorithm are not in quantitative agreement with the RS analytical predictions (with a value of $T$ and $\theta$ chosen to fit the numerical ones). A possible explanation for this discrepancy is that the greedy algorithm is a very ``out-of-equilibrium'' algorithm, hence the configurations it reaches are not the typical ones of the ``equilibrium'' measure (\ref{eq_eta_us}).

\begin{figure}
\includegraphics[width=9cm]{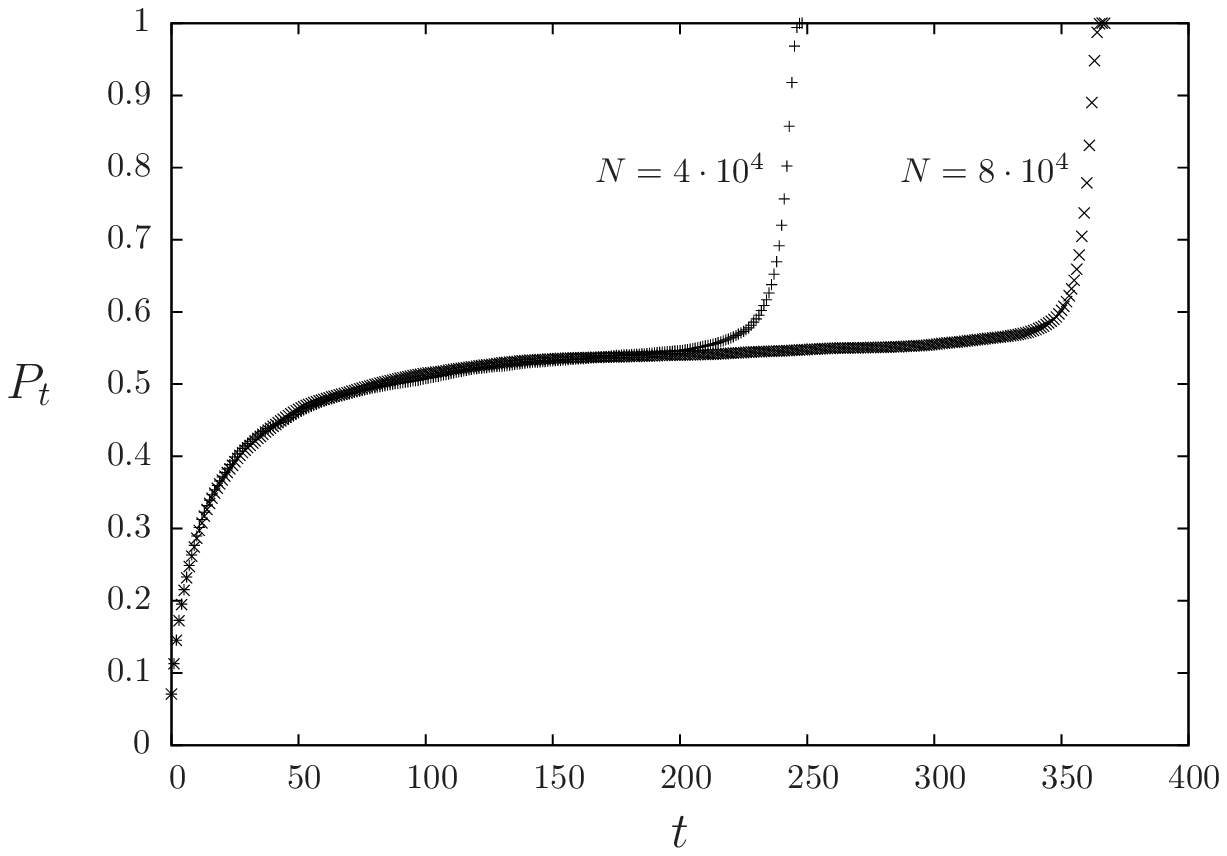}
\caption{The distribution of activation times for the least dense activating initial configurations found by the greedy algorithm for $T=\infty$, for $k=3$, $l=2$, and two different sizes $N$ of the graph. For $N=4 \cdot 10^4$ the complete activation took place after $T=248$ steps, while for $N=8 \cdot 10^4$ it occured at $T=367$.}
\label{fig_tact_num}
\end{figure}

\subsection{Survey propagation}
\label{sec_single_sample_mp}

The second algorithmic procedure we investigated is based on the insight provided by the statistical mechanics analysis on the structure of the configuration space of the problem; it corresponds indeed to the Survey Propagation algorithm introduced in~\cite{MeZe} for the analysis of random satisfiability problem (and more precisely to its variant introduced in~\cite{SPy} for the energy minimization in the unsatisfiable phase of such problems). An idealized thought experiment for the construction of minimal contagious sets would be to sequentially assign the values of the $\s_i$ according to their marginal probabilities in the law (\ref{eq_eta_us}), with $\epsilon=+\infty$ and $\mu=-\infty$; the exact determination of such marginals is in general a very hard computational tasks, and in practice one has to content oneself with approximations provided for instance by message passing procedures. This is the road we have followed here, by implementing the single-sample energetic 1RSB equations (\ref{eq_G_1RSBy}), i.e. assigning to each directed edge $i\to j$ of the graph under study a vector $P_{i\to j}$ of $2T$ probabilities. At each step $\tau$ of the algorithm the equations (\ref{eq_G_1RSBy}) are iterated several times to look for a global solution of these equations; the presence of $\tau$ active (decimated) vertices in the current configuration $\us(\tau)$ is implemented as a boundary condition in these equations, easily seen to be $P_{i\to j}(h)=\delta(h-B_0)$ for the outgoing messages from an activated vertex $i$. The information contained in such a solution of the 1RSB equations can be a priori exploited in several ways; we chose to compute, for each vertex $i$ not yet activated, the quantity
\beq
W_i = 1 - \frac{\partial}{\partial y}\ln \Zsite(\{ P_{j\to i} \}_{j \in \di} )+ \frac{1}{2} \sum_{j \in \di} \frac{\partial}{\partial y} \ln \Zedge(P_{i \to j},P_{j\to i}) \ ,
\eeq
i.e. the contribution of the site $i$ to the derivative of the potential $\Phi_{\rm e}$ given in Eq.~(\ref{eq_Phie_single_sample}). This number measures indeed the tendency of $i$ to be active in all configurations belonging to the clusters considered in the energetic 1RSB formalism. Accordingly we choose the vertex $i$ with the largest value of $W_i$ to be the new active vertice to be added to $\us(\tau)$ in order to form $\us(\tau+1)$. For simplicity we fixed the value of $y$ in the whole procedure to the value $y_{\rm s}$ determined analytically, that leads to a vanishing complexity before the decimation; we also tried to recompute this value of $y$ during the course of the decimation but did not obtain significant improvement of the performances in the cases considered.

The values of the density of the percolating initial configurations we managed to construct in this way are presented in Table \ref{tab_single_sample_finiteT} for the two cases $k=l=2$ and $k=3$, $l=2$, for several (relatively small) values of $T$. The results are better than the simple greedy algorithm, and in most of the cases also than the maxsum replica-symmetric algorithm~\cite{Torino1,Torino2,private}, but in some cases deviate significantly from the prediction $\tmino$ for the density of minimal contagious sets. An analytical understanding of the performances of such decimation procedures is actually a challenging open problem (see~\cite{RiSe09,Coja11} for partial results in the simpler case of the Belief-Propagation guided decimation). We did not study much larger values of $T$ because we faced in this case convergence issues for the iterations of the equations (\ref{eq_G_1RSBy}), that a simple damping did not seem to alleviate efficiently. A pragmatic, even if not completely satisfactory, position we adopted for the results at $T\ge 4$ for the case $k=3$, $l=2$, was to ignore somehow the convergence problems, stopping the iterations of (\ref{eq_G_1RSBy}) after a time fixed beforehand, and computing the value of $W_i$ from these unconverged messages. As Table \ref{tab_single_sample_finiteT} demonstrates this attitude is not unreasonable as the densities reached are still better than the one of the greedy algorithm (yet can get worse than the maxsum procedure~\cite{Torino1,Torino2,private}).

\section{Conclusions and perspectives}
\label{sec_conclu}

In this paper we have continued the study initiated in~\cite{Torino1,Torino2} of the minimal contagious sets for the bootstrap percolation (or threshold model) dynamics on random graphs. We have shown the importance of taking into account the phenomenon of replica symmetry breaking in the determination of the minimal density $\tmin$ of active vertices in percolating initial conditions, and could simplify analytically the equations determining $\tmin$ in the limit $T\to\infty$ where the constraint on the time to reach a complete activation of the graph disappears. Reformulating the problem as the minimal number of vertices to be removed in a graph in order to destroy some specific subgraphs (its cycles or more generically its $q$-core) we recovered a previously known result for the decycling number of 3-regular random graphs~\cite{decycling} as well as a conjecture for 4-regular ones~\cite{decycling}, and proposed new quantitative conjectures for the sizes of the minimal ``de-coring'' sets for all pairs of degree of the graph and minimal degree of the targeted core. These take a particularly simple rational form for the removal of the 3-core in 5- and 6- regular random graphs.

Let us sketch now some possible directions for future study. A first project would be to test the stability of the 1RSB ansatz we used to compute $\tmino$, to assess for which values of $(k,l)$ this number should be expected to be the exact value $\tmin$ and not only a lowerbound. This computation should be doable following the techniques of~\cite{stab1,stab2,bm} for all finite $T$, and might even be simplified in the large $T$ limit. By analogy with the independent set problem which is a marginal case of the problem investigated here one could surmise to find that the 1RSB ansatz is stable for large enough values of the degree $k$ (and maybe also of the threshold $l$). This is also the regime where one can hope to see a mathematically rigorous proof of these predictions, as recently obtained for the independent sets in~\cite{is_Sly}. Asymptotic expansions of $\tminz(k,l)$ and $\tmino(k,l)$ in the large $k$ limit for $k>l$ should also be performed, considering either $l$ fixed in this limit, $l$ proportional to $k$, or $k-l$ fixed.

For the sake of concreteness and simplicity we presented explicit results only for regular random graphs, however we gave the intermediate equations of the RS and 1RSB cavity method under a form that can be directly applied to any sparse random graph ensembles with arbitrary prescribed degree distribution, and possibly fluctuating thresholds for activation. The latter could naturally be correlated with the degree of the vertices, triggering for instance the activation if the fraction of active neighbours reaches some fixed proportion (instead of a fixed number). It would be interesting to see how the results presented here are qualitatively modified by the local fluctuations in the graph structure, which would be particularly severe in the case of power-law tails in the degree distribution.

We also concentrated exclusively in this paper on the problem of optimizing the number of initially active vertices, imposing that all vertices are active at a later time. The variant of this problem where one puts a constraint on the maximal number of active vertices allowed in the initial configuration and try to maximize the level of activation at a later time is also relevant, in particular for applications to real-world situations. At the RS level we have sketched how to do this by controlling the parameter $\epsilon$ (the cost to be paid for finally inactive vertices) that we kept arbitrary in the first steps of the computations, a systematic study and the inclusion of the effects of replica symmetry breaking remains to be done.

Finally we believe that the message passing procedure inspired by the energetic 1RSB equations presented in Sec.~\ref{sec_single_sample_mp} would be worth investigated further. One should try to study (and cure) the convergence issues that arise for larger values of $T$, maybe changing the way the information provided by the messages is used. One could in particular exploit them in a softer way by implementing a reinforcement technique~\cite{Torino1,Torino2} instead of a direct decimation. A more extensive comparison with the maxsum message passing procedure studied in~\cite{Torino1,Torino2} could also be interesting.

\acknowledgments
We warmly thank
Fabrizio Altarelli,
Victor Bapst,
Alfredo Braunstein,
Amin Coja-Oghlan,
Luca Dall'Asta,
Svante Janson,
Marc Lelarge and
Riccardo Zecchina
for useful discussions, and in particular FA, AB, LDA and RZ for sharing with us the unpublished numerical results~\cite{private} on their maxsum algorithm, and SJ for a useful correspondence and for pointing out the reference \cite{decycling}.

The authors acknowledge the support of the French Agence Nationale de la Recherche (ANR) under reference ANR-11-JS02-005-01 (GAP project) and of the People
Programme (Marie Curie Actions) of the European Union's Seventh Framework Programme
FP7/2007-2013/ under REA grant agreement no 290038.

\appendix

\section{The limit $\mu\to-\infty$ of the fields recursion}
\label{app_eqs_WP}

We justify here the equation (\ref{eq_g_ABt}) for the recursion $h=g(h_1,\dots,h_k)$ between ``hard fields'' $h_i\in \{A_0,A_1,\dots,A_{T-1},A_T=B_T,B_{T-1},\dots,B_1,B_0 \}$. We can first notice that in Eqs.~(\ref{eq_at_WP},\ref{eq_bt_WP})  the (constrained) maximum over the partitions $I,J,K$ of $\S_t$ is always reached for $|I|+|J|$ and $|I|$ as small as possible (because $a_t^{(i)} \ge b_{t-1}^{(i)} \ge b_{t-2}^{(i)} $), which allows to rewrite 
\bea
a_t &=& \max\left(0, \max_{t'\in[1,T]} \max_{\substack{J,K \\ |J| = l - \ind(t' \ge t+1)} } \S_{t'}(h_1,\dots,h_k;\emptyset ,J,K) \right) \ , \\
b_t &=& \max\left(0, \max_{t'\in[1,t]} 
\max_{\substack{J,K \\ |J| = l}}
\S_{t'}(h_1,\dots,h_k;\emptyset,J,K) \right) \ ,
\eea
where $J,K$ forms a partition of $\{1,\dots,k\}$. In addition one realizes that
\beq
\max_{\substack{J,K \\ |J| = l}}
\S_t (h_1,\dots,h_k;\emptyset,J,K) = 1 \Leftrightarrow\left( \sum_{i=1}^k \ind(h_i \in \{A_0,\dots,A_{t-1} \} ) = 0 \ \ \text{and} \ \
\sum_{i=1}^k \ind(h_i \in \{B_0,\dots,B_{t-1} \} ) \ge l \right)\ ,
\eeq
which by logical negation leads to
\beq
\max_{\substack{J,K \\ |J| = l}}
\S_t (h_1,\dots,h_k;\emptyset,J,K) \le 0 \Leftrightarrow \left( \sum_{i=1}^k \ind(h_i \in \{A_0,\dots,A_{t-1} \} ) \ge 1 \ \ \text{or} \ \
\sum_{i=1}^k \ind(h_i \in \{B_0,\dots,B_{t-1} \} ) \le l-1 \right) \ .
\eeq
Combining these logical rules leads after a short reasoning to
\bea
g(h_1,\dots,h_k)=A_t &\Leftrightarrow& 
\left( a_t =1 \ \text{and} \ a_{t+1}=0 \right) \label{eq_g_At}\\
&\Leftrightarrow& 
\begin{cases}
& \sum_{i=1}^k \ind(h_i \in \{B_0,\dots,B_t \}) = l-1 \\
\text{and}& \sum_{i=1}^k \ind(h_i \in \{A_0,\dots,A_t \})=0 \\
\text{and}& \sum_{i=1}^k \ind(h_i=A_{t+1}) \ge 1
\end{cases}
 \ ,
\eea
and
\bea
g(h_1,\dots,h_k)=B_t &\Leftrightarrow& 
\left( b_t =1 \ \text{and} \ b_{t-1}=0 \right) \label{eq_g_Bt}\\
&\Leftrightarrow& 
\begin{cases}
& \sum_{i=1}^k \ind(h_i \in \{B_0,\dots,B_{t-1} \}) \ge l \\
\text{and}& \sum_{i=1}^k \ind(h_i \in \{B_0,\dots,B_{t-2} \})\le l-1 \\ 
\text{and}& \sum_{i=1}^k \ind(h_i \in \{A_0,\dots,A_{t-1} \})=0
\end{cases}
\ .
\eea
Considering the various possible cases leading to a field of type $A_t$ or $B_t$ yields finally (\ref{eq_g_ABt}).

\section{Technical details on the resolution of the factorized RS and energetic 1RSB equations}
\label{sec_app}

We shall present in this Appendix the details of the RS and energetic 1RSB cavity equations in the particular case of random $k+1$ regular graphs with an uniform threshold $l$ of activations. It turns out that despite their different interpretations these two version of the cavity method can be treated in an unified way. We thus begin by introducing this common formulation, then we unveil the simplifications that arise in the case $l=k$, before finally discussing the limit $T\to\infty$, both in the case $l=k$ and $l<k$.

\subsection{Common formulation}

\subsubsection{RS cavity method}

Consider the fixed-point RS equation $h=g(h,\dots,h)$, with $g$ defined in Eq.~(\ref{eq_g}); alternatively we saw in Eqs.~(\ref{eq_diff_aT},\ref{eq_diff_at}) an expression for the differences $e^{-\mu a_t} - e^{-\mu a_{t+1}}$. Setting $h_i=h$ in the right-hand sides of these equations, and using the identity
\beq
\sum_{\substack{I,J,K \\  |I| \le l-1 \\ |I|+|J| \ge l}} f(I,J,K) =
\sum_{\substack{I,J,K \\ |I|+|J| \ge l}} f(I,J,K) -
\sum_{\substack{I,J,K \\ |I| \ge l}} f(I,J,K) \ ,
\eeq
for any function $f$ of a partition $I,J,K$, allows to show the equivalence of the fixed-point equation on $h=(a_0,\dots,a_T,b_{T-1},\dots,b_1)$ with:
\bea
e^{-\mu a_t}- e^{-\mu a_{t+1}} &=& e^{-\mu + \mu k a_0} \binom{k}{l-1} e^{-\mu (l-1) b_t} \left[ 
\left( e^{-\mu a_{t+1}} - e^{-\mu b_t } \right)^{k-l+1}  -
\left( e^{-\mu a_{t+2}} - e^{-\mu b_t } \right)^{k-l+1}
\right] \ ,
\\
e^{-\mu b_{t+1}}- e^{-\mu b_t} &=& e^{-\mu + \mu k a_0} \sum_{p=l}^k \binom{k}{p} \left[ 
e^{-\mu p b_t} \left(e^{-\mu a_{t+1}} - e^{-\mu b_t} \right)^{k-p}
- e^{-\mu p b_{t-1}} \left(e^{-\mu a_{t+1}} - e^{-\mu b_{t-1}} \right)^{k-p}
\right] \ .
\eea
These equations are valid for $t\in\{0,\dots,T-1\}$, with the boundary conditions $e^{-\mu b_{-1}}=0$, $b_0=1$, $a_T=b_T$, $a_{T+1}=b_{T-1}$. The thermodynamic quantities can also be simplified in this factorized case, the site contribution to the RS free-entropy reading from Eq.~(\ref{eq_zsite}):
\beq
\zsite = 1 + e^{-\mu + \mu (k+1) a_0} \sum_{t=1}^T 
\sum_{p=l}^{k+1} \binom{k+1}{p} \left[ 
e^{-\mu p b_{t-1}} \left(e^{-\mu a_t} - e^{-\mu b_{t-1}} \right)^{k+1-p}
- e^{-\mu p b_{t-2}} \left(e^{-\mu a_t} - e^{-\mu b_{t-2}} \right)^{k+1-p}
\right] \ ,
\label{eq_zsite_reg}
\eeq
while the edge contribution of Eq.~(\ref{eq_zedge}) becomes
\beq
\zedge = e^{2 \mu a_0} \left[
e^{- 2 \mu a_T} + 2 \sum_{t=0}^{T-1}
 \left( e^{-\mu a_t} - e^{-\mu a_{t+1}}\right) e^{-\mu b_t}
\right] \ .
\label{eq_zedge_reg}
\eeq

Let us introduce some new notations and define a change of parameters on the unknowns $a_t,b_t$, as $u_t=e^{-\mu a_t}$, $v_t=e^{-\mu b_t}$. We also define a new parameter $\lambda$, with $\lambda=e^{-\mu + \mu k a_0}$. In terms of these new quantities the above set of equations becomes
\bea
u_t - u_{t+1} &=& D(u_{t+1},v_t) - D(u_{t+2},v_t) \ , \label{eq_u} \\
v_{t+1} - v_t &=& S(u_{t+1},v_t) - S(u_{t+1},v_{t-1}) \ , \label{eq_v}
\eea
with $v_{-1}=0$, $v_0=1$, $u_T=v_T$, $u_{T+1}=v_{T-1}$, and
\beq
D(u,v) = \lambda \binom{k}{l-1} v^{l-1} (u-v)^{k-l+1} \ , \qquad
S(u,v) = \lambda \sum_{p=l}^k \binom{k}{p} v^p (u-v)^{k-p} \ .
\label{eq_DS}
\eeq
In other words the $u$'s and $v$'s are solutions of a set of polynomial equations, and as such should be viewed as a function of $\lambda$ and $T$ (and of course of $k$ and $l$). They also obey, on top of the boundary conditions, the inequalities $u_0 \ge u_1 \ge \dots \ge u_T = v_T \ge v_{T-1} \ge \dots v_1 \ge v_0=1$. The chemical potential $\mu$ has disappeared from this set of equations, but actually it is now implicitly a function of $\lambda$ and $T$, as from the definition of $\lambda$ one recovers $\mu$ with $\mu = -\ln(\lambda u_0^k)$.

For future use we emphasize here an identity between the derivatives of $D$ and $S$ and introduce a new function $C(u,v)$:
\beq
C(u,v) = \frac{\partial D}{\partial u} = \frac{\partial S}{\partial v} = \lambda l \binom{k}{l} v^{l-1} (u-v)^{k-l} \ .
\label{eq_C}
\eeq

Let us also rewrite the thermodynamic quantities in terms of these new variables. The expressions (\ref{eq_zsite_reg}) and (\ref{eq_zedge_reg}) become
\beq
\zsite = 1 + \Fsite \ , \qquad \zedge = \frac{1}{u_0} \Fedge \ , 
\label{eq_rs_zofF}
\eeq
where we introduced the two functions
\bea
\Fsite(\lambda,T) &=& \frac{\lambda}{u_0} \sum_{t=1}^T 
\sum_{p=l}^{k+1} \binom{k+1}{p} \left[ v_{t-1}^p (u_t - v_{t-1})^{k+1-p}
- v_{t-2}^p (u_t - v_{t-2})^{k+1-p}
\right] \ ,\label{eq_Fsite}\\
\Fedge(\lambda,T) &=& \frac{1}{u_0} 
\left[ v_T^2 + 2 \sum_{t=0}^{T-1} (u_t -u_{t+1}) v_t \right] \ .
\label{eq_Fedge}
\eea
We emphasize here the dependency on $\lambda$ and $T$, which was kept implicit in the $u_t$ and $v_t$'s. One has then the final expressions of all RS thermodynamic quantities as:
\beq
\phi = \mu + \ln(\zsite) - \frac{k+1}{2} \ln(\zedge) \ , \qquad \mu = -\ln(\lambda u_0^k) \ , \qquad s = \phi - \mu \theta \ , \qquad \theta = \frac{1}{\zsite} \ .
\label{eq_RS_thermo}
\eeq
One can also express the probability distribution of the activation times in terms of these new variables. Denoting $P_t$ the cumulative distribution, i.e. the probability that the activation time of one vertex is smaller or equal than $t$, one has from Eq.~(\ref{eq_etasite}): 
\beq
P_t = \frac{1}{\zsite}\left[ 1 + \Fsite(\lambda,T,t) \right] \ , 
\label{eq_Pt}
\eeq
where we defined
\beq
\Fsite(\lambda,T,t) =\frac{\lambda}{u_0} \sum_{t'=1}^t 
\sum_{p=l}^{k+1} \binom{k+1}{p} \left[ v_{t'-1}^p (u_{t'} - v_{t'-1})^{k+1-p}
- v_{t'-2}^p (u_{t'} - v_{t'-2})^{k+1-p}\right] \ .
\label{eq_Fsite_t}
\eeq
One can check that, as it should, $P_0=\theta$ the fraction of initially active sites (summations over empty sets being equal to zero by convention), and $P_T=1$ (as $\epsilon=+\infty$ all vertices are active at the final time).

\subsubsection{Energetic 1RSB cavity method}

We now turn to a similar study of the energetic 1RSB equations in the factorized case, namely the determination of the normalized vector of probabilities $P=(p_0,\dots,p_{T-1},q_T,\dots,q_0)$, solution of the fixed-point equation $P=G(P,\dots,P)$, with the mapping $G$ defined in Eq.~(\ref{eq_G_1RSBy}).

Let us first note that in general the normalization $Z[P_1,\dots,P_k]$ of (\ref{eq_G_1RSBy}) can be expressed in terms of $q_0$,
\beq
Z = 1 + (e^y-1) (1-Z q_0) \ \Rightarrow \ \ \frac{e^y}{Z} = 1 + q_0 (e^y -1) 
\ .
\eeq
This remark allows to rewrite the fixed-point equation $P=G(P,\dots,P)$ as

\bea
p_t &=&(1 + q_0 (e^y -1)) \binom{k}{l-1}\left(\sum_{t'=0}^t q_{t'}\right)^{l-1}
\left[  \left(\sum_{t'=t+1}^T q_{t'} + \sum_{t'=t+1}^{T-1} p_{t'}\right)^{k-l+1}
-  \left(\sum_{t'=t+1}^T q_{t'} + \sum_{t'=t+2}^{T-1} p_{t'}\right)^{k-l+1}
\right] \ ,\nonumber\\
q_t &=& (1 + q_0 (e^y -1)) \sum_{p=l}^k \binom{k}{p}\left[
\left(\sum_{t'=0}^{t-1} q_{t'}\right)^p
\left(\sum_{t'=t}^T q_{t'} + \sum_{t'=t}^{T-1} p_{t'}\right)^{k-p}
- \left(\sum_{t'=0}^{t-2} q_{t'}\right)^p
\left(\sum_{t'=t-1}^T q_{t'} + \sum_{t'=t}^{T-1} p_{t'}\right)^{k-p}
\right]\ , \nonumber
\eea
where in the first line $t\in\{0,\dots,T-1\}$ and in the second  $t\in\{1,\dots,T\}$. These two sets of equations are supplemented by the normalization condition $q_0+\dots+q_T+p_{T-1}+\dots+p_0=1$.

The site and edge contributions of the energetic 1RSB potential, defined in
(\ref{eq_Zsite_y},\ref{eq_Zedge_y}), become in the factorized case:
\bea
\Zsite &=& 1+(e^y-1)\sum_{t=1}^T
\sum_{p=l}^{k+1} \binom{k+1}{p}\left[
\left(\sum_{t'=0}^{t-1} q_{t'}\right)^p
\left(\sum_{t'=t}^T q_{t'} + \sum_{t'=t}^{T-1} p_{t'}\right)^{k+1-p}
- \left(\sum_{t'=0}^{t-2} q_{t'}\right)^p
\left(\sum_{t'=t-1}^T q_{t'} + \sum_{t'=t}^{T-1} p_{t'}\right)^{k+1-p}
\right]
\ , \nonumber\\
\Zedge &=& e^{-y}+(1-e^{-y}) \left[ 
\left(\sum_{t=0}^T q_t \right)^2
+ 2 \sum_{t=0}^{T-1} p_t \sum_{t'=0}^t q_{t'}
\right]  \ .\nonumber
\eea

Now let us change variables and trade the unknowns $p_t,q_t$ for some variables $u_t$, $v_t$, and the parameter $y$ for some parameter $\lambda$, according to 
\beq
u_t = \frac{1}{q_0} \left(\sum_{t'=0}^T q_{t'} + \sum_{t'=t}^{T-1} p_{t'}
\right) \ , \qquad
v_t = \frac{1}{q_0} \sum_{t'=0}^t q_{t'} \ , \qquad
\lambda= (1 + q_0 (e^y -1)) q_0^{k-1} \ .
\eeq
Inserting these definitions in the above equations one realizes that the quantities $u_t$ and $v_t$ are solutions of exactly the same set of equations (\ref{eq_u},\ref{eq_v}) defined in the RS case, and obey the same boundary conditions and inequalities. From the solution of these equations, for a given value of the parameter  $\lambda$, one recovers the parameter $y$ noting that by the normalization condition one has $u_0=1/q_0$, hence $y = \ln(\lambda u_0^k - u_0 +1)$. The expressions of $\Zsite$ and $\Zedge$ within this parametrization are easily obtained from the above equations and read:
\beq
\Zsite = 1 + \left(1 - \frac{1}{\lambda u_0^{k-1}} \right) \Fsite \ , \qquad
\Zedge = \frac{1+ (\lambda u_0^{k-1} - 1) \Fedge}{\lambda u_0^k - u_0 +1} \ ,
\eeq
with the same functions $\Fsite$ and $\Fedge$ defined in Eqs.~(\ref{eq_Fsite},\ref{eq_Fedge}) for the RS case. One has finally an expression for the thermodynamic quantities of the energetic 1RSB formalism as
\beq
\Phi_{\rm e} = - y + \ln \Zsite - \frac{k+1}{2} \ln \Zedge \ , \quad
 y = \ln(\lambda u_0^k - u_0 +1) \ , \quad
\Sigma_{\rm e} = \Phi_{\rm e} + y \theta \ ,
\label{eq_1RSBy_thermo}
\eeq
where $\theta$ is here the opposite of the derivative of $\Phi_{\rm e}$ with respect to $y$, which after a short computation reads
\bea
\theta &=& 1- \frac{e^y}{e^y-1} \frac{\Zsite - 1}{\Zsite} 
- \frac{k+1}{2} \frac{1}{e^y-1} \frac{1-\Zedge}{\Zedge} 
\label{eq_1RSBy_theta} \\
&=&
\frac{1-\frac{1}{\lambda u_0^k} \Fsite}{1 + \left(1 - \frac{1}{\lambda u_0^{k-1}} \right) \Fsite} 
- \frac{k+1}{2} \frac{1-\frac{1}{u_0} \Fedge}{1+(\lambda u_0^{k-1}-1) \Fedge} \ .
\nonumber
\eea

\subsubsection{Simplifications for $l=k$}
\label{sec_uv_klequal}

In the case $l=k$ further simplifications arise. Indeed the function $S(u,v)$ defined in (\ref{eq_DS}) is in this case independent of $u$, and the equations (\ref{eq_u},\ref{eq_v}) can be rewritten as:
\bea
v_0 &=& 1 \ , \\
v_t &=& 1 + \lambda \, v_{t-1}^k \qquad \text{for} \ t \in \{1,\dots,T\} \ , \label{eq_vt_klequal}\\
u_{T-1} &=& v_T + \lambda k \, v_{T-1}^{k-1} \, (v_T - v_{T-1}) \ , \\
u_t &=& u_{t+1} + \lambda k \, v_t^{k-1} \, (u_{t+1} - u_{t+2}) \qquad \text{for} \ t \in \{0,\dots,T-2\} \label{eq_ut_klequal}
\ .
\eea
This set of equations is particularly simple to solve, and admits a single solution for each value of $\lambda$. One can indeed compute by recurrence the value of the $v_t$ for increasing values of $t$ from $0$ to $T$, then deduce the value of $u_{T-1}$, and finally by a downward recurrence the values of $u_t$ for $t$ from $T-2$ to $0$. The thermodynamic observables are then deduced from (\ref{eq_RS_thermo}) in the RS case or (\ref{eq_1RSBy_thermo}) in the energetic 1RSB case, where the site contributions can be simplified from (\ref{eq_Fsite}), yielding
\beq
\Fsite(\lambda,T)=\frac{\lambda}{u_0}\left[v_{T-1}^{k+1} + (k+1) \sum_{t=1}^T (u_t-u_{t+1}) v_{t-1}^k \right] \ .
\label{eq_Fsite_klequal}
\eeq
These simplifications can also be performed for the function (\ref{eq_Fsite_t}) giving the distribution of activation times, which reads in the case $k=l$:
\beq
\Fsite(\lambda,T,t)=\frac{\lambda}{u_0}\left[v_{t-1}^{k+1} + (k+1) v_{t-1}^k (u_{t+1}-v_{t-1})+(k+1) \sum_{t'=1}^t (u_{t'}-u_{t'+1}) v_{t'-1}^k \right] \ .
\label{eq_Fsite_t_klequal}
\eeq

\subsubsection{Numerical resolution for $l<k$}

In the case $l<k$ we did not find a simple change of variables on the unknowns $u_t,v_t$ that would put the system of equations (\ref{eq_u},\ref{eq_v}) in the triangular form that appeared naturally when $k=l$ and led to a direct resolution by successive substitutions. We therefore resorted to the Newton-Raphson iterative method for solving (\ref{eq_u},\ref{eq_v}), taking care of choosing a good initial condition for the iterations to be convergent. This guess on the solution was provided by analytical asymptotic expansions, either in the limit $\lambda \to 0$ or with $T\to\infty$ (see next paragraph). Depending on the values of $\lambda$ and $T$ we found either 0, 1 or 2 relevant solutions of (\ref{eq_u},\ref{eq_v}), but this multi valuedness has no physical meaning and comes only from the arbitrary choice of the parametrization in terms of $\lambda$. Indeed there is a single solution for each value of the chemical potential $\mu$ (or $y$ in the energetic 1RSB formalism).

\subsection{The large $T$ limit}
\label{sec_app_Tinfty}

In the rest of this Appendix we shall justify analytically the claims made in Sec.~\ref{sec_kequall_Tinfty} and \ref{sec_lmk_Tinfty} on the behaviour of the RS and energetic 1RSB solutions as $T$ goes to infinity.

\subsubsection{The trivial solution}
\label{sec_app_Tinfty_trivial}

As anticipated in Sec.~\ref{sec_results_rrg}, in the large $T$ limit the portion of the curve $s(\theta)$ corresponding to $\theta>\tr$ should coincide with the entropy $-\theta \ln \theta - (1-\theta) \ln(1-\theta)$ counting all configurations with a fraction $\theta$ of initially active sites, as such configurations are typically activating (see the reminder on random initial configurations of Sec.~\ref{sec_reminder_random}). Let us see how to prove this statement. A moment of thought, considering for instance the form of the RS equations at $\epsilon=0$, reveals that this situation should correspond to a solution of (\ref{eq_u},\ref{eq_v}) with $u_t=\tu$, independently of $t$. This ansatz is indeed consistent with Eq.~(\ref{eq_u}), and with this substitution Eq.~(\ref{eq_v}) becomes
\beq
v_{t+1} = 1+ S(\tu,v_t) \ .
\label{eq_vt_trivial}
\eeq
This last equation is a simple recursion on the $v$'s, with the initial value $v_0=1$. For the boundary condition $u_T=v_T$, $u_{T+1}=v_{T-1}$ to be asymptotically (when $T\to\infty$) verified one has to impose the values of $\tu$ and $\lambda$ such that the $v_t$ solution of (\ref{eq_vt_trivial}) converge to $\tu$ when $t \to \infty$, in other words that the smallest fixed point solution $v \ge 1$ of $v=1+S(\tu,v)$ is precisely equal to $\tu$. The condition $\tu=1+S(\tu,\tu)$ imposes the following relationship between $\tu$ and $\lambda$, $\tu=1+\lambda \tu^k$. Using this condition one can then rewrite (\ref{eq_vt_trivial}) as
\beq
\frac{v_{t+1}}{\tu} = \frac{1}{\tu} + \left(1-\frac{1}{\tu}\right) 
\sum_{p=l}^k \binom{k}{p} \left(\frac{v_t}{\tu}\right)^p \left(1-\frac{v_t}{\tu}\right)^{k-p} \ .
\eeq
Comparing this equation with (\ref{eq_random_tx}) one realizes that by definition of $\tr$, all the values of $\tu$ in the interval $[1,1/\tr[$ are such that the condition $v_t \to \tu$ is fulfilled (with the value of $\lambda$ fixed by $\tu=1+\lambda \tu^k$). Let us now compute the RS thermodynamic quantities associated with this solution. As the $u_t$ are independent of $t$ the summation in Eq.~(\ref{eq_Fsite}) can be performed with a telescopic identity, and yields after a short computation $\Fsite=\tu-1$. Similarly one sees easily from (\ref{eq_Fedge}) that $\Fedge=\tu$ for this solution. This gives indeed the function $s(\theta)=-\theta \ln \theta - (1-\theta) \ln(1-\theta)$ for $\theta > \tr$ upon replacing in the expression of the RS thermodynamic potential (cf. Eq.~(\ref{eq_RS_thermo})). In addition the cumulative distribution $P_t$ of activation times defined in Eq.~(\ref{eq_Pt}) coincides on this solution with the series $x_t$ of Eq.~(\ref{eq_random_x}) obtained as the activation time cumulative distribution of a random initial condition.

In the following we shall describe the non-trivial part of the resolution of the RS and energetic 1RSB equations in the large $T$ limit, i.e. in the RS case the part of the curve $s(\theta)$ for $\theta < \tr$. The cases $l=k$ and $l<k$ are technically rather different, we shall thus divide the discussion according to this distinction.

\subsubsection{Asymptotics for $l=k$}
\label{sec_app_Tinfty_klequal}

\begin{figure}
\includegraphics[width=8.5cm]{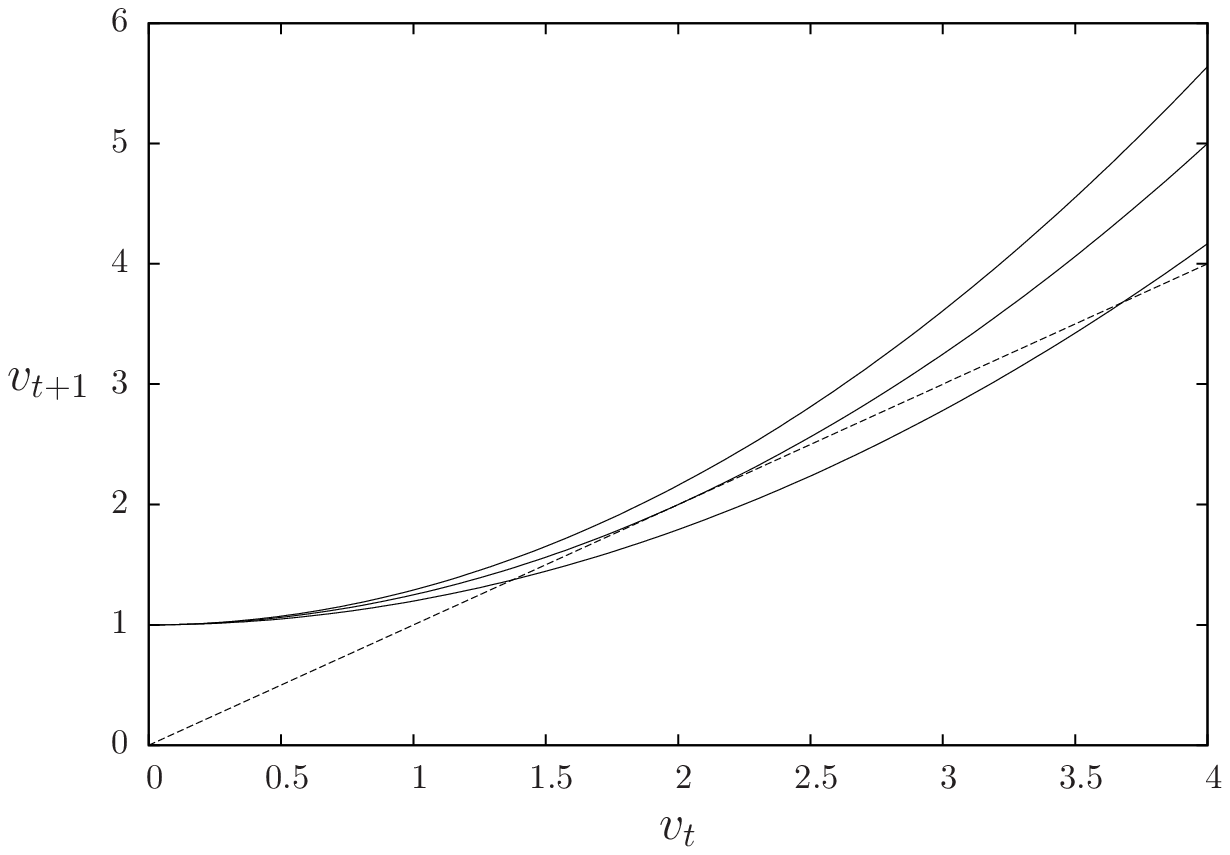}
\caption{A graphical representation of the recursion $v_{t+1}=1+\lambda v_t^k$ (here for $k=2$). The dashed straight line corresponds to $v_{t+1}=v_t$, the three solid curves are, from bottom to top, for $\lambda < \lambda_{\rm c}$, $\lambda = \lambda_{\rm c}$ and $\lambda > \lambda_{\rm c}$.}
\label{fig_recurs_klequal}
\end{figure}

As explained in Sec.~\ref{sec_uv_klequal} in the case $l=k$ the equations on $v_t$ decouple, these quantities become independent of $T$ and are solutions of the recurrence $v_{t+1}=1+\lambda v_t^k$. A straightforward study of this equation (see Fig.~\ref{fig_recurs_klequal} for an illustration) reveals the existence of a critical value $\lambda_{\rm c}$ such that $v_t$ converges to a finite value when $t\to\infty$ if $\lambda \le \lambda_{\rm c}$, while it diverges when $\lambda > \lambda_{\rm c}$. This critical parameter and the associated fixed-point $v_{\rm c}$ of the recurrence are solution of the equations:
\beq
v_{\rm c} = 1+ \lambda_{\rm c} v_c^k \ , \qquad 1 = \lambda_{\rm c} k \, v_{\rm c}^{k-1} \ ,
\eeq
which are easily solved and yield $\lambda_{\rm c}=\frac{(k-1)^{k-1}}{k^k}$, $v_{\rm c}=\frac{k}{k-1}$.

The case  $\lambda < \lambda_{\rm c}$ corresponds actually to the trivial solution already discussed above, let us thus consider the alternative situation, $\lambda > \lambda_{\rm c}$. The divergence of $v_t$ is then actually very steep, with a double exponential form. Indeed when $v_t \gg 1$ the recurrence becomes approximately $v_{t+1} \approx \lambda v_t^k$, which reveals that $(\ln \ln v_t)/t$ converges to $\ln k$. As $u_0 \ge v_T$ one also has a divergence of $u_0$ with $T$ in this regime; from (\ref{eq_RS_thermo}) (resp. (\ref{eq_1RSBy_thermo})) this implies that the chemical potential $\mu$ of the RS formalism (resp. the parameter $y$ of the energetic 1RSB one) go to $-\infty$ (resp. $+\infty$), i.e. that the parametric curve $s(\theta)$ (resp. $\Sigma_{\rm e}(\theta)$) has a vertical tangent in this regime. Furthermore we shall prove now that the corresponding density $\theta$ of initially active sites converges to $(k-1)/(2k)$ (both in the RS and energetic 1RSB cases), hence this branch corresponds to a vertical segment. This is actually a consequence of the following statement on the behaviour of the functions $\Fsite$ and $\Fedge$ of Eqs.~(\ref{eq_Fsite_klequal},\ref{eq_Fedge}):
\beq
\forall \lambda > \lambda_{\rm c} \ , \quad
\lim_{T\to\infty} \Fsite(\lambda,T) = \frac{k+1}{k-1} \ , \qquad \lim_{T\to\infty} \Fedge(\lambda,T) = \frac{2k}{k-1} \ ,
\label{eq_claim_Fs}
\eeq
as can be easily deduced from the expressions of $\theta$ given in (\ref{eq_rs_zofF},\ref{eq_RS_thermo}) and (\ref{eq_1RSBy_theta}), along with the divergence of $u_0$ in the latter case. To prove the claim of Eq.~(\ref{eq_claim_Fs}), let us first note that, iterating (\ref{eq_ut_klequal}), one obtains
\bea
u_t-u_{t+1} &=& (u_0-u_1) \frac{1}{k^t} \frac{1}{(\lambda v_0^{k-1}) (\lambda v_1^{k-1}) \dots (\lambda v_{t-1}^{k-1})} \\
&=& (u_0-u_1) \frac{1}{k^t} \frac{v_1 v_2 \dots v_{t-1}}{(v_1-1)(v_2-1) \dots (v_t -1)} \ ,
\eea
where we used (\ref{eq_vt_klequal}) to go from the first to the second line. We can thus write
\beq
u_t-u_{t+1} = (u_0-u_1) \frac{1}{k^t} \alpha_t \frac{1}{v_t} \ ,
\eeq
where we introduced the sequence $\alpha_t$ (note its independence on $T$) as
\beq
\alpha_t = \prod_{t'=1}^t \frac{v_{t'}}{v_{t'}-1} \ , \qquad \alpha_0=1 \ .
\eeq
We also have, in terms of this series,
\beq
u_0-u_1 = k^T \frac{1}{\alpha_T} v_T (v_T - v_{T-1}) \ .
\eeq
Using these relations, along with the representation $u_0 = v_T + \sum_{t=0}^{T-1} (u_t -u_{t+1})$, allows to rewrite the definition of (\ref{eq_Fedge}) as:
\beq
\Fedge= \frac{ \alpha_T \frac{1}{k^T} \frac{v_T}{v_T-v_{T-1}} + 2 \overset{T-1}{\underset{t=0}{\sum}} \alpha_t \frac{1}{k^t} }{\alpha_T \frac{1}{k^T} \frac{1}{v_T-v_{T-1}} + \overset{T-1}{\underset{t=0}{\sum}} \frac{\alpha_t}{v_t} \frac{1}{k^t}} \ .
\eeq
The sum in the denominator can be transformed by noting that, from the definition of $\alpha_t$, $\alpha_t /v_t = \alpha_t - \alpha_{t-1}$. This yields
\beq
\Fedge= \frac{ \alpha_T \frac{1}{k^T} \frac{v_T}{v_T-v_{T-1}} + 2 \overset{T-1}{\underset{t=0}{\sum}} \alpha_t \frac{1}{k^t} }{\alpha_T \frac{1}{k^T} \frac{1}{v_T-v_{T-1}} + \frac{1}{k^T} \alpha_{T-1} + \frac{k-1}{k}\overset{T-1}{\underset{t=0}{\sum}} \alpha_t \frac{1}{k^t}} \ .
\eeq
Notice now that $\alpha_t$ has a finite limit when $t \to \infty$, thanks to the divergence of $v_t$ (for the limit of $\alpha_t$ to exists it is actually enough that $v_t \gg t$). Hence the summations in the above equation converge when $T\to \infty$ thanks to the exponentially decaying factor $1/k^t$, and all other terms in the numerator and denominator are neglectible in this limit. This proves the limit $2k/(k-1)$ for $\Fedge$ (one could also compute the main correction, of order $k^{-T}$, from this expression). The statement on $\Fsite$ is proved with similar manipulations, that brings from (\ref{eq_Fsite_klequal}) to the expression (exact for all $T$),
\beq
\Fsite= \frac{ \alpha_T \frac{1}{k^T} \frac{v_{T-1}(v_T-1)}{v_T (v_T-v_{T-1})} + \frac{k+1}{k} \overset{T-1}{\underset{t=0}{\sum}} \alpha_t \frac{1}{k^t} }{\alpha_T \frac{1}{k^T} \frac{1}{v_T-v_{T-1}} + \frac{1}{k^T} \alpha_{T-1} + \frac{k-1}{k}\overset{T-1}{\underset{t=0}{\sum}} \alpha_t \frac{1}{k^t}} \ .
\eeq
As above the limit $T\to \infty$ can now be taken safely, the converging summations being the only non-vanishing terms of the numerator and denominator, hence the convergence of $\Fsite$ to $(k+1)/(k-1)$, with corrections of order $k^{-T}$. These corrections actually contribute to the non-trivial dependence on $\lambda$ of $s$ and $\Sigma_{\rm e}$ (which are both finite) in this regime; we did not push their determination further, and merely observe here that their order $k^{-T}$ explains the statement on the finite $T$ corrections to $\tmin$ for $k=l=2$ and $k=l=3$ made in Sec.~\ref{sec_kequall_Tinfty}.

We have just seen that in the $T \to \infty$ limit the cases $\lambda < \lambda_{\rm c}$ and $\lambda > \lambda_{\rm c}$ describe, respectively, the trivial branch $\theta > \tr$ of the RS entropy and its vertical segment at $\tr/2$. To describe the range $[\tr/2,\tr]$ of non-trivial densities of initially active sites one has thus to investigate a regime where $\lambda$ is in a $T$-dependent scaling window around $\lambda_{\rm c}$.

Let us denote $\tv_t$ the solution of the recursion right at the critical point, i.e. $\tv_{t+1}= 1 + \lambda_{\rm c}\tv_t^k$, with $\tv_0=1$. This series converges to $v_{\rm c}$, with an asymptotic behaviour which is easily found to be
\beq
\tv_t = v_{\rm c} - \frac{2k}{(k-1)^2} \frac{1}{t} + O\left( \frac{1}{t^2} \right) \ .
\label{eq_tvt_asymptot}
\eeq
Now if $\lambda = \lambda_{\rm c} + \delta$, with an infinitesimal positive value of $\delta$, the solution $v_t$ of the recursion $v_{t+1}=1+\lambda v_t^k$ spends a time of order $\delta^{-1/2}$ around the avoided fixed-point $v_{\rm c}$ before crossing over to the doubly exponentially growing regime investigated above (this is a general feature of such recursive equations in the neighbourhood of a bifurcation, see for instance~\cite{bifurcation}). It is thus natural to investigate the scaling window parametrized by $\hlambda$ as
\beq
\lambda = \lambda_{\rm c} + 2\pi^2 \frac{(k-1)^{k-2}}{k^{k-1}}\frac{\hlambda^2}{T^2} \ ,
\eeq
the numerical prefactor and the square on $\hlambda$ being chosen to simplify the following expressions. One can then look for a solution of the recurrence equation under the form $v_t = v_{\rm c} + \frac{1}{T} V(t/T)$, with $V(s)$ a scaling function. Expanding at the leading order in $T$ one obtains a differential equation on $V$,
\beq
V'(s)=\frac{2\pi^2 k \hlambda^2}{(k-1)^2} + \frac{(k-1)^2}{2 k} V(s)^2 \ .
\eeq
The latter can be integrated into
\beq
V(s) = - \frac{2 k}{(k-1)^2} \frac{\pi \hlambda}{\tan(\pi \hlambda s)} \ ,
\eeq
the constant in the solution of the differential equation being obtained by a matching argument between the regime $s\to 0$ and the large $t$ asymptotics of the critical series $\tv_t$ given in (\ref{eq_tvt_asymptot}). Note that this form is only valid for $\hlambda < 1$, otherwise one enters the regime where $v_T$ diverges with $T$. One can furthermore assume a similar scaling ansatz for the $u_t$, introducing a scaling function $U(s)$ under the form $u_t = v_{\rm c} + U(t/T)$. Inserting these forms in Eq.~(\ref{eq_ut_klequal}) yields a differential equation on $U$, 
\beq
\frac{U''(s)}{U'(s)}=-\frac{(k-1)^2}{k} \, V(s) \ ,
\eeq
which is integrated in
\beq
U'(s) = B \sin^2(\pi \hlambda s) \ , \qquad U(s)=A + \frac{B}{2} \left( s - \frac{\sin(2 \pi \hlambda s)}{2 \pi \hlambda} \right) \ ,
\eeq
with $A$ and $B$ two constants of integration. These can be fixed by imposing the boundary conditions $u_T=v_T$ and $u_{T+1}=v_{T-1}$, which translates here in $U(1)=V(1)/T$ and $U'(1)=-V'(1)/T$. Solving these equations yield $A$ and $B$; considering in particular $u_0=v_{\rm c} + U(0)$ one obtains, at the leading order in a large $T$ expansion,
\beq
u_0 = v_{\rm c} + \frac{1}{T} \frac{\hlambda^2}{\sin^4(\pi \hlambda)} \left(1-\frac{\sin(2\pi \hlambda)}{2 \pi \hlambda} \right) \frac{\pi^2 k}{(k-1)^2} - \frac{1}{T} \frac{\hlambda}{\tan(\pi \hlambda)} \frac{2 \pi k}{(k-1)^2} \ .
\label{eq_u0_hlambda}
\eeq
One realizes at this point that for any fixed $\hlambda < 1$, the limit of $u_0$ coincides with $v_{\rm c}$, in other words we are describing in this regime the end of the trivial branch, with $\theta \approx \tr$. To describe the non-trivial regime of densities $[\tr/2,\tr]$ one has thus to further refine the scaling window, taking now $\hlambda$ approaching $1$ in a $T$-dependent way. The inspection of (\ref{eq_u0_hlambda}) reveals that the correct scaling that allows to obtain a non-trivial limit of $u_0$ corresponds to $\hlambda = 1 - O(T^{-1/4})$. We shall thus set
\beq
\hlambda = 1 - \frac{1}{\sqrt{\pi}} \left(\frac{\tlambda}{(k-1) T} \right)^\frac{1}{4} \ , 
\eeq
with $\tlambda>0$ the new parameter describing this scale, the numerical prefactor being chosen for convenience. After a short computation one obtains the limit as $T \to \infty$ of the thermodynamic quantities in this scaling regime of $\lambda$ as
\beq
u_0(\tlambda) = \frac{k}{k-1}\, \frac{1+ \tlambda}{\tlambda} \ , \qquad
\Fsite(\tlambda) = \frac{1}{k-1}\, \frac{k+1+ \tlambda}{1 + \tlambda} \ , \qquad
\Fedge(\tlambda) = \frac{k}{k-1}\, \frac{2 + \tlambda}{1 + \tlambda} \ ,
\eeq
the last two expressions being obtained by inserting the scaling ansatz on $u_t$ and $v_t$ in the definitions (\ref{eq_Fedge},\ref{eq_Fsite_klequal}); at the lowest order one can actually replace the $v_t$'s by $v_{\rm c}$ there. This yields a parametric representation of the thermodynamic quantities of the RS (resp. energetic 1RSB) formalism in terms of $\tlambda$, by inserting these last results in Eq.~(\ref{eq_RS_thermo}) (resp. (\ref{eq_1RSBy_thermo},\ref{eq_1RSBy_theta})). In the RS case one can check that $\tlambda \to 0$ corresponds to $\theta \to \tr/2$, while $\tlambda \to \infty$ yields $\theta \to \tr$, hence this scaling regime allows to cover the desired range $[\tr/2,\tr]$ for the densities of initially active sites. It is furthermore possible to invert the relation $\theta(\tlambda)$, which yields finally the formula (\ref{eq_s_klequal_Tinfty}) announced in the main text for the entropy of activating initial configurations of density in the non-trivial interval $[\tr/2,\tr]$. In the energetic 1RSB case this last step does not seem possible and the final result (\ref{eq_1RSBy_klequal_Tinfty}) is presented in a form parametrized by $\tlambda$. We did not embark in a systematic study of the finite $T$ corrections in this regime, it is however clear that they are polynomially small in $T$, which justifies the statement made in Sec.~\ref{sec_kequall_Tinfty} on the corrections to $\tmin(T)$ for $k=l\ge 4$.

Let us finally justify the results presented at the end of Sec.~\ref{sec_kequall_Tinfty} on the distribution of activation times. Assuming a finite value of $t$, the expression of (\ref{eq_Fsite_t_klequal}) becomes in the regime parametrized by $\tlambda$:
\beq
\Fsite(\tlambda,t) = \frac{\lambda_{\rm c}}{u_0(\tlambda)} \left[
\tv_{t-1}^{k+1} + (k+1) \tv_{t-1}^k (u_0(\tlambda) - \tv_{t-1}) \right] \ ,
\eeq
the last summation in (\ref{eq_Fsite_t_klequal}) yielding a subdominant correction of order $1/T$. Note that $\Fsite(\tlambda,t)$ tends to $\Fsite(\tlambda)$ as $t\to\infty$, which means that the support of the distribution of the activation times does not scale with $T$ in this regime. The expression (\ref{eq_Pt_klequal_Tinfty}) for the cumulative distribution of activation times follows then easily from its generic definition given in Eq.~(\ref{eq_Pt}), upon expressing all the quantities depending on $\tlambda$ as a function of the corresponding $\theta$. In the main text we introduced for clarity the series $w_t = \tr \tv_t$, to allow for an easier comparison with the distribution of activation times from a random initial condition.

\subsubsection{Asymptotics for $l<k$}
\label{sec_app_Tinfty_lmk}

Let us now discuss the solution of the set of equations (\ref{eq_u},\ref{eq_v}) in the limit $T\to\infty$, in the case $l<k$, and justify the statements made in Sec.~\ref{sec_lmk_Tinfty}; as we shall see their behaviour and the method of study is qualitatively different compared to the case $l=k$.

We shall first rephrase Eqs.~(\ref{eq_u},\ref{eq_v}) as a single recursive equation, by introducing a four-dimensional vector $w_t$ defined by
\beq
w_t=\begin{pmatrix} u_{t\phantom{+1}} \\ u_{t+1} \\ v_{t\phantom{+1}} \\ v_{t-1} \end{pmatrix} \ .
\eeq
The recursive equations (\ref{eq_u},\ref{eq_v}) on the $u_t$'s and $v_t$'s become a single recursion on $w_t$, of the form $w_{t+1} = R(w_t)$ where the function $R$ is given by
\beq
R \begin{pmatrix} u_{\phantom +} \\ u_+ \\ v_{\phantom +} \\ v_- \end{pmatrix} =
\begin{pmatrix} u_+ \\ E(u,u_+,v) \\ v+S(u_+,v)-S(u_+,v_-) \\ v \end{pmatrix}
\ .
\eeq
The function $S$ was defined in (\ref{eq_DS}), while $E(u,u_+,v)$ is given implicitly as $D(E(u,u_+,v),v)=D(u_+,v)+u_+-u$, with the function $D$ of (\ref{eq_DS}). Inverting this relation one obtains an explicit expression of $E$:
\beq
E(u,u_+,v) = v+\left( (u_+-v)^{k-l+1} + \frac{1}{\lambda \binom{k}{l-1}} \frac{u_+-u}{v^{l-1}}  \right)^\frac{1}{k-l+1} \ .
\eeq

We have thus a representation of the time evolution of $w$ as the flow of a discrete dynamical system in a four-dimensional space. The boundary conditions on the $u_t$'s and $v_t$'s translate into conditions on the allowed values of $w_0$ and $w_T$. The former must indeed lie in the two-dimensional manifold with $v=1$ and $v_-=0$, while the latter is restricted to the two-dimensional manifold defined by $u=v$ and $u_+=v_-$. When $T\to\infty$, for a fixed value of $\lambda$, the solution $w_t$ of the recursion $w_{t+1}=R(w_t)$ must find a way to go infinitely slowly from the first manifold at $t=0$ to the second one at $t=T\to\infty$. It must in consequence remains as close as possible to the fixed points of the evolution map $R$.

\begin{figure}
\begin{center}
\includegraphics[width=7cm]{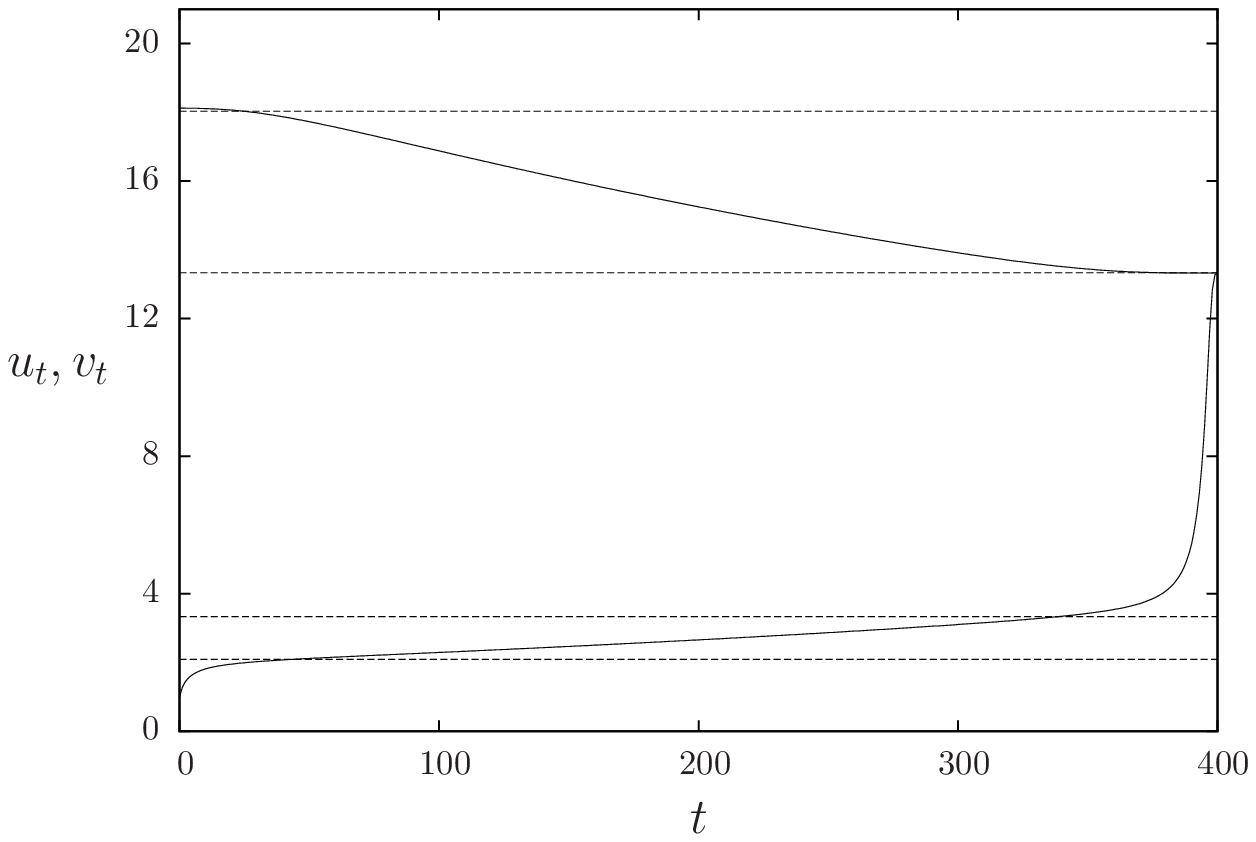}
\hspace{1cm}
\includegraphics[width=7cm]{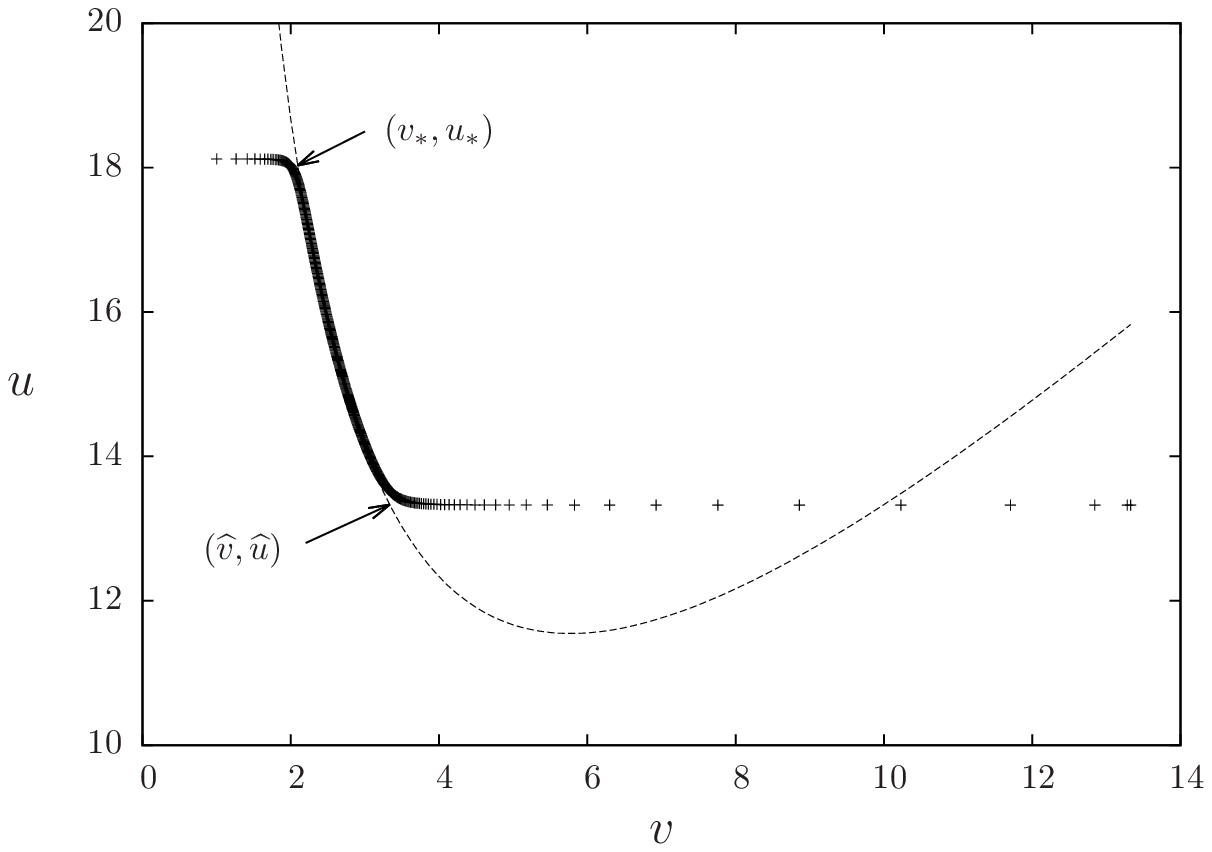}
\end{center}
\caption{The solution of the equations (\ref{eq_u},\ref{eq_v}) for $k=3$, $l=2$, with $\lambda=0.005$ and $T=400$. Left panel: the solid curves are $u_t$ (top) and $v_t$ (bottom) as functions of $t$; the dashed horizontal lines correspond, from top to bottom, to $u_*$, $\hu$, $\hv$ and $v_*$, solutions of (\ref{eq_fixedpoint_star},\ref{eq_fixedpoint_hat}). Right panel: parametric plot of the same data, with symbols instead of lines to appreciate the discreteness in $t$. Dashed line is the solution of the equation $C(u,v)=1$, almost superimposed with most of the points $(v_t,u_t)$. The arrows point to the beginning $(v_*,u_*)$ and end $(\hv,\hu)$ of the scaling regime along the curve $C(u,v)=1$.}
\label{fig_k3l2_uv}
\end{figure}

The study of the equation $w=R(w)$ is very simple and shows that these fixed points span the two-dimensional subspace with $u=u_+$, $v=v_-$. One can then compute the Jacobian matrix of $R$ on such a fixed-point, and realizes that this matrix has two eigenvalues equal to 1 (corresponding to the invariance of the fixed-point subspace under $u\to u+\delta u$ and $v \to v+\delta v$), and two eigenvalues $C(u,v)$ and $1/C(u,v)$, where $C$ is the function defined in (\ref{eq_C}). All the fixed points have thus an unstable direction, except the one-dimensional set of fixed points obeying the further condition $C(u,v)=1$, which constitutes a line of marginal fixed points. In the $T\to\infty$ limit the solution $w_t$ is thus expected to remain close to this line, otherwise the flow along the unstable directions forbid to go from one boundary manifold at $t=0$ to the other one at $t=T \gg 1$. This analysis is corroborated by the numerical results presented in Fig.~\ref{fig_k3l2_uv}, where we show the solution $u_t,v_t$ determined numerically for some large but finite value of $T$. In particular the right panel demonstrate that for most values of $t$ (i.e. excluding both $t$ finite and $T-t$ finite in the large $T$ limit), the couple $(u_t,v_t)$ falls on the marginal fixed-point line $C(u,v)=1$.

\begin{figure}
\begin{center}
\includegraphics[width=8cm]{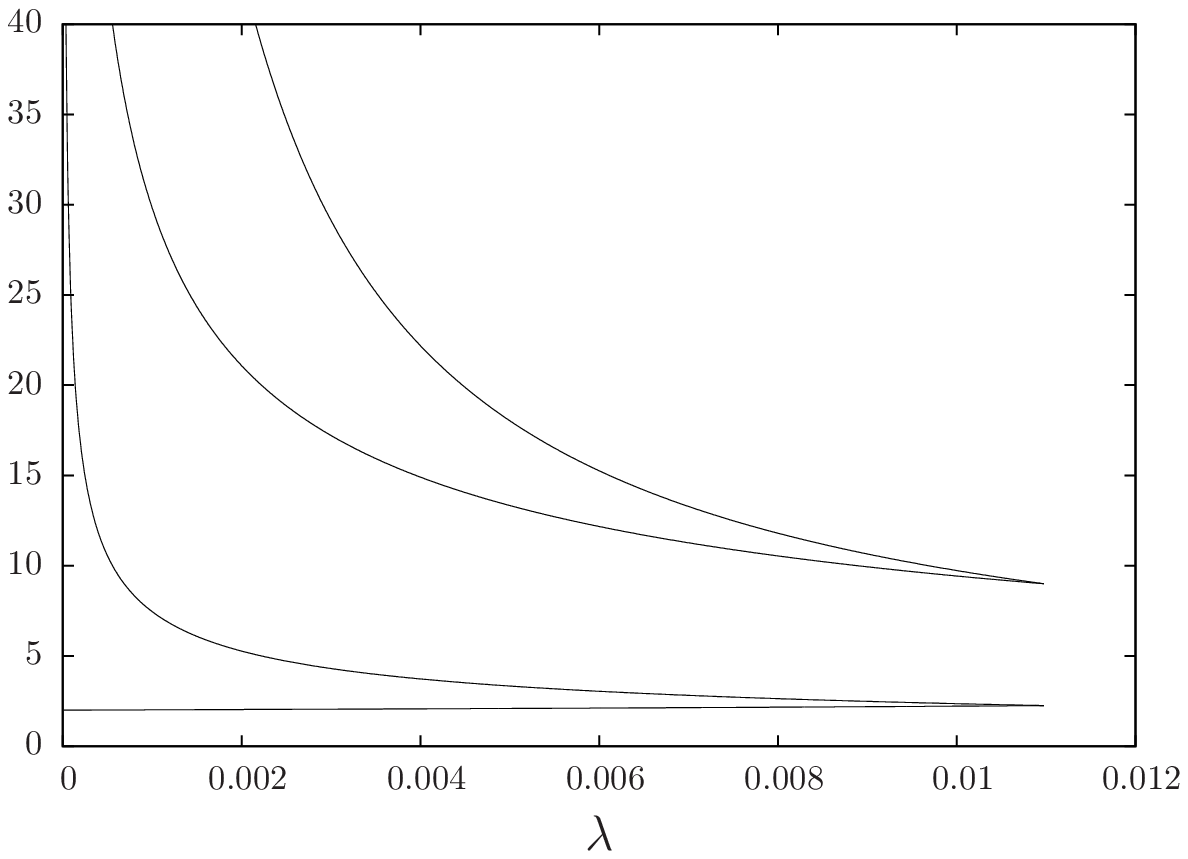}
\end{center}
\caption{The functions $u_*$, $\hu$, $\hv$ and $v_*$ (from top to bottom) solutions of Eqs.~(\ref{eq_fixedpoint_star},\ref{eq_fixedpoint_hat}) as a function of $\lambda$ for $k=3$, $l=2$. The upper two and lower two curves meet in $\lambda=\lambda_{\rm r}$. When $\lambda \to 0$ the upper three curves diverge, while $v_*$ converges to $l/(l-1)$.}
\label{fig_k3l2_fixedpoints}
\end{figure}

More precisely, the solution $u_t,v_t$ can be described in the large $T$ limit by two scaling functions $U(s)$ and $V(s)$, function of a rescaled time $s=t/T \in ]0,1[$, such that at the leading order,
\beq
u_t = U\left(\frac{t}{T}\right) \ , \qquad v_t = V\left(\frac{t}{T}\right) \ .
\eeq
Inserting this ansatz in the equations (\ref{eq_u},\ref{eq_v}), one realizes that the condition $C(U(s),V(s))=1$, that we obtained intuitively above, is indeed precisely what is needed to enforce (\ref{eq_u},\ref{eq_v}) at the leading order in the large $T$ limit. Note that the explicit dependency of $U$ and $V$ on $s$ can be determined from the sub-dominant corrections in this limit; however we shall not need it in what follows. It will indeed be enough to compute the value of $U$ and $V$ for $t$ small and $t$ close to $T$, i.e. for $s$ around 0 and 1. As revealed by the numerical data presented in Fig.~\ref{fig_k3l2_uv}, the matching between the scaling regime described by the functions $U,V$ (i.e. for $s$ strictly between 0 and 1) and the boundary conditions at $t=0$ and $t=T$ affects the series $v_t$ but not $u_t$. In other words, for $t$ finite while $T\to\infty$ one has $u_t\to u_*=U(0)$ independently of $t$, where $u_*$ is some ($\lambda$ dependent) constant still to be determined, while $v_t$ converges to the solution of the recursion $v_{t+1}=v_t+S(u_*,v_t)-S(u_*,v_{t-1})$ obtained from (\ref{eq_v}) by replacing $u_t$ by its limit $u_*$. Equivalently one has in this regime $v_{t+1}=1+S(u_*,v_t)$. When $t\to\infty$ (after the large $T$ limit) this series $v_t$ converges to $v_*=V(0)$, the smallest fixed-point solution of this recursion on $v$; for this behaviour to match the beginning of the scaling regime (i.e. $s\to 0$) one must impose simultaneously
\beq
C(u_*,v_*)= 1 \ , \qquad \text{and} \ \ \ v_*=1+S(u_*,v_*) \ .
\label{eq_fixedpoint_star}
\eeq
The first equation allows to express $u_*$ as a function of $v_*$; replacing in the second one leads to the single equation on $v_*$ given in Eq.~(\ref{eq_vstar}), while (\ref{eq_ustar}) is nothing but an explicit version of the condition $C(u_*,v_*)= 1$. A similar reasoning in the regime $T-t$ finite reveals that $U(1)=\hu$ and $V(1)=\hv$ have to obey
\beq
C(\hu,\hv) =1 \ , \qquad \text{and} \ \ \ 
\hv = S(\hu,\hv) + \hu - S(\hu,\hu) \ .
\label{eq_fixedpoint_hat}
\eeq
It is easy to check that the expressions of $\hu$ and $\hv$ given in (\ref{eq_uvhat}) are indeed solutions of these two equations, using the equations on $\tr$ and $\txr$ of Eq.~(\ref{eq_tandx_r}). By definition for $\lambda \in ]0,\lambda_{\rm r}]$ one has $u_* \ge \hu \ge \hv \ge v_*$, see Fig.~\ref{fig_k3l2_fixedpoints} for a representation of the solution of the equations (\ref{eq_fixedpoint_star},\ref{eq_fixedpoint_hat}) as a function of $\lambda$. In $\lambda_{\rm r}$, where one recovers the trivial solution studied in App.~\ref{sec_app_Tinfty_trivial}, one has $u_*=\hu=1/\tr$ and $v_*=\hv=\txr/\tr$.

Let us now deduce the value of $\Fsite$ and $\Fedge$ in the large $T$ limit from the above characterization of the behaviour of the $u_t$'s and $v_t$'s. From Eq.~(\ref{eq_Fedge}) one has in this limit
\beq
\lim_{T \to \infty}\Fedge(\lambda,T) = \frac{1}{u_*}\left[\hu^2 - 2 \int_0^1 \dd s \, U'(s) V(s) \right] \ ,
\eeq
the matching regimes of $t$ finite and $T-t$ finite having neglectible contributions to the summation. The integral above can be computed even if we have not determined the time-dependency of the scaling functions $U(s)$ and $V(s)$: using $\dd s \, U'(s) = \dd u$ and the condition $C(U(s),V(s))=1$, one has
\beq
-\int_0^1 \dd s \, U'(s) V(s) = \int_{\hu}^{u_*} \dd u \ v(u) = u_* v_* - \hu\, \hv + \int_{v_*}^{\hv} \dd v \ u(v) \ , 
\label{eq_intFedge}
\eeq
where $u(v)$ (resp. $v(u)$) is the solution of $C(u(v),v)=1$ (resp. $C(u,v(u))=1$). The equation $C(u(v),v)=1$ can be explicitly solved into
\beq
u(v) = v + \left(\lambda l \binom{k}{l} \right)^{-\frac{1}{k-l}} v^{- \frac{l-1}{k-l}} \ .
\label{eq_uofv}
\eeq
This allows to compute the integral in (\ref{eq_intFedge}) and to obtain (\ref{eq_Fedge_lmk}).

We shall now compute similarly the limit of $\Fsite$ that was defined in Eq.~(\ref{eq_Fsite}). In that equation we shall exploit the fact that $u_t-u_{t+1}$ is of order $1/T$ to perform the approximation 
\beq
(u_t-v_{t-2})^{k+1-p} = (u_{t-1}-v_{t-2})^{k+1-p}+(k+1-p) (u_t-u_{t-1})(u_{t-1}-v_{t-2})^{k-p} + O\left(\frac{1}{T^2} \right) \ .
\eeq
Within this approximation the first term leads to a telescopic summation, we then get
\beq
\Fsite \sim \frac{\lambda}{u_0} 
\sum_{p=l}^{k+1} \binom{k+1}{p} \left[ v_{T-1}^p (u_T - v_{T-1})^{k+1-p} -
(k+1-p)
\sum_{t=1}^T v_{t-2}^p  (u_t-u_{t-1})(u_{t-1}-v_{t-2})^{k-p}
\right]
\eeq
As $u_T=v_{T-1}+O(1/T)$ in the first summation only the term $p=k+1$ survives; the second term can be rearranged as above in terms of integrals of the scaling functions, namely
\bea
\lim_{T\to\infty} \Fsite(\lambda,T)&=&\frac{\lambda}{u_*}\left[
\hu^{k+1} - (k+1) \sum_{p=l}^k \binom{k}{p} 
\int_0^1 \dd s \ U'(s) V(s)^p (U(s)-V(s))^{k-p} \right] \\
&=&\frac{\lambda}{u_*}\left[
\hu^{k+1} + (k+1) \sum_{p=l}^k \binom{k}{p} \int_{\hu}^{u_*} \dd u \
v(u)^p (u-v(u))^{k-p}
\right] \\
&=&\frac{\lambda}{u_*}\left[
\hu^{k+1} + (k+1) \sum_{p=l}^k \binom{k}{p} \int_{v_*}^{\hv} \dd v \ (-u'(v))
v^p (u(v)-v)^{k-p}
\right]
\eea
Inserting the expression of $u(v)$ given in Eq.~(\ref{eq_uofv}) yields easily to the value of $\Fsite$ written in (\ref{eq_Fsite_lmk}). The parametric representations of $s(\theta)$ and $\Sigma_{\rm e}(\theta)$ given in Sec.~\ref{sec_lmk_Tinfty} are then direct consequences of Eqs.~(\ref{eq_RS_thermo},\ref{eq_1RSBy_thermo},\ref{eq_1RSBy_theta}).

For what concerns the distribution of activation times, one has in the regime $t=sT$ with $s\in]0,1[$ the following limit for the function $\Fsite$ defined in (\ref{eq_Fsite_t}):
\bea
\lim_{T\to\infty} \Fsite(\lambda,T,t=sT)=&&\frac{\lambda}{u_*} \left[ \sum_{p=l}^{k+1} \binom{k+1}{p} V(s)^p (U(s) - V(s))^{k+1-p} \right. \nonumber \\ && \hspace{1.5cm} \left.
- (k+1) \sum_{p=l}^k \binom{k}{p} 
\int_0^s \dd s' \ U'(s') V(s')^p (U(s')-V(s'))^{k-p}
\right] \ .
\eea
Studying the limit $s\to 0+$ and $s\to 1^-$ of this expression leads to the expressions (\ref{eq_Pslimits}) for the fraction of vertices which activate at the very beginning and at the very end of the process.

\bibliographystyle{h-physrev}
\bibliography{biblio}

\end{document}